\documentclass[11pt]{article}
\usepackage{pgfplots}
\usepackage{enumerate}
\usepackage[OT1]{fontenc}
\usepackage{amsmath, amssymb, mathtools}
\usepackage{dsfont}
\usepackage{pgfplots}
\usepackage{smile}
\usepackage{multirow}
\usepackage{rotating}
\usepackage{enumerate}
\usepackage{esvect}
\usepackage{pdflscape}
\usepackage{tikz}
\usetikzlibrary{tikzmark,decorations.pathreplacing,calc}
\usetikzlibrary{patterns}
\usetikzlibrary{arrows}
\usetikzlibrary{bayesnet}
\usepackage{authblk}
\usepackage{caption}
\usepackage{bbold}
\usepackage{setspace}
\usepackage[colorlinks,
            linkcolor=red,
            anchorcolor=blue,
            citecolor=blue
            ]{hyperref}
\usepackage{algorithm}
\usepackage{algorithmic}

\def\supp{\mathop{\text{supp}}}

\long\def\comment#1{}

\def\cS{{\mathcal{S}}}

\newcommand{\bel}{\begin{eqnarray}\label}
\newcommand{\eel}{\end{eqnarray}}
\newcommand{\bes}{\begin{eqnarray*}}
\newcommand{\ees}{\end{eqnarray*}}

\let\hat\widehat
\let\tilde\widetilde
\def\mid{\,|\,}

\def\E{{\mathbb E}}

\def\tca{{\textsc{T-CACE}}}

\def\supp{\mathop{\text{supp}\kern.2ex}}
\def\argmin{\mathop{\text{\rm arg\,min}}}

\def\given{{\,|\,}}

\def\supp{\mathop{\text{supp}}}

\def\prob{{\mathbb{P}}}

\theoremstyle{plain}

\usepackage{mathrsfs}
\usepackage{fullpage}

\usepackage{hyperref}
\usepackage[protrusion=false,expansion=true]{microtype}
\def\##1\#{\begin{align}#1\end{align}}
\def\$#1\${\begin{align*}#1\end{align*}}

\usepackage{undertilde}

\newtheorem{customassumption}{Assumption}

\theoremstyle{mytheoremstyle}

\def\nend{\nonumber\\}

\newcommand{\RNum}[1]{\uppercase\expandafter{\romannumeral #1\relax}}

\usepackage{subcaption}
\usetikzlibrary{shapes, decorations, arrows, calc, arrows.meta, fit, positioning}
\tikzset{
    -Latex, auto, node distance =1 cm and 1 cm,semithick,
    state/.style ={circle, draw, minimum width = 1 cm},
    missingstate/.style ={circle, draw, minimum width = 1 cm, fill=lightgray},
    hiddenstate/.style ={circle, draw, minimum width = 1 cm, fill=lightgray, text=black},
    point/.style = {circle, draw, inner sep=0.04cm,fill,node contents={}},
    bidirected/.style={Latex-Latex,dashed},
    el/.style = {inner sep=2pt, align=left, sloped},
    normal/.style={-stealth', line width=1},
    double/.style={stealth'-stealth', line width=1},
}
\DeclareUnicodeCharacter{FF0C}{,}

\title{\huge Generalizing causal effects with noncompliance: Application to deep canvassing experiments\footnote{The authors would like to thank Erin Hartman, Emily Flanagan, and participants at the American Causal Inference Conference for helpful feedback.}}
\author[1]{Zhongren Chen\thanks{Email: zhongren.chen@yale.edu}}

\author[2,1]{Melody Huang\thanks{Email: melody.huang@yale.edu, Website: \texttt{www.melodyyhuang.com}}}

\affil[1]{
\small
\textit{Department of Statistics and Data Science, Yale University}}
\affil[2]{
\small
\textit{Department of Political Science, Yale University}}
\date{}
\pgfplotsset{compat=1.17}

\begin{document}
\maketitle
\vspace{-25pt}

\begin{abstract}
Standard approaches in generalizability often focus on generalizing the intent-to-treat (ITT). However, in practice, a more policy-relevant quantity is the generalized impact of an intervention across compliers. While instrumental variable (IV) methods are commonly used to estimate the complier average causal effect (CACE) within samples, standard approaches cannot be applied to a target population with a different distribution from the study sample. This paper makes several key contributions. First, we introduce a new set of identifying assumptions in the form of a population-level exclusion restriction that allows for identification of the target complier average causal effect (T-CACE) in both randomized experiments and observational studies. This allows researchers to identify the T-CACE without relying on standard principal ignorability assumptions. Second, we propose a class of inverse-weighted estimators for the T-CACE and derive their asymptotic properties. We provide extensions for settings in which researchers have access to auxiliary compliance information across the target population. Finally, we introduce a sensitivity analysis for researchers to evaluate the robustness of the estimators in the presence of unmeasured confounding and extend existing tests to evaluate instrument validity in this context. We illustrate our proposed method through extensive simulations and a study evaluating the impact of deep canvassing on reducing exclusionary attitudes.
\end{abstract}
\clearpage 

\doublespacing
\section{Introduction} \label{sec: introduction}

Recent work in political science has clarified and formalized the challenges in generalizing the treatment effect from an experimental sample to a target population \citep[e.g.,][]{egami2023elements, findley2021external}. However, many existing statistical approaches focus explicitly on generalizing the intent-to-treat effect, ignoring the presence of non-compliance in both the experiment and the target population \citep[e.g.,][]{hotz1999predicting, hartman2015sample, egami2023elements,  huang2023leveraging, cole2010generalizing, kern2016assessing, zhang2024minimax}. In practice, political scientists are often interested in understanding the impact of interventions across individuals who actually comply with the treatment. As such, a more policy-relevant interest is the \textit{target complier average causal effect} (T-CACE), which considers the generalized impact of a treatment, across the subset of individuals who would comply. Unfortunately, in the presence of non-compliance, standard assumptions used to generalize treatment effects are no longer sufficient to identify the T-CACE \citep{dahabreh2022generalizing}. 

In this paper, we introduce a framework for political scientists to generalize the \textit{complier average causal effect} (CACE), also commonly referred to as the local average treatment effect (LATE) \citep{imbens1994identification}, from a study sample to a target population. The problem presents a unique challenge, as non-compliers are generally unidentifiable in both the study sample and the target population. Furthermore, individuals who are likely to select into an experimental study may also be more likely to comply, resulting in differential rates of compliance between the experiment and the target population. We propose an instrumental variable (IV) approach \citep{angrist1996identification} to address the problem of generalizing non-compliance. We invoke a population-level exclusion restriction, which allows us to identify the T-CACE as the ratio of the conditional expectation of the difference in outcomes to the conditional expectation of the difference in treatment received, both evaluated in the target population. Our identification strategy is distinct from a concurrent study by \citet{clark2024transportability}, who introduce principal ignorability assumptions to identify the T-CACE.\footnote{This assumption posits that the conditional mean of potential outcomes is consistent across different compliance types, which often fails to hold in practice.} 

From our identification result, we propose a family of weighted estimators that enable consistent estimation of the T-CACE, along with a derivation of their asymptotic properties. We also propose a multiply robust estimator that allows researchers to leverage outcome models in the estimation stage. Notably, our identification strategy and estimators are applicable to both randomized controlled trials (RCTs) and observational studies.

We then provide several extensions of the framework to address practical considerations that arise in political science. First, we provide ways for researchers to integrate partial, or proxy, information about compliance across the target population into the estimation procedure. Second, we introduce an optimization-based sensitivity analysis framework to evaluate the robustness of the estimators in the presence of unmeasured confounding \citep[e.g.,][]{rosenbaum1987sensitivity, aronow2013interval, zhao2019sensitivity, huang2024sensitivity, huang2025relative}. Finally, we extend existing IV tests for researchers to evaluate the validity of the instrument used in this generalizability context.

The paper is organized as follows. Section \ref{sec: background} introduces the notational framework and formally defines the T-CACE. Section \ref{sec: complete randomization} examines the assumptions required to derive the identification results. Section \ref{sec: causal estimation} proposes the weighted estimator, the weighted least squares estimator, and the multiply robust estimator, along with their asymptotic statistical properties. Section \ref{sec: practical considerations} discusses practical considerations on how to incorporate additional compliance information and evaluate potential violations of key assumptions. Section \ref{sec: simulation study} evaluates the performance of the proposed estimators in a simulation study. Section \ref{sec: application exclusionary attitudes} applies the proposed methods to a dataset from a canvassing study that employs techniques designed to reduce exclusionary attitudes. Section \ref{sec: conclusion} concludes the paper.

\subsection{Motivating Example: Deep canvassing to reduce exclusionary attitudes}
To illustrate the proposed methodological framework, we introduce a motivating example on generalizing the impact of deep canvassing on reducing exclusionary attitudes, as originally studied in \citet{kalla2020reducing}. Deep canvassing is used to persuade individuals regarding their underlying political beliefs. Deep canvassing employs non-judgmental narrative exchange and interpersonal conversation to persuade others by sharing personal experiences and listening without judgment \citep[e.g,][]{kalla2020reducing, chen2025framework, offer2026deep}.

We focus on the randomized experiment from \citet{kalla2020reducing}, where the authors study whether deep canvassing can effectively reduce prejudice against unauthorized immigrants in a series of field experiments. Initially, 217,600 registered voters were recruited by mail. Of these, 7,870 responded and were randomly assigned to one of three groups: 2,624 to persuasion with narrative exchange, and 2,623 to persuasion without narrative exchange, and 2,623 to a control group.\footnote{For simplicity, we focus on participants assigned to the narrative exchange group and the control group, excluding those in the persuasion-without-narrative group.} Voters were randomly assigned at the household level, ensuring that voters who completed the pre-survey within the same household were always assigned to the same treatment. Assignment was conducted within matched blocks of households. Following \citet{kalla2020reducing}, we define compliance as the setting in which canvassers indicated that they engaged in an effective conversation with the participants. 

One week after the conversations between the canvassers and the participants, 1079 participants were reached to complete a follow-up survey, corresponding to an attrition rate of \(79.4\%\). From this survey, an overall index measuring support for immigration-related policies and unauthorized immigrants was constructed and used as the experiment’s outcome variable. \citet{kalla2020reducing} reported a within-sample intent-to-treat effect of 8.8$\%$ increase in support for unauthorized immigrants and a complier average causal effect of 13.2$\%$ increase, indicating that deep canvassing effectively reduced prejudice amongst individuals who selected into the experiment and successfully completed the follow-up survey. A policy-relevant question is whether we would expect the impact of deep canvassing to be just as effective across compliers who did not complete the follow-up survey or select into the experiment. In what follows, we propose a suite of methods to address this question. 

\section{Background} \label{sec: background}
We define a study sample with $n$ units, where each unit is drawn i.i.d. from an infinite super-population.  Let $Z_i \in \{0, 1\}$ denote the assigned treatment to each unit, where $Z_i = 1$ implies unit $i$ is assigned to treatment and $Z_i = 0$ for control. Let $D_i$ be an indicator of treatment \textit{received}, where $D_i=1$ if the unit $i$ received treatment and $D_i=0$ if the unit $i$ did not receive the treatment. We can similarly write the observed treatment received indicator as $D_i := D_i(1) Z_i + D_i(0) (1-Z_i)$, where $D_i(1)$ and $D_i(0)$ represent the potential treatment received under treatment and control, respectively. Define the potential outcomes under treatment and control as $Y_i(1), Y_i(0)$, respectively. Finally, we define an indicator $C_i \in\{0,1\}$ for whether an individual complies with treatment assignment (i.e., $D_i(1) = 1$, $D_i(0) = 0$).

We define a \textit{target population} of interest, with $N$ units, where each unit is drawn i.i.d. at random from an infinite super-population. Let $S_i$ be the indicator such that $S_i = 1$ indicates the unit $i$ is in the study sample, while $S_i = 0$ indicates that the unit $i$ is in the target population. We assume that for all units in the study sample and the target population, we have observed a set of pre-treatment covariates $X_i \in \cX.$ However, across the target population, we do not have access to treatment assignment, treatment receipt, or outcome information. 

Throughout the paper, we will denote $p$ the density of continuous random variables and $\prob$ the probability of an event or probability mass function of a discrete random variable. We denote $\cS$ the index set of the experiment units: $\cS := \{i: S_i = 1\}$ and $\cT$ as the index set of the target units: $\cT := \{i: S_i = 0\}.$

To begin, we assume that within the study sample, treatment assignment is conditionally ignorable.
\begin{customassumption}[Treatment Ignorability] \label{asp: randomization}
$$Z \indep \sbr{Y(1), Y(0), D(1), D(0)} \given X, S=1.$$

\end{customassumption}
\noindent Assumption \ref{asp: randomization} will hold by construction in settings when researchers conduct a randomized experiment. In observational settings, Assumption \ref{asp: randomization} assumes that, given a set of pre-treatment covariates $X$, the treatment assignment and the potential outcomes are conditionally ignorable.

Under Assumption \ref{asp: randomization}, we can identify the \textit{intent-to-treat effect} (ITT) within the study: 
$$\tau_{\textsc{S-ITT}} = \E\sbr{Y(1) - Y(0) \mid S = 1}.$$
Existing literature in external validity has largely focused on generalizing the intent-to-treat effect \citep[e.g.,][]{buchanan2018generalizing, cole2010generalizing, hartman2015sample, egami2023elements,huang2023leveraging, ross2026transporting}, where the target estimand is the ITT across the target population:
$$\tau_{\textsc{T-ITT}} = \E\sbr{Y(1) - Y(0) \mid S = 0}.$$

Identifying the intent-to-treat effect across the target population often relies on an assumption of mean exchangeability of selection and treatment effect heterogeneity.
\begin{customassumption}[Mean Exchangeability of Selection and Treatment Effect Heterogeneity] \label{asp: cond_ign_selection}
$$\E\sbr{Y(1) - Y(0)  \mid X, S = 1} = \E\sbr{Y(1) - Y(0)  \mid X, S = 0}.$$    
\end{customassumption}
\noindent Existing methods include reweighting the data in the study sample to balance the distribution of the target population \citep[e.g.,][]{cole2010generalizing, buchanan2018generalizing}, or modeling the treatment effect heterogeneity and projecting to the target population \citep[e.g.,][]{kern2016assessing}.

In settings where there is perfect compliance (i.e., everyone who is encouraged to take the treatment receives the treatment), the generalized ITT effect will be equivalent to the average treatment effect. However, in settings when there is non-compliance, a more relevant estimand of interest is the average treatment effect across compliers---i.e., individuals who would receive the treatment if encouraged: 

$$\tau_{\textsc{T-CACE}} = \E \big[ Y(1) - Y(0) \mid C = 1, S =0 \big].$$

Much of the existing literature has focused on identifying and estimating the \textit{within-study} complier average causal effect \citep[e.g.,][]{angrist1996identification, sovey2011instrumental, jo2002estimation, aronow2013sharp, jo2009use, feller2017principal, ding2017principal}. In particular, \citet{angrist2010extrapolate, aronow2013beyond, mogstad2018using, gulotty2025must} examine how the IV estimate can be generalized to other subgroups or causal estimands \textit{within} the study sample. However, limited work exists on generalizing the complier average causal effect to a target population. In addition to having to account for selection bias, researchers must additionally account for the fact that the selection bias from entering a particular study is likely confounded with the propensity for compliance. It is this particular setting that we focus on in the paper.

Most closely related to our proposed framework are \citet{rudolph2017robust} and \citet{clark2024transportability}. \citet{rudolph2017robust} propose targeted maximum likelihood estimators (TMLEs) for transported encouragement-design effects, including the T-CACE. Their setting differs from ours in the observed data structure and identification problem. In their setup, treatment assignment and treatment received are observed in the target population, while the outcome is missing. In contrast, we assume that researchers only have access to pre-treatment covariates across the target population. \citet{clark2024transportability} adopt a different set of identification assumptions (i.e., principal ignorability), which we compare in detail through simulation studies in \S F.2.

Our framework is also related to the economics literature on marginal treatment effects (MTEs) and policy-relevant treatment effects (PRTEs) \citep{heckman2001policy, heckman2005structural, mogstad2018identification, mogstad2024instrumental, shea2023ivmte, blandhol2022tsls}. This literature represents CACE and other policy-relevant treatment effects as weighted averages of marginal treatment response (MTR) functions, allowing researchers to extrapolate away from the compliers and therefore examine the robustness of CACE by expanding or contracting the complier subpopulations. Our goal is different. We study a two-population generalizability problem in which extrapolating from study-sample compliers to other latent complier groups within the same population would not address the central challenge: the target population may differ from the study sample in both its covariate distribution and compliance process, and may contain only pre-treatment covariates.

\section{Identifying the Target Complier Average Causal Effect} \label{sec: complete randomization}
This section outlines the necessary conditions for deriving the identification results for $\tau_\tca.$ Our identification result builds on existing instrumental variable approaches. Informally, we leverage the randomization of treatment assignment \textit{within} the study as an instrument for treatment received. Within the context of an experimental setting, exogeneity of treatment assignment can be directly controlled by design. We then introduce a new assumption, which allows researchers to generalize the information about the instrument from the study to the target population.

We invoke standard assumptions from instrumental variables.

\begin{customassumption}[Monotonicity and Valid Instrument] \label{asp: IV} Assume monotonicity and that the treatment encouragement $Z$ is a valid instrument. More formally: 
\begin{enumerate} 
\item[(a)] Monotonicity (i.e., no defiers)
$$\prob\rbr{D(1) < D(0)} = 0$$
\item[(b)] Exclusion Restriction:  $Y(z,D(z)) = Y(z',D(z'))$ for all $z,z'$ such that $D(z) = D(z')$

\item[(c)] Instrument Relevance 
    $$\EE\sbr{D(1) - D(0) \mid S = 0} \neq 0.$$
\end{enumerate} 
\end{customassumption}

Assumption \ref{asp: IV}-(a) effectively rules out the existence of \textit{defiers} (i.e., individuals who receive the treatment when assigned to control, and refuse it when assigned to treatment) in both the study sample and the target population. In the context of the canvassing via persuasion example, this implies that individuals who are never visited by a canvasser would not engage in a conversation with one. Following \citet{imbens1994identification}, Assumption \ref{asp: IV}-(a) should be read as a sign-uniformity restriction on the choice response to the instrument. This condition is needed so that the target Wald ratio in Theorem \ref{thm: randomization identification tcace} is interpreted as a T-CACE rather than a net contrast of compliers and defiers. Assumption \ref{asp: IV}-(b) states that the instrument $Z$ can only affect the outcome through $D$---in other words, the outcome only depends on whether treatment is received, not treatment assignment. This allows us to simplify the potential outcomes to be a function of just $D$. Assumption \ref{asp: IV}-(c) says that the expected difference in potential treatment received is nonzero.

In the context of persuasion through canvassing, Assumption \ref{asp: IV}-(b) implies that being assigned to a persuasion influences the attitude of a voter only through the act of persuasion itself, without any other direct effect. Assumption \ref{asp: IV}-(c) implies that there is a nonzero probability that an individual in the target population whose receipt of the persuasion intervention depends on whether they were assigned to it. We note that Assumption \ref{asp: IV} (a)-(b) must hold at the population level, meaning they apply to both the study sample and target population. In Section \ref{sec: practical considerations}, we discuss guidelines for sharp tests of Assumption~\ref{asp: IV} \citep{kitagawa2015test, mourifie2017testing, yu2025binary}. Following \citet{deaton2010instruments,swanson2014think,swanson2018partial}, we view these diagnostic procedures as falsification, rather than as mechanical validation of the identifying assumptions.

\begin{customassumption} [Mean Exchangeability of the First Stage] \label{asp: randomized treatment received exchangeability}
 \begin{align*}
     \EE_{X \given S=0} \cbr{\EE\sbr{D(1) - D(0) \given S=0, X}} = \EE_{X \given S=0} \cbr{\EE\sbr{D(1) - D(0) \given S=1, X}}.
 \end{align*}
\end{customassumption}
\noindent Assumption \ref{asp: randomized treatment received exchangeability} implies that conditioned on pre-treatment covariates, the average difference in treatment received is the same across the study sample and the target population. 

Assumptions \ref{asp: cond_ign_selection} and \ref{asp: randomized treatment received exchangeability} together imply \textit{the conditional exchangeability of compliance between the study sample and the target population}. This means that conditioned on a set of covariates $X$, the latent compliance patterns across both the study sample and the target population will be identical. In the context of the motivating example, this assumes individuals with the same demographic information and pre-intervention support for immigration are willing to engage in conversation with a canvasser with the same probability, regardless of if they are in the study sample or the target population. Crucially, the mean exchangeability assumption assumes away a fixed effect on the probability of compliance from selecting into the study sample. 

These mean exchangeability assumptions are most plausible when researchers have access to a rich set of covariates that can plausibly account for potential confounding from selecting into the sample. Unlike standard generalizability contexts, the pre-treatment covariates $X$ not only must capture potential treatment effect heterogeneity, but also account for potential heterogeneity in treatment \textit{receipt}. In other words, the mean exchangeability assumptions are most credible in settings when the covariates are prognostic of \textit{both} treatment effect heterogeneity, as well as variation in treatment receipt. To help researchers evaluate the sensitivity of their estimates to potential violations in the exchangeability assumptions, we propose a sensitivity analysis in Section 5. 

Alternative approaches have used principal stratification to identify the T-CACE \citep[e.g.,][]{clark2024transportability, ottoboni2020estimating}. However, these approaches require researchers to model the compliance patterns. In contrast, our approach leverages the exogeneity from a randomized instrument within the study to account for the compliance patterns, which is more feasible in many contexts. For example, in the deep canvassing study, principal stratification requires that, conditional on pre-treatment covariates, compliers and non-compliers exhibit the same treatment effect. However, non-compliers---those who refuse to engage in a full conversation with the canvassers---may also be more resistant to persuasion, suggesting a violation of this assumption. Given that the pre-treatment covariates capture only limited demographic information, it is unrealistic to expect the assumptions underlying principal stratification to hold. In contrast, because the experimenters have full control over the RCT, the treatment assignment is likely to serve as a valid IV.

With Assumptions \ref{asp: randomization}-\ref{asp: randomized treatment received exchangeability}, we can directly identify $\tau_\tca$.

\begin{theorem}[Causal Identification for T-CACE]\label{thm: randomization identification tcace}
Let $\mu_{yz}(x) := \EE\sbr{Y \given Z = z, S=1, X=x}$ and $\mu_{dz}(x) := \EE\sbr{D \given Z = z, S=1, X=x}.$
Under Assumptions \ref{asp: randomization}-\ref{asp: randomized treatment received exchangeability}, the T-CACE can be identified as
\begin{align}
    \tau_{\tca} &= \EE \sbr{Y(1) - Y(0) \mid C = 1, S =0} \nonumber \\
    &=\frac{\EE\sbr{Y (1) - Y(0) \mid S = 0}}{\EE\sbr{D(1) - D(0) \mid S = 0}} \label{eqn:wald} \\
    &= \frac{\EE_{X \given S=0} \sbr{\mu_{y1}(x) - \mu_{y0}(x)}}{\EE_{X \given S=0} \sbr{\mu_{d1}(x) - \mu_{d0}(x)}}. \label{eqn:id_result}
\end{align}
\end{theorem}

\eqref{eqn:wald} arises from leveraging the validity of the instrument across the target population (i.e., Assumption \ref{asp: IV}). This allows us to rewrite $\tau_\tca$ as a Wald ratio across the target population. \eqref{eqn:id_result} arises from mean exchangeability of treatment effect heterogeneity (i.e., Assumption \ref{asp: cond_ign_selection}) and mean exchangeability of the first stage (i.e., Assumption \ref{asp: randomized treatment received exchangeability}). The identification result can be interpreted as a weighted version of the traditional instrumental variables (IV) estimator, adjusting for distributional differences in $X$ across the study sample and the target population.

The identification result in Theorem \ref{thm: randomization identification tcace} highlights that researchers should, when possible, collect covariates that are prognostic of \textit{both} selection into the study and compliance. We consider a reweighting approach that adjusts for the differences in covariate distributions between the two populations. Specifically, we reweight the observed data in the study sample to match the covariate distribution of the target population. This enables consistent estimation of the T-CACE, even when the covariate distributions differ substantially. In the next section, we formalize this approach and introduce a family of estimators of T-CACE.
\section{Estimating the T-CACE} \label{sec: causal estimation}
In the following section, we leverage the identification results from Theorem \ref{thm: randomization identification tcace} to introduce a class of estimators to estimate the T-CACE. We propose three estimators: (1) a weighting-based estimator, which allows researchers to adjust for the distributional differences across the study sample and target population; (2) a weighted least squares estimator, which allows researchers to offset potential efficiency loss from weighting by using an agnostic outcome model to explain variation within the study sample; and (3) a multiply robust estimator, which allows researchers to augment a traditional weighting estimator with a treatment effect heterogeneity model. We demonstrate that all proposed estimators are consistent for T-CACE and derive their asymptotic distributions. 

For all of the estimation approaches, we require overlap in both the treatment assignment as well as the sample selection process.
\begin{customassumption}[Overlap] \label{asp: cre overlap} The following relationships hold almost surely:
    \begin{align*}
    0 < \prob\rbr{S = 1 \given X} \text{ and } 0 < \mathbb{P}\rbr{Z=1 \given S=1,X} < 1.
\end{align*}
\end{customassumption}
Assumption \ref{asp: cre overlap} is a technical condition necessary for the validity of most IPW methods \citep{rosenbaum1983assessing} and ensures that each unit has a non-zero probability of being selected into the study sample and assigned to the treatment group within that population. Recent research has proposed various methods to either assess robustness to violations of overlap or mitigate such violations. For details, see \citet{huang2024overlap, li2019addressing, crump2009dealing}. This Assumption is most likely to be violated when the experimental sample is drawn after a screening process of the target population or there is a contextual shift between the experimental sample and the target population (geographical regions, institutional differences, etc) \citep{huang2024overlap}. Consider, for example, a cash transfer experiment (e.g., the Youth Opportunities Program in Northern Uganda, as studied in \citealp{blattman2016returns, egami2021covariate, huang2024overlap}). The cash-transfer treatment is randomized only among groups that submitted applications and passed screening. Uninformed unemployed adults outside that application process would therefore have zero probability of inclusion in the experimental sample \citep{huang2024overlap}. This concern is less direct in our canvassing application because participants were not screened through any application process. 

\subsection{Weighted Estimator} \label{subsec:weighted}
The identification result in Theorem \ref{thm: randomization identification tcace} expresses the T-CACE as a ratio of two terms: (1) the generalized intent-to-treat effect, and (2) the generalized first stage. In this subsection, we construct a weighted estimator for $\tau_{\tca}$ by deriving separate weighted estimators for both the numerator and the denominator. We show that the weighted estimator is a consistent estimator of the T-CACE and derive its asymptotic distribution.

We begin by defining the following weights: 
\begin{equation}
    w_z(X) := \frac{\prob(S = 0 \given X)}{\prob(S = 1 \given X)} \cdot \frac{1}{\mathbb{P}(Z = z \given S = 1, X)},
    \label{eqn:weights}
\end{equation}
where $\mathbb{P}(S=0|X)/{\mathbb{P}(S=1|X)}$ accounts for distribution shifts in the observed covariates between the study sample and the target population, and $\mathbb{P}(Z=z|S=1, X)$ accounts for imbalance in the treatment and control groups. In the context of a RCT, the probability of treatment is known by the researcher. The weights in \eqref{eqn:weights} are equivalent to the standard generalization weights introduced in \citet{cole2010generalizing}. A common approach to estimating the weights is to estimate a propensity score model for the underlying sample selection process. Alternatively, researchers can employ balancing approaches, which directly target the distributional differences in the underlying covariates without directly fitting a parametric model \citep[e.g.,][]{hainmueller2012entropy, imai2014covariate}. 

We define the re-weighted intent-to-treat effect as: 
\begin{align} \label{eq: randomization Hajek outcome}
    \hat{\tau}_{w}^{Y} =  \frac{\sum_{i:S_i=1} \hat{w}_1(X_i) Z_i Y_i} {\sum_{i:S_i=1} \hat{w}_1(X_i) Z_i} - \frac{\sum_{i:S_i=1} \hat{w}_0(X_i) (1-Z_i)Y_i} {\sum_{i:S_i=1} \hat{w}_0(X_i) (1-Z_i)},
\end{align}
where $\hat w_1(X_i)$ and $\hat w_0(X_i)$ are estimates of the true generalization weights in \eqref{eqn:weights}. This is the standard weighted estimator used in the generalizability literature, which ignores non-compliance.

We also define the weighted estimator for the treatment received:

\begin{align} \label{eq: randomization Hajek treatment received}
    \hat{\tau}_{w}^{D} =  \frac{\sum_{i:S_i=1} \hat{w}_1(X_i) Z_i D_i} {\sum_{i:S_i=1} \hat{w}_1(X_i) Z_i} - \frac{ \sum_{i:S_i=1} \hat{w}_0(X_i) (1-Z_i)D_i} {\sum_{i:S_i=1} \hat{w}_0(X_i) (1-Z_i)}.
\end{align}
$\hat{\tau}_{w}^{D}$ is analogous to $\hat{\tau}_{w}^{Y}$ in that both share the same weights to account for the distribution shift. However, $\hat{\tau}_{w}^{D}$ pertains to treatment received rather than the outcome. Notably, $\hat{\tau}_{w}^{D}$ effectively estimates the denominator of the identification result of Theorem \ref{thm: randomization identification tcace}.

The weighted estimator of $\tau_\tca$ is given by the ratio of these two quantities: 
\begin{align} \label{eq: Hajek estimator}
    \hat{\tau}_{w} = \frac{\hat{\tau}_{w}^{Y}}{\hat{\tau}_{w}^{D}}.
\end{align}

If the estimated weights $\hat w_1(X)$ and $\hat w_0(X)$ are correctly specified, then the weighted estimator is a consistent estimator for $\tau_\tca$.

\begin{theorem}[Consistency of the Weighted Estimator] \label{thm: randomization Hajek Estimator consistency}
Assume that $\mathbb{P}(S = 1) = \frac{n}{n+N}$ is fixed, so that the ratio between $n$ and $N$ remains constant as $n \to \infty$ and $N \to \infty$. Assume $\sup_{x \in \cX} \abr{\hat{w}_1(x) - w_1(x)} = o_p(1)$ and $\sup_{x \in \cX} \abr{\hat{w}_0(x) - w_0(x)} = o_p(1).$ Then, under Assumptions \ref{asp: randomization}-\ref{asp: randomized treatment received exchangeability}:

\begin{align*}
    \hat{\tau}_{w} \overset{p}{\rightarrow} \tau_\tca.
\end{align*}
    \begin{proof}
        See \S\ref{pro: randomization Hajek Estimator consistency} for a detailed proof.
    \end{proof}
\end{theorem}
The fixed ratio condition is used only as an asymptotic device and ensures that as the study sample grows towards infinity, the corresponding target population size also grows towards infinity. In settings when researchers are interested in finite-population settings, we refer readers to \citet{li2017general}. To characterize the asymptotic distribution of the weighted estimator, we begin by first assuming that the selection model and the treatment assignment model are known. Then, the weighted estimator will be asymptotically normally distributed. 

\begin{theorem}[Asymptotic Distribution of the Weighted Estimator]\label{thm: cre known model asymptotics Hajek}
Assume the selection model (i.e., $\mathbb{P}(S = 1 \mid X)$) and the treatment assignment model (i.e., $\mathbb{P}(Z = 1 \mid S=1, X)$) are both known. Let $g(\cdot):\mathbb{R}^6\rightarrow \mathbb{R}$ be defined as $g(\theta) = (\frac{\theta_1}{\theta_3} - \frac{\theta_2}{\theta_4})/(\frac{\theta_5}{\theta_3} - \frac{\theta_6}{\theta_4}).$ Denote $\nabla g$ as its gradient vector. Then, under the same assumptions as Theorem \ref{thm: randomization Hajek Estimator consistency} and the regularity conditions $\EE \sbr{w_z(X)^2 Y^2} < \infty,$

\begin{align*}
    \sqrt{n + N}(\hat{\tau}_{w} - \tau_\tca) \overset{d}{\rightarrow} N(0, \nabla g ^T \Sigma_{\theta^*} \nabla g), 
\end{align*}
    where we define the covariance matrix $\Sigma_{\theta^*}$ in \eqref{eq: covariance matrix known weights} of \S\ref{pro: randomization Hajek Estimator asymptotic}.
    \begin{proof}
        See \S\ref{pro: randomization Hajek Estimator asymptotic} for a detailed proof.
    \end{proof}
\end{theorem}
The result of Theorem \ref{thm: cre known model asymptotics Hajek} follows from establishing the asymptotic normality of each component of $\hat{\tau}_{w}$ by leveraging the theory of estimating equations. In particular, we write $\hat{\tau}_{w}$ as the function $g$ of an M-estimator \citep{huber1992robust} and then apply the Delta method to $g$. The covariance matrix, $\Sigma_{\theta^*},$ though unknown, can be consistently estimated using its empirical counterpart (see \S\ref{app: cre confidence intervals} for details). 

In practice, researchers do not know the true probability of selecting into the study sample. 
When the selection mechanism is unknown, we make an assumption of the model specification. In Assumption \ref{asp: cre model specification} and Theorem \ref{thm: cre logistic model asymptotics Hajek} it follows, we assume the covariate matrix $X$ includes a vector of ones as its first column.

\begin{customassumption}[Model Specification for the Selection Mechanism] \label{asp: cre model specification}
We denote the logistic function $\sigma(\beta^T X) := \rbr{1 + \exp(-\beta^T X)}^{-1}.$ We assume that the study sampling process follows a logistic regression model: $\prob(S = 1 \given X) = \sigma(\beta'^T X)$ for some $\beta' \in \mathbb{R}^{\text{dim}(X)}.$ In addition, we assume that $\prob(Z = 1 \given S = 1, X)$ is known. 
\end{customassumption}

Assumption \ref{asp: cre model specification} assumes that the study sample selection process follows a logistic model. Logistic regression is arguably the most popular method for modeling sample selection \citep{kern2016assessing, buchanan2018generalizing}. The condition that $\prob(Z = 1 \given S = 1, X)$ is known holds for all RCTs. Theorem \ref{thm: cre logistic model asymptotics Hajek} below gives the $\sqrt{n + N}$-asymptotic distribution of $\hat{\tau}_{w}.$
\begin{theorem}[Asymptotic Distribution of the Weighted Estimator for Unknown Selection Mechanism] \label{thm: cre logistic model asymptotics Hajek}
Suppose that all the assumptions in Theorem \ref{thm: randomization Hajek Estimator consistency} hold. Suppose that model specification for the selection mechanism (i.e., Assumption \ref{asp: cre model specification}) holds. If we set $\hat{\prob}(S = 1 \given X) = \sigma(\hat{\beta}^T X)$ in $\hat{\tau}_{w},$ where $\hat{\beta}$ is the maximum likelihood estimator of $\beta,$ then under the regularity conditions $\EE \sbr{||X||_2 ^2} < \infty,$ $\EE \sbr{w_z(X)^2 Y^2} < \infty,$ and $\EE\sbr{XX^T}$ being full rank,
\begin{align*}
    \sqrt{n + N}\rbr{\hat{\tau}_{w} - \tau_\tca} \overset{d}{\rightarrow} N(0, \nabla g ^T \Sigma_{\theta^*, \beta^*} \nabla g), 
\end{align*}
    where we define the covariance matrix $\Sigma_{\theta^*, \beta^*}$ in \eqref{eq: weighted covariance matrix unknown weights} of \S\ref{pro: randomization Hajek Estimator asymptotic unknown}.
\end{theorem}

\begin{proof}
    See \S\ref{pro: randomization Hajek Estimator asymptotic unknown} for a detailed proof.
\end{proof}
Theorem \ref{thm: cre logistic model asymptotics Hajek} derives the asymptotic variance of $\hat{\tau}_{w}$ when $\hat{\prob}(S = 1 \given X)$ follows a logistic regression. The asymptotic variance of $\hat{\tau}_{w}$, given by $\nabla g ^T \Sigma_{\theta^*, \beta^*} \nabla g,$ differs from the asymptotic variance $\nabla g ^T \Sigma_{\theta^*} \nabla g$ in Theorem \ref{thm: cre known model asymptotics Hajek}. This difference arises because, when the study sample selection mechanism is unknown, we must account for the additional variance introduced by estimating the logistic regression model. In practice, we can construct Wald-type confidence intervals for $\tau_\tca$ by replacing the covariance matrices in Theorem \ref{thm: cre known model asymptotics Hajek} and Theorem \ref{thm: cre logistic model asymptotics Hajek} with their empirical counterparts. The details of the construction of the empirical sandwich-type variance estimator and the Wald-type confidence intervals are given in \S\ref{app: cre confidence intervals}. As a result, Theorem \ref{thm: cre logistic model asymptotics Hajek} allows us to perform inference for $\hat \tau_{w}$, while accounting for the variance in estimating the selection probabilities.

\subsection{Weighted Least Squares} \label{subsec: wls estimation}
While Section \ref{subsec:weighted} introduces an estimator that allows us to consistently recover the T-CACE, a well-known drawback to weighted estimators is the potential efficiency loss that occurs from reweighting \citep{miratrix2018worth, huang2023leveraging}. This concern is exacerbated in the instrumental variables setting, where the combination of reweighting, as well as a potentially weak instrument, can result in large amounts of variance inflation \citep{hartman2023improving}. To combat some of the losses in efficiency, we propose a weighted least squares approach to estimating the T-CACE.

More formally, we define the covariate-adjusted weighted least squares estimator as:

\begin{align} \label{eq: wls}
    \hat{\tau}_{\text{wls}} := \frac{\hat{\tau}^{Y}_{\text{wls}}}{\hat{\tau}^{D}_{\text{wls}}},
\end{align}
where $\hat{\tau}^{Y}_{\text{wls}}$ and $\hat{\tau}^{D}_{\text{wls}}$ are given by
\begin{align} \label{eq: wls objective}
    (\hat{\tau}_{\text{wls}}^{Y}, \hat{\gamma}^Y) := \argmin_{\tau, \gamma}\sum_{i:S_i=1}\hat{w}_i(X_i) \left\{ Y_i - (\tau Z_i + \gamma^T X_i) \right\}^2, \nend
    (\hat{\tau}_{\text{wls}}^{D}, \hat{\gamma}^D) := \argmin_{\tau, \gamma}\sum_{i:S_i=1}\hat{w}_i(X_i) \left\{ D_i - (\tau Z_i + \gamma^T X_i)\right\}^2.
\end{align}
Intuitively, the weighted least squares estimator reweights the residual variance of the outcome and treatment received, after controlling for covariates $X$. If the pre-treatment covariates $X$ are predictive of $Y$ and $D,$ incorporating $X$ into the model will reduce the weighted sample residual variance, leading to a more precise estimate. Furthermore, while the estimated weights depend on covariates that are measured across both the study sample and the target population, the regression uses only observations across the study sample. Using the weighted least squares approach allows researchers to leverage additional covariates, measured across just the study sample, to help improve efficiency.

We can view the standard weighted estimator $\hat{\tau}_{w}$ as a special case of $\hat{\tau}_{\text{wls}},$ in which no covariates $X$ are included except for the intercept. We show in Theorem~\ref{thm: cre logistic model asymptotics distribution wls} that, under the same conditions as the weighted estimator, $\hat{\tau}_{\text{wls}}$ will be a consistent estimator and will be asymptotically normally distributed.

\subsection{Multiply Robust Estimator}
We now propose a multiply robust estimator that allows researchers to simultaneously model the sample selection process, as well as both the treatment effect heterogeneity and the treatment received. We demonstrate that the proposed multiply robust estimators remain consistent for T-CACE as long as either the sample selection model is correctly specified or both the outcome and treatment received models are correctly specified \citep{robins1994estimation}.

To construct the multiply robust estimator, we augment both $\hat{\tau}_{w}^{Y}$ and $\hat{\tau}_{w}^{D}$ with outcome-based estimators for both $Y$ and $D$. As such, the multiply robust estimator will be defined as the ratio of the two augmented models. More specifically, the multiply robust estimator for T-CACE is defined as:
\begin{align*}
    \hat{\tau}_{\text{mr}}:= \frac{\hat{\tau}_{\text{mr}}^{Y}}{\hat{\tau}_{\text{mr}}^{D}},
\end{align*}
where the numerator is an augmented weighted estimator for the ITT, and is defined as: 
\begin{align*}
    \hat{\tau}_{\text{mr}}^{Y}:= \underbrace{\frac{\sum_{i:S_i=1} \hat{w}_1(X_i) Z_i(Y_i - \hat{\mu}_{y1}(X_i))} {\sum_{i:S_i=1} \hat{w}_1(X_i) Z_i} - \frac{\sum_{i:S_i=1} \hat{w}_0(X_i) (1-Z_i)(Y_i - \hat{\mu}_{y0}(X_i))} {\sum_{i:S_i=1} \hat{w}_0(X_i) (1-Z_i)}}_{\text{Weighting-based estimator using outcome residuals}} \nend
    + \underbrace{\frac{\sum_{i:S_i=0} (\hat{\mu}_{y1}(X_i) - \hat{\mu}_{y0}(X_i))} {N}}_{\text{Outcome-based estimator}},
\end{align*}
and the denominator is an augmented weighted estimator for the treatment received: 
\begin{align*}
    \hat{\tau}_{\text{mr}}^{D}:= \underbrace{\frac{\sum_{i:S_i=1} \hat{w}_1(X_i) Z_i(D_i - \hat{\mu}_{d1}(X_i))} {\sum_{i:S_i=1} \hat{w}_1(X_i) Z_i} - \frac{\sum_{i:S_i=1} \hat{w}_0(X_i) (1-Z_i)(D_i - \hat{\mu}_{d0}(X_i))} {\sum_{i:S_i=1} \hat{w}_0(X_i) (1-Z_i)}}_{\text{Weighting-based estimator using treatment received residuals}} \nend
    + \underbrace{\frac{\sum_{i:S_i=0} (\hat{\mu}_{d1}(X_i) - \hat{\mu}_{d0}(X_i))} {N}}_{\text{Treatment-received based estimator}},
\end{align*}
where $\hat{\mu}_{yz}(X)$ is an estimator of $\EE[Y \given Z = z, S=1, X=x]$, and $\hat{\mu}_{dz}(X)$ is an estimator of $\EE[D \given Z = z, S=1, X=x].$ $\hat \mu_{yz}(X)$ and $\hat \mu_{dz}(X)$ represent additional nuisance functions that researchers must estimate. In practice, researchers can use linear regression, or more flexible, black-box approaches (e.g., \citealp{rudolph2017robust, chernozhukov2018double, athey2019estimating, liu2024encoding, huang2025distilling}) to construct estimates of $\hat \mu_{yz}(X)$ and $\hat \mu_{dz}(X)$.

The proposed multiply robust estimator will be a consistent estimator of the T-CACE, so long as the weights for the sample selection process are correctly specified, or both the outcome and treatment received models are correctly specified. We present the formal theoretical properties of $\hat \tau_{mr}$ in \S\ref{thm: consistency doubly robust estimator}. 

\section{Practical Considerations} \label{sec: practical considerations}
In the following section, we consider several extensions of the framework to address practical challenges. First, we provide ways for researchers to integrate partial, or proxy, information about compliance across the target population into the estimation procedure. Second, we introduce a sensitivity analysis framework to evaluate the robustness of the estimators when the underlying mean exchangeability assumptions are violated. Finally, we extend existing IV tests for researchers to evaluate the validity of the instrument used in this generalizability context.

\subsection{Incorporating Partial Compliance Information from Target Population} \label{subsec: partial compliance information}
Throughout Sections \ref{sec: complete randomization} and \ref{sec: causal estimation}, we have assumed that researchers have minimal knowledge of the compliance patterns in the target population. As a result, they can only utilize compliance information from the study sample. In practice, researchers often have additional information about compliance in the target population. For example, in door-to-door canvassing studies, an organization’s administrative records may indicate which residents in the target area responded to previous survey attempts. Similarly, in a medical setting, electronic health records may indicate whether patients have historically followed certain treatment regimens. 

In this subsection, we consider two settings: (1) researchers have observed compliance status for a subset of individuals in the target population; (2) researchers have access to a proxy measure of compliance for individuals across the target population. We show that in settings when researchers have access to partial compliance information across the target population, they can reduce the reliance on the IV assumptions of the compliance pattern (i.e., Assumption \ref{asp: IV}), and use this information to evaluate the validity of the underlying identifying assumptions (i.e., mean exchangeability of the first stage---Assumption \ref{asp: randomized treatment received exchangeability}).  \\ 

\noindent \textit{Case 1: Observed Compliance Status for a Subset of Units.}
In some settings, researchers can directly observe the compliance status for a subset of individuals in the target population. For example, some participants assigned to the treatment group during the experiment fail to complete the follow-up survey. As such, the outcome in the units is not observed; however, we \textit{do} know whether or not they complied with the treatment.

To estimate the T-CACE for the non-follow-up group, we leverage the partially observed compliance information to construct a variant of $\hat \tau_w$: 
\begin{align} \label{eq: Hajek over time}
      \hat{\tau}_{\text{w-pc}} := \hat{\tau}_{w}^{Y} \times \frac{\sum_{i: S_i =0}Z_i} {\sum_{i: S_i =0}D_i}.
\end{align}

By incorporating information about treatment received across the target population, $\hat{\tau}_{\text{w-pc}}$ simplifies the form of $\hat{\tau}_{w}.$ In particular, $\hat{\tau}_{w}^{D}$ is replaced with the fraction of compliers among the treated individuals in the target population. Under this framework, $\hat{\tau}_{\text{w-pc}}$ consistently recovers the T-CACE without requiring mean exchangeability in the first stage (i.e., Assumption \ref{asp: randomized treatment received exchangeability}), provided there are no always-takers and treatment assignment is completely random. For the formal assumptions and proof, refer to Assumption \ref{asp: partial compliance information} and Theorem \ref{thm: consistency partial information}.\\

\noindent \textit{Case 2: Proxy Measure of Compliance}
We now consider a setting in which researchers do not have a direct measure of compliance across the target population, but instead have a proxy measure of whether or not individuals may comply. For example, consider a medical trial in which researchers wish to evaluate the efficacy of flu vaccination. While researchers would not have access to an individual's compliance for this year's flu vaccination, they could have a measure of whether or not individuals took the flu shot in previous years. While we cannot directly generalize compliance patterns using the proxy measure of compliance to our current study without further assumptions, we can use the proxy measure as a way to evaluate the validity of our underlying identifying assumptions. 

More formally, let $\tilde C$ represent a proxy measure of compliance. Assuming no always takers, we can directly compute $\prob(\tilde C = 1 \mid S = 0).$ For example, this could represent the proportion of compliers based on the previous year's vaccine behavior. We can then compare $\prob(\tilde C = 1 \mid S = 0)$ to our estimate of the overall proportion of compliers in the target population, represented by $\EE_{X \given S=0}[\mu_{d1}(X)-\mu_{d0}(X)].$ While a large discrepancy between the two does not necessarily imply the violation of the underlying identifying assumptions, it does serve as a potential signal that the underlying pre-treatment covariates $X$ may not be sufficiently generalizing the underlying compliance patterns. 

\subsection{Sensitivity Analysis} \label{subsec: sensitivity analysis}
In order to identify the T-CACE, we introduced two different mean exchangeability assumptions (i.e., Assumption \ref{asp: cond_ign_selection} and Assumption \ref{asp: randomized treatment received exchangeability}). This requires that researchers not only have a sufficiently rich set of pre-treatment covariates that can explain the potential confounding effects of selection on treatment effect heterogeneity but also a set of covariates that can explain away differential compliance patterns between the study sample and the target population. In practice, this can be a tenuous assumption to leverage, as the set of covariates researchers tend to have access to across \textit{both} the study sample and the target population can be relatively limited. Furthermore, it can be implausible to evaluate whether these assumptions hold in practice. In this subsection, we propose a sensitivity analysis framework for researchers to evaluate the underlying sensitivity in their T-CACE estimates to potential omitted variables.

To begin, we define a latent or unmeasured covariate $U$ such that, \textit{if} researchers additionally accounted for $U$, mean exchangeability of both treatment effect heterogeneity and the first stage would hold.
\begin{customassumption} [Mean Exchangeability of Selection and Treatment Effect Heterogeneity with Unmeasured Confounders] \label{asp: cond_ign_selection unmeasured covariates}
\begin{align*}
   \EE_{X, U \given S=0} \cbr{\EE\sbr{Y(1) - Y(0) \given S=0, X, U}} = \EE_{X, U \given S=0} \cbr{\EE\sbr{Y(1) - Y(0) \given S=1, X, U}}. 
\end{align*}
\end{customassumption}
Similarly, we assume the same for the treatment received.
\begin{customassumption} [Mean Exchangeability of the First Stage with Unmeasured Confounders] \label{asp: randomized treatment received exchangeability unmeasured covariates}
\begin{align*}
   \EE_{X, U \given S=0} \cbr{\EE\sbr{D(1) - D(0) \given S=0, X, U}} = \EE_{X, U \given S=0} \sbr{\EE\sbr{D(1) - D(0) \given S=1, X, U}}. 
\end{align*}

\end{customassumption}
Assumptions \ref{asp: cond_ign_selection unmeasured covariates} and \ref{asp: randomized treatment received exchangeability unmeasured covariates} do not aid in constructing a weighted estimator for identifying T-CACE, as the correct weights cannot be estimated due to the unobserved $U.$  In what follows, we propose a sensitivity analysis procedure to assess the robustness of the weighted estimators to the presence of an omitted $U$. For simplicity, we focus on the weighted estimator, although the procedure can be easily extended to the weighted least squares estimator. Throughout the section, we assume that Assumption \ref{asp: cond_ign_selection unmeasured covariates} and Assumption \ref{asp: randomized treatment received exchangeability unmeasured covariates} hold. 

To begin, define the \textit{oracle} weights as: 
$$w^*_z(X,U) :=\frac{\prob(S = 0 \given X, U)}{\prob(S = 1 \given X, U)\prob(Z = z \given S=1, X, U)}, \text{ for } z \in \{0,1\}.$$
The magnitude of the gap between the estimable weights ($w_z(X)$) and the oracle weights will depend directly on how imbalanced $U$ is between the study sample and target population \citep{huang2025variance}. 
When there is a greater degree of imbalance in $U$ between the study sample and the target population, the difference between $w_z(X)$ and $w^*_z(X,U)$ will be larger. The bias in the overall T-CACE estimate will depend on how the imbalance in $U$ is related to both the treatment effect heterogeneity, as well as the first stage (i.e., differences in treatment received---$D(1) - D(0)$).

Following \citet{zhao2019sensitivity}, we assume that the gap between the oracle weights and the estimable weights can be constrained by some constant $\Gamma \geq 1$.

\begin{customassumption} [Marginal Sensitivity Model] \label{asp: marginal sensitivity model} For some constant $\Gamma \geq 1$, 
$$\varepsilon(\Gamma) := \left\{  w^* ~ : ~ \Gamma^{-1} \leq \frac{w^*(x,u)}{w(x)} \leq \Gamma \text{ for all } x \in \mathcal{X}, u \in \mathcal{U}\right\}.$$
\end{customassumption}

The marginal sensitivity model essentially assumes that the worst-case error that occurs from omitting $U$ in the weights can be bounded \citep{nie2021covariate, zhao2019sensitivity}. A larger $\Gamma$ implies a greater degree of potential confounding. For a fixed $\Gamma$ value, we can bound the range of possible values of the T-CACE by minimizing and maximizing the estimand with respect to the oracle weights $w^*$ over $\varepsilon(\Gamma)$. However, this is a high-dimensional non-convex optimization problem, which is NP-hard and computationally intractable \citep{snoek2015scalable}. We instead solve a computationally tractable problem that provides a conservative bound of the original problem. See \S\ref{app: sensitivity} for details on the estimation problem.

In practice, we recommend researchers recompute the range of possible T-CACE estimates for increasing $\Gamma \geq 1$ values to find the threshold $\Gamma^*$ value that results in an interval that includes zero. This implies that for a $\Gamma = \Gamma^*$ value, the unmeasured confounders $U$ could be sufficiently strong to result in a statistically insignificant T-CACE. A relatively small $\Gamma^*$ (i.e., close to 1) implies that the estimated T-CACE is highly sensitive to potential omitted variables. We provide a benchmarking approach in \S\ref{app: illustrating sensitivity analysis}, which allows researchers to use observed covariate data to calibrate plausible $\Gamma$ values. 

\subsection{Evaluating validity of the IV}
The sensitivity analysis proposed in the previous subsection allows researchers to consider potential violations in the mean exchangeability assumptions. To help researchers evaluate the other necessary assumptions (i.e., Assumption 3), we extend existing IV tests for the external validity context. More specifically, there is a growing literature \citep[e.g.,][]{kitagawa2015test, mourifie2017testing, yu2025binary} on evaluating instrument validity. We focus specifically on an approach proposed by \citet{kitagawa2015test}.  \citet{kitagawa2015test} derives a testable observable implication that holds when the IV is valid in the study sample and shows that this implication is sharp--i.e., no other feature of the study-sample data can provide additional information for ruling out invalid instruments. In \S G.2, we provide a sample assignment ignorability condition under which the study-sample observable implications can be transported to the target population. Under this condition, we can apply the approach in \citet{kitagawa2015test} to test Assumption~\ref{asp: IV}. We then discuss directions for researchers in cases where these tests fail in \S G.3. We emphasize that the proposed IV tests are falsification tests, in the sense that they can provide evidence that an IV is not valid; however, failure to reject the tests should not be interpreted as evidence that the IV \textit{is} valid.
\section{Simulation Study} \label{sec: simulation study}
We now illustrate the performance of the proposed estimators in a series of simulations. We show that when mean exchangeability of treatment effect heterogeneity and mean exchangeability of the first stage hold, all three proposed weighted estimators will be consistent estimators for the T-CACE. Furthermore, our proposed variance estimators provide nominal coverage across the different simulation settings. Compared to the weighted estimator, the WLS and the multiply robust estimators exhibit lower variance, demonstrating their potential utility in practice for efficiently estimating the T-CACE.

\paragraph{Simulation Set-Up.} We provide an overview of the simulation set-up, with details in \S\ref{app: detailed standard simulation setup}. We generate the sample selection indicator as $S_i\given X_i \sim \text{Bernoulli}\cbr{r' \cdot \sigma\rbr{\sum_{j=1}^{10}X^{j}_i}},$ where $r'$ is a hyperparameter that controls the ratio of the study sample size to the combined size of the study sample and target population, and the covariates $X$ follow a Uniform distribution (i.e., $X_i \in \RR^{10}$, where $X^{j}_i \overset{i.i.d.} {\sim} \text{Unif(-0.3, 0.5)}$ for $j \in \{1, ..., 10\}$). For each $n + N \in \{1500, 5000,  10000\},$ we set $r' \in \{1, 0.2, -0.4, -1.5\},$ so that the ratio of the study sample size to the combined size of the study sample and target population is approximately $0.71,$ $0.55,$ $0.40$ and $0.23$.  We generate the compliance type $C_i$ for each unit $i$ from a multinomial logit model on $X_i$.

Finally, the outcome is a linear combination of the treatment received, the covariates, and interaction terms: 
$$Y_i = 2D_i + \sum_{j=1}^{10}X^{j}_i + D_i \times \sum_{j=1}^{10}X^{j}_i +  \epsilon_i, \text{ where } \epsilon_i \sim N(0, 0.5).$$ 

Throughout this section, we consider a completely randomized treatment indicator $Z_i \sim \text{Bernoulli}(0.5)$. We consider an observational setting in \S\ref{subsec: sim obs study}. For each pair of $(n + N, r')$, we estimate the following estimators: the weighted estimator $\hat{\tau}_{w},$ the WLS estimator $\hat{\tau}_{\text{wls}},$ the multiply robust estimator $\hat{\tau}_{\text{mr}},$ and the weighted ITT estimator that does not account for compliance (i.e., $\hat{\tau}_{w}^Y$). We repeat each scenario for 1,000 total iterations. Table \ref{table:combined_coverage} displays the results for the four estimators when $N + n = 5000$, under two settings where $n/(n + N) \approx 0.71$ or $n/(n + N) \approx 0.23$. 

\paragraph{Simulation Results.} Overall, we see that the mean errors of $\hat{\tau}_{w},$ $\hat{\tau}_{\text{wls}},$ and $\hat{\tau}_{\text{mr}}$ are negligible, highlighting the consistency of the underlying estimators. The weighted ITT estimator that does not account for compliance is biased downward, suggesting that the causal effect is stronger among the compliers in the target population. Furthermore, the WLS and multiply robust estimators exhibit substantially smaller variances than the weighted estimator. Interestingly, the weighted ITT estimator is not only downward biased but also exhibits greater variance than the three T-CACE estimators. We conjecture that the higher variance of the ITT estimator arises from its inclusion of both compliers and non-compliers, which increases outcome heterogeneity. In contrast, T-CACE estimators focus exclusively on compliers, thereby eliminating this heterogeneity and reducing variance.

\begin{table} 
\centering
\begin{tabular}{lcccc} \toprule 
& \multicolumn{2}{c}{Scenario 1} & \multicolumn{2}{c}{Scenario 2} \\
& Bias & s.d. & Bias & s.d.  \\ \midrule 
Weighted ITT Estimator &-1.15 &0.27 &-1.37 &0.40 \\ 
Weighted Estimator &-0.00 &0.12 &0.00 &0.30 \\ 
Weighted Least Squares &0.00 &0.06 &-0.02 &0.13 \\ 
Multiply Robust &0.00 &0.06 &-0.01 &0.13 \\ \bottomrule 
\end{tabular} 
\caption{The bias and standard deviations of the weighted estimator, WLS estimator, multiply robust estimator, and the weighted ITT estimator. The results are based on 1,000 trials with a total sample size of $N + n = 5000$. Scenario 1 represents $n/(n + N) \approx 0.71$, while scenario 2 corresponds to $n/(n + N) \approx 0.23$.}
\label{table:combined_coverage}
\end{table}

Finally, we evaluate the performance of the different estimators under more complex settings. \S\ref{subsec: IV vs PS} compares the performance of the principal stratification estimators defined in Theorem 2 of \citet{clark2024transportability} with the WLS estimators. We find that in scenarios where principal ignorability is violated due to unmeasured confounders affecting both treatment received and outcome, the WLS estimator remains robust as the dimensionality of the unobserved variables increases, while the principal stratification estimator suffers from systematic bias. In \S \ref{subsec: sim obs study}, we conduct extensive experiments under the setting of an observational study. We further evaluate the proposed sensitivity analysis framework in \S\ref{subsec: sensitivity analysis simulation}.

\section{Empirical Application: Reducing Exclusionary Attitudes through 
Deep Canvassing} \label{sec: application exclusionary attitudes}
We now return to the motivating example of reducing exclusionary attitudes through deep canvassing. We consider two different settings. The first generalizes the complier average causal effect to participants who failed to complete the follow-up survey, but initially opted into the experiment. The second considers generalizing to the set of voters who did not opt into the experiment to begin with. Since voters were randomized at the household level, we adjust the variance estimator to account for within-household dependence (see \S\ref{app: empirical addtional tables} for details).

\subsection{Setting 1: Generalizing to Participants Who Did Not Take the Follow-up Survey} \label{sec: application not follow up}
We begin by generalizing the results to the set of participants who did not take the follow-up survey. While we have a measure of compliance for some of the individuals in the target group, we do not have access to their outcomes. For each voter, we have access to 33 total pre-treatment covariates that includes demographic information, such as age, gender, past electoral participation, as well as 25 survey items measuring immigrant-related opinions prior to treatment. We apply our weighted estimator and WLS estimator to generalize the CACE from the study sample to these participants.

The estimated T-CACEs range from 12-20\%. Notably, the results for the weighted and WLS T-CACE estimators are statistically significant at the 5\% level. We also include the principal stratification estimator (weighted (PS)) introduced in \S\ref{subsec: IV vs PS}, which is based on the principal ignorability assumption. It provides a comparable estimate to our methods: $(12.4\% ; [2.8, 23.3]).$ Taken together, these estimates provide robust evidence that deep canvassing positively influenced the target index, with some variation in magnitude across estimation methods. The generalized CACE in this context is larger than the generalized ITT $(10.3\% ; [2.4, 18.2])$ and comparable to the within-sample CACE estimate $13.2\% ; [6.2,20.1]).$ This suggests that the estimated treatment effect on support for unauthorized immigrants from deep canvassing generalizes well to the population of individuals who failed to participate in the follow-up survey. \\

\subsection{Setting 2: Generalizing to Voters Who Did not Participate in the Experiment} \label{sec: application whole}
We now consider the harder task of generalizing to the set of individuals who did not respond to the initial recruitment survey. This consists of 209,730 registered voters. Unlike the first setting, we are restricted to basic demographics, as the individuals who did not participate in the experiment failed to complete the baseline survey. As a result, we can only adjust for age, gender, location, and past electoral participation history.

In contrast to the first setting, we find that, except for the PS estimator, T-CACE estimates are all statistically insignificant at the 5\% level. We conclude that among the 209,730 voters who did not participate in the experiment, the impact of deep canvassing remains unclear at the 5\% significance level. The larger intervals are likely due to the greater degree of imbalance between the study sample's units and the target. This is in line with what we would expect intuitively, as there likely is a greater degree of distribution shift between individuals who did not participate at all in the experiment than those who initially participated, but failed to participate in the follow-up survey (i.e., Setitng 1). 

\begin{figure}[htbp]
\centering
\includegraphics[width=0.8\textwidth]{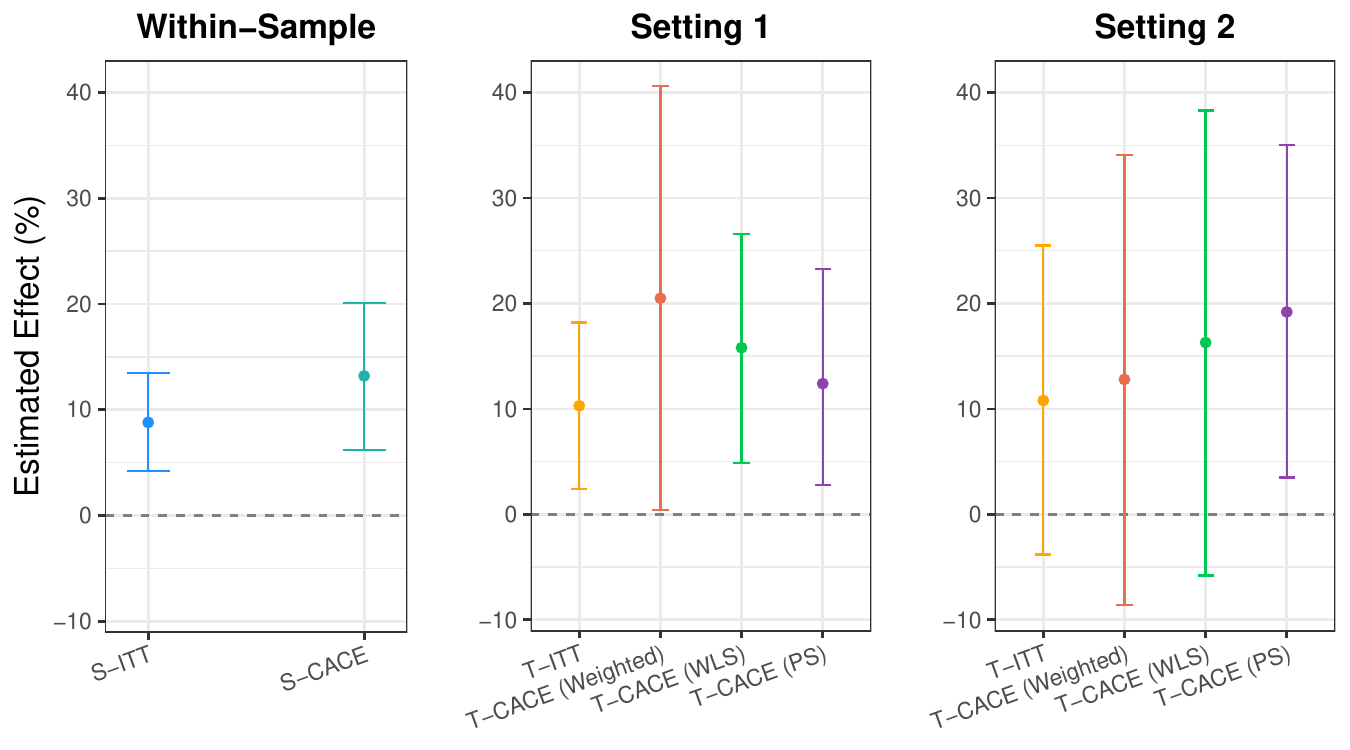}
\caption{Plot of various estimates and 95\% confidence intervals for the deep canvassing application. The left-most panel shows the within-sample effects. The middle panel displays results for participants who did not take the follow-up survey. The right-most panel presents estimates for participants who did not participate in the experiment at all. We use the prefix “S-” to denote within-sample estimates and “T-” to denote the corresponding target quantities of interest.}
\label{fig: real data estimator plot}
\end{figure}

Figure \ref{fig: real data estimator plot} summarizes the results for these two settings, as well as the estimated within-sample effects, with numerical results in Table 8 of \S\ref{app: empirical addtional tables}. We provide an illustration of the sensitivity analysis in \S\ref{app: illustrating sensitivity analysis}. \\ 

\section{Conclusion} \label{sec: conclusion}
In this paper, we address the critical challenge of generalizing the complier average causal effect (CACE) from a study sample to a target population. Our contributions are threefold. First, we derive the identifying assumptions to identify the target complier average causal effect (T-CACE) using instrumental variables. Second, we propose a set of estimators that allow researchers to efficiently estimate the T-CACE. We extend the framework to consider settings in which researchers have access to auxiliary compliance information across the target population. Finally, we introduce a sensitivity analysis to evaluate the robustness of the T-CACE estimators to potential violations in the underlying identification assumptions. We apply our results to a deep canvassing experiment, and show that there is statistical evidence that the efficacy of deep canvassing on reducing prejudice generalizes to the set of individuals who could not be reached in a follow-up survey. 

There are several interesting avenues of future research. First, \citet{clark2024transportability} introduces an alternative set of identification assumptions using principal ignorability, which requires that researchers can feasibly model the compliance pattern across both the study sample and the target population. In contrast, our paper has largely focused on an instrumental variables approach. Future work could consider how to reason about the plausibility of both sets of assumptions in practice.

Second, in settings when there are high rates of non-compliance, treatment assignment will become a weak instrument \citep[e.g.,][]{hartman2023improving}, which results in large amounts of efficiency loss, even in estimating the within-sample CACE. Concerns about weak instruments are amplified in generalization contexts, where reweighting the study sample can result in further efficiency loss \citep[e.g.,][]{miratrix2018worth}. An interesting avenue of future work could consider how to mitigate concerns about weak instruments, specifically in the context of generalizing the CACE.

\newpage
\bibliographystyle{ims}
\bibliography{reference}

\newpage 
\appendix

\section*{\centering Supplementary Materials: Generalization of Causal Effects with Noncompliance}

\section{Additional Discussion}
\subsection{Theoretical Properties of the Weighted Least Squares Estimator} \label{app:wls}
In the following subsection, we provide the formal theoretical results for $\hat{\tau}_{\text{wls}}$. 
\begin{theorem}[Consistency of the Weighted Least Squares Estimator] \label{thm: randomization wls Estimator consistency}
Suppose that all the assumptions for Theorem 3.1 hold. Suppose $\sup_{x \in \cX} |\hat{w}_1(x) - w_1(x)| = o_p(1)$ and $\sup_{x \in \cX} |\hat{w}_0(x) - w_0(x)| = o_p(1).$ 
Then

\begin{align*}
    \hat{\tau}_{\text{wls}} \overset{p}{\rightarrow} \tau_\tca.
\end{align*}
    \begin{proof}
        See \S\ref{pro: randomization wls Estimator consistency} for a detailed proof.
    \end{proof}
\end{theorem}

Theorem \ref{thm: randomization wls Estimator consistency} states that $\hat{\tau}_{\text{wls}}$ is consistent under the same set of conditions required for the consistency of $\hat{\tau}_{w}.$ Moreover, using standard estimating equation theory, we show that $\hat{\tau}_{\text{wls}}$ is $\sqrt{n + N}$-consistent. The proof proceeds similarly to that of Theorem \ref{thm: cre logistic model asymptotics Hajek}, except that one must additionally account for the variability due to the estimation of the nuisance parameters $\hat{\gamma}^Y$ and $\hat{\gamma}^D.$

\begin{theorem}[Asymptotic Distribution of the Weighted Least Squares Estimator for Unknown Selection Mechanism] \label{thm: cre logistic model asymptotics distribution wls}
Suppose that all the assumptions in Theorem \ref{thm: randomization wls Estimator consistency} hold. Suppose the model specification for the selection mechanism (i.e., Assumption \ref{asp: cre model specification}) holds. Then, provided that $\EE[||X||_2^4] < \infty$, $\EE[Y^2] < \infty$, and the design matrices for both the selection model ($\EE[XX^T]$) and the weighted outcome regression ($\EE[S w_z(X) \tilde{X} \tilde{X}^T]$) are full rank,
\begin{align*}
    \sqrt{n + N}(\hat{\tau}_{\text{wls}} - \tau_\tca) \overset{d}{\rightarrow} N(0, \nabla g_{\text{wls}} ^T \Sigma_{\theta_{\text{wls}}^*, \beta^*} \nabla g_{\text{wls}}), 
\end{align*}
    where we define $g_{\text{wls}}$ and the covariance matrix $\Sigma_{\theta_{\text{wls}}^*, \beta^*}$ in the proof.
    \begin{proof}
        See \S\ref{pro: cre logistic model asymptotics distribution wls} for a detailed proof.
    \end{proof}    
\end{theorem}

As a result, we can derive a sandwich-type variance estimator, analogous to those in Theorems~\ref{thm: cre known model asymptotics Hajek} and \ref{thm: cre logistic model asymptotics Hajek}, and construct a Wald-type confidence interval for the WLS estimator.

\subsection{Theoretical Properties of Multiply Robust} \label{app:mr}

\begin{theorem}[Consistency Properties of the Multiply Robust Estimator] \label{thm: consistency doubly robust estimator}
Suppose that all the assumptions for Theorem~\ref{thm: randomization identification tcace} hold. Assume 
$\sup_{x \in \cX} \abr{\hat{w}_z(x) - \tilde{w}_z(x)} = o_p(1),$ 
for $z \in \{0,1\}$, for some functions $\tilde{w}_1(x)$ and $\tilde{w}_0(x)$, $\sup_{x \in \cX} \abr{\hat{\mu}_{yz}(x) - \tilde{\mu}_{yz}(x)} = o_p(1)$ for $z \in \{0,1\}$ and some functions $\tilde{\mu}_{yz}(x),$ and $\sup_{x \in \cX} \abr{\hat{\mu}_{dz}(x) - \tilde{\mu}_{dz}(x)} = o_p(1)$ for $z \in \{0,1\}$ and some functions $\tilde{\mu}_{dz}(x)$. If at least one of the following conditions holds, then $\hat{\tau}_{\text{mr}}\overset{p}{\rightarrow} \tau_\tca:$

\begin{itemize}
    \item [(i).] $\tilde{w}_1(x) = w_1(x)$ and $\tilde{w}_0(x) = w_0(x).$ 
    \item [(ii).] For any $z\in \{0,1\},$ $\tilde{\mu}_{yz}(x) = \mu_{yz}(x)$ and $\tilde{\mu}_{dz}(x) = \mu_{dz}(x).$ 
\end{itemize}

\begin{proof}
    See \S \ref{pro: consistency doubly robust estimator} for a detailed proof.
\end{proof}
\end{theorem}

Because $\hat{\tau}_{\text{mr}}$ is a smooth function of two standard doubly robust estimators, researchers can apply a nonparametric bootstrap to estimate its standard error \citep{funk2011doubly}.

\section{Proofs in Section 3}
\subsection{Theorem~\ref{thm: randomization identification tcace} (Outcome-Based Identification)} 
\begin{proof} \label{pro: randomization identification tcace}
We start by noticing that 
\begin{align} \label{eq: randomization outcome identification decomposition}
    \EE[Y(1) - Y(0) \given S = 0] & = \sum_{k\in \{0, 1\}}\EE\sbr{Y(1) - Y(0)\given S=0, C = k}\prob(C=k \given S=0) \nend
    &= \EE\sbr{Y(1) - Y(0)\given S=0, C = 1}\prob(C=1 \given S=0).
\end{align}
where we apply the law of total probability to partition over the compliance types in the first equality. The second equality holds by monotonicity (i.e., Assumption~\ref{asp: IV}-(a)) and exclusion restriction (i.e., Assumption~\ref{asp: IV}-(b)). We have

\begin{align} \label{eq: randomization treatment received identification decomposition}
    \EE[D(1) - D(0) \given S = 0] & = \sum_{k\in \{0, 1\}}\EE\sbr{D(1) - D(0)\given S=0, C = k}\prob(C=k \given S=0) \nend
    &= \prob(C=1 \given S=0),
\end{align}
where the first equality holds by the law of total probability, and the second equality holds by monotonicity (i.e., Assumption~\ref{asp: IV}-(a)) and exclusion restriction (i.e., Assumption~\ref{asp: IV}-(b)). Then by instrument relevance (i.e., Assumption~\ref{asp: IV}-(c)), we arrive at an equation for $\tau_\tca:$
\begin{align} \label{eq: randomization tcace ratio}
    \EE[Y(1) - Y(0)\given C = 1, S = 0] = \frac{\EE[Y (1) - Y(0) \mid S = 0]}{\EE[D(1) - D(0) \mid S = 0]}.
\end{align}
We finish the proof by showing the identification formulae for the numerator and the denominator for \eqref{eq: randomization tcace ratio}. For the numerator,
\begin{align*}
    \EE[Y(1) - Y(0)\given S = 0] &= \EE_{X \given S=0}\{\EE[Y(1) - Y(0) \given S = 0, X]\} \nend
    &= \EE_{X \given S=0}\{\EE[Y(1) - Y(0) \given S = 1, X]\} \nend
    &= \EE_{X \given S=0}\{\EE[Y \given Z = 1, S = 1, X] - \EE[Y \given Z = 0, S=1, X]\},
\end{align*}
where the first equality holds by the law of total expectation, the second equality holds by mean exchangeability of selection and treatment effect heterogeneity (i.e., Assumption~\ref{asp: cond_ign_selection}), and the last equality holds by treatment ignorability (i.e., Assumption~\ref{asp: randomization}). Similarly for the denominator,
\begin{align*}
    \EE[D(1) - D(0)\given S = 0] &= \EE_{X \given S=0}\{\EE[D(1) - D(0) \given S = 0, X]\} \nend
    &= \EE_{X \given S=0}\{\EE[D(1) - D(0) \given S = 1, X]\} \nend
    &= \EE_{X \given S=0}\{\EE[D \given Z = 1, S = 1, X] - \EE[D \given Z = 0, S=1, X]\},
\end{align*}
where the first equality holds by the law of total expectation, the second equality holds by mean exchangeability of the first stage (i.e., Assumption~\ref{asp: randomized treatment received exchangeability}), and the last equality holds by treatment ignorability (i.e., Assumption~\ref{asp: randomization}). We complete the proof.

\end{proof}

\section{Proofs in Section 4}
\subsection{IPW-Based Identification} 
\begin{corollary}
[An Equivalent Form of T-CACE]\label{thm: randomization identification tcace IPW}
Suppose that treatment ignorability (i.e., Assumption~\ref{asp: randomization}), the IV assumptions (i.e., Assumption~\ref{asp: IV}), mean exchangeability (i.e., Assumption~\ref{asp: cond_ign_selection} and \ref{asp: randomized treatment received exchangeability}, and overlap (i.e., Assumption~\ref{asp: cre overlap}) hold. Let

\begin{align*} 
    \tau_{w}^{Y} = \frac{\mathbb{E}\left[ w_1(X)ZSY \right]}{\mathbb{E}\left[ w_1(X)ZS \right]}  -   \frac{\mathbb{E}\left[ w_0(X)(1-Z)SY \right]}{\mathbb{E}\left[ w_0(X)(1-Z)S \right]}.
\end{align*}
We define similarly that
\begin{align*} 
    \tau_{w}^{D} = \frac{\mathbb{E}\left[ w_1(X)ZSD \right]}{\mathbb{E}\left[ w_1(X)ZS \right]}  -   \frac{\mathbb{E}\left[ w_0(X)(1-Z)SD \right]}{\mathbb{E}\left[ w_0(X)(1-Z)S \right]}.
\end{align*}

Then the T-CACE can be written as 
\begin{align*}
    \tau_\tca = \frac{\tau_{w}^{Y}}{\tau_{w}^{D}}.
\end{align*}

\begin{proof} \label{pro: randomization identification tcace IPW}
    We prove the theorem by using Theorem~\ref{thm: randomization identification tcace} after showing that 
\begin{align} \label{eq: ipw outcome identification}
    \frac{\EE_{X \given S=0}[\mu_{y1}(X)-\mu_{y0}(X)]}{\EE_{X \given S=0}[\mu_{d1}(X)-\mu_{d0}(X)]} = \frac{\tau_{w}^{Y}}{\tau_{w}^{D}}.
\end{align}
    
We start by establishing the relationship: 
\begin{align} \label{eq: ipw outcome identification 1}
    \mathbb{E} \left[ \frac{\mathbb{P}(S=0|X)}{\mathbb{P}(S=1|X)} \cdot \frac{ZSY}{\mathbb{P}(Z=1|S=1,X)} \right] = \mathbb{P}(S=0) \cdot \mathbb{E}_{X|S=0}\{\mathbb{E} [ Y\given Z=1, S=1, X]\}.
\end{align}
We write the left-hand side as
\begin{align*}
    & \mathbb{E} \sbr{\frac{\mathbb{P}(S=0|X)}{\mathbb{P}(S=1|X)} \cdot \frac{ZSY}{\mathbb{P}(Z=1|S=1,X)}} \nend
    & = \int_{\cX} \mathbb{E} \left[ \frac{\mathbb{P}(S=0|X)}{\mathbb{P}(S=1|X)} \cdot \frac{ZSY}{\mathbb{P}(Z=1|S=1,X)} \Bigg| X=x \right] p_X(x) \, dx \nend
    & = \int_{\cX} \frac{\mathbb{P}(S=0|X=x)}{\mathbb{P}(S=1|X=x)} \cdot \mathbb{E} \left[ \frac{ZSY}{\mathbb{P}(Z=1|S=1,X)} \Bigg| X=x \right] p_X(x) \, dx \nend
    & = \int_{\cX} \frac{\mathbb{P}(S=0|X=x)}{\mathbb{P}(S=1|X=x)} \cdot \mathbb{E} \left[ \frac{ZSY}{\mathbb{P}(Z=1|S=1,X)} \Bigg| X=x, S=1 \right] \mathbb{P}(S=1|X) \cdot p_X(x) \, dx,
\end{align*}
where the first equality holds by the Tower property, and the third equality holds by the law of total expectation. Proceeding from above,
\begin{align*}
    &=\int_{\cX} \mathbb{P}(S=0|X=x) \cdot p_X(x) \cdot \mathbb{E} \sbr{\frac{ZY}{\mathbb{P}(Z=1|S=1,X=x)} \Bigg| X=x, S=1} \, dx \nend
    & = \int_{\cX} p_{X|S=0}(x) \cdot \mathbb{P}(S=0) \cdot \mathbb{E} \left[ \frac{Y}{\mathbb{P}(Z=1|S=1,X=x)} \Bigg| Z=1, S=1, X=x \right] \mathbb{P}(Z=1|S=1, X=x) dx \\
    & = \mathbb{P}(S=0) \int_{\cX} p_{X|S=0}(x) \cdot \mathbb{E} [ Y\given Z=1, S=1 X=x] dx \\
    & = \mathbb{P}(S=0) \cdot \mathbb{E}_{X|S=0}\{\mathbb{E} [ Y\given Z=1, S=1, X]\},
\end{align*}
where the second equality holds by Bayes' rule and the law of total expectation. Similarly, we can show that 
\begin{align} \label{eq: ipw outcome identification 2}
    \mathbb{E} \left[ \frac{\mathbb{P}(S=0|X)}{\mathbb{P}(S=1|X)} \cdot \frac{ZS}{\mathbb{P}(Z=1|S=1,X)} \right] &= \mathbb{E} \left[ \frac{\mathbb{P}(S=0|X)}{\mathbb{P}(S=1|X)} \cdot \frac{(1-Z)S}{\mathbb{P}(Z=0|S=1,X)} \right] =  \mathbb{P}(S=0), \\
    \mathbb{E} \left[ \frac{\mathbb{P}(S=0|X)}{\mathbb{P}(S=1|X)} \cdot \frac{(1-Z)SY}{\mathbb{P}(Z=1|S=1,X)} \right] &= \mathbb{P}(S=0) \cdot \mathbb{E}_{X|S=0}\{\mathbb{E} [ Y\given Z=0, S=1, X]\}, \label{eq: ipw outcome identification 3}\\
    \mathbb{E} \left[ \frac{\mathbb{P}(S=0|X)}{\mathbb{P}(S=1|X)} \cdot \frac{ZSD}{\mathbb{P}(Z=1|S=1,X)} \right] &= \mathbb{P}(S=0) \cdot \mathbb{E}_{X|S=0} \left\{ \mathbb{E} \left[ D \given Z=1, S=1, X \right] \right\}, \text{ and} \label{eq: ipw outcome identification 4}\\
    \mathbb{E} \left[ \frac{\mathbb{P}(S=0|X)}{\mathbb{P}(S=1|X)} \cdot \frac{(1-Z)SD}{\mathbb{P}(Z=0|S=1,X)} \right] &= \mathbb{P}(S=0) \cdot \mathbb{E}_{X|S=0} \left\{\mathbb{E} \left[ D \given Z=0, S=1, X \right] \right\}. \label{eq: ipw outcome identification 5}
\end{align}
Therefore, by combining \eqref{eq: ipw outcome identification 1}, \eqref{eq: ipw outcome identification 2}, \eqref{eq: ipw outcome identification 3}, \eqref{eq: ipw outcome identification 4} and \eqref{eq: ipw outcome identification 5}, we establish \eqref{eq: ipw outcome identification}. We complete the proof by Theorem~\ref{thm: randomization identification tcace}.
\end{proof}
\end{corollary}

\subsection{Theorem~\ref{thm: randomization Hajek Estimator consistency} (Consistency of the Weighted Estimator)} 
\begin{proof} \label{pro: randomization Hajek Estimator consistency}
    Note that $\hat{\tau}_{w}$ is of the form $a_1/a_2 - a_3/a_4,$ where the four terms $(a_1, a_2, a_3, a_4)$ represent different arithmetic means. We first prove that each of the four terms in $\hat{\tau}_{w}$ is consistent to its expectation counterpart in the identification formula of $\tau_\tca$ in Corollary \ref{thm: randomization identification tcace IPW}. We then use the continuous mapping theorem to combine the four terms and obtain the overall consistency. For example, we start by proving that the first term in $\hat{\tau}_{w}^{Y}$ is consistent to 
    
\begin{align} \label{eq: randomization Hajek Estimator consistency}
    \frac{\mathbb{E}\left[ \frac{\mathbb{P}(S=0|X)}{\mathbb{P}(S=1|X)} \cdot \frac{ZSY}{\mathbb{P}(Z=1|S=1,X)} \right]}{\mathbb{E}\left[ \frac{\mathbb{P}(S=0|X)}{\mathbb{P}(S=1|X)} \cdot \frac{ZS}{\mathbb{P}(Z=1|S=1,X)} \right]}.
\end{align}    
For the numerator of the first term of $\hat{\tau}_{w}^{Y},$ we have by the convergence condition of $(\hat{w}_1(x), \hat{w}_0(x)),$ 
\begin{align*}
    \frac{1}{n + N}\sum_{i=1}^{n + N} \frac{\hat{\prob}(S_i = 0 \given X_i)}{\hat{\prob}(S_i = 1 \given X_i)} \frac{Z_iS_iY_i}{\hat{\prob}(Z_i = 1 \given S_i=1, X_i)} - \frac{1}{n + N}\sum_{i=1}^{n + N} \frac{\prob(S_i = 0 \given X_i)}{\prob(S_i = 1 \given X_i)} \frac{Z_iS_iY_i}{\hat{\prob}(Z_i = 1 \given S_i=1, X_i)} = o_p(1).
\end{align*}
Then, combined with the weak law of large numbers, we derive that 
\begin{align*}
    \frac{1}{n + N}\sum_{i=1}^{n + N} \frac{\hat{\prob}(S_i = 0 \given X_i)}{\hat{\prob}(S_i = 1 \given X_i)} \frac{Z_iS_iY_i}{\hat{\prob}(Z_i = 1 \given S_i=1, X_i)} \overset{p}{\rightarrow} \mathbb{E}\left[ \frac{\mathbb{P}(S=0|X)}{\mathbb{P}(S=1|X)} \cdot \frac{ZSY}{\mathbb{P}(Z=1|S=1,X)} \right].
\end{align*}
Similarly, we can show that the limit of the denominator of the first term of $\hat{\tau}_{w}^{Y}$ is equal to $\mathbb{P}(S=0):$
\begin{align*}
    {\frac{1}{n + N}\sum_{i=1}^{n + N} \frac{\hat{\prob}(S_i = 0 \given X_i)}{\hat{\prob}(S_i = 1 \given X_i)} \frac{Z_iS_i}{\hat{\prob}(Z_i = 1 \given S_i=1, X_i)}} \overset{p}{\rightarrow} \mathbb{P}(S=0).
\end{align*}
Therefore, by the continuous mapping theorem, the first term of $\hat{\tau}_{w}^{Y}$ is consistent to \eqref{eq: randomization Hajek Estimator consistency}. By applying a similar proof to each of the three remaining terms and using the continuous mapping theorem again, we complete the proof.
\end{proof}

\subsection{Theorem~\ref{thm: cre known model asymptotics Hajek} (Asymptotic Distribution of the Weighted Estimator)} 
\begin{proof} \label{pro: randomization Hajek Estimator asymptotic}
    Denote $\hat{\theta_1} = \frac{1}{n + N}\sum_{i=1}^{n + N} \frac{\hat{\prob}(S_i = 0 \given X_i)}{\hat{\prob}(S_i = 1 \given X_i)} \frac{Z_iS_iY_i}{\prob(Z_i = 1 \given S_i=1, X_i)},$ $\hat{\theta_3} = \frac{1}{n + N}\sum_{i=1}^{n + N} \frac{\hat{\prob}(S_i = 0 \given X_i)}{\hat{\prob}(S_i = 1 \given X_i)} \frac{Z_iS_i}{\prob(Z_i = 1 \given S_i=1, X_i)},$ etc., up to $\hat{\theta_6},$ such that $\hat{\tau}_{w} = g(\hat{\theta}).$ We also denote $\hat{\theta_i}^{(j)}$ to be the $j-$th term of the summation of $\hat{\theta_i}.$ For example, $\hat{\theta_1}^{(2)} = \frac{\hat{\prob}(S_2 = 0 \given X_2)}{\hat{\prob}(S_2 = 1 \given X_2)} \frac{Z_2S_2Y_2}{\prob(Z_2 = 1 \given S_2=1, X_2)}.$ We plan to derive the asymptotic properties of $\hat{\theta}$ using the estimating equation method and then apply the delta method to $g(\hat{\theta}).$ We establish the estimating equation as:

\begin{align*}
    \frac{1}{n + N}\sum_{i=1}^{n + N} \Psi \left(\theta; Z_i, X_i, Y_i, D_i, S_i \right) = \frac{1}{n + N} \left(\begin{array}{c}
    \sum_{j=1}^{n + N}\rbr{\hat{\theta_1}^{(j)}-\theta_1}\\ 
    \sum_{j=1}^{n + N}\rbr{\hat{\theta_2}^{(j)}-\theta_2}\\ 
    \sum_{j=1}^{n + N}\rbr{\hat{\theta_3}^{(j)}-\theta_3}\\ 
    \sum_{j=1}^{n + N}\rbr{\hat{\theta_4}^{(j)}-\theta_4}\\ 
    \sum_{j=1}^{n + N}\rbr{\hat{\theta_5}^{(j)}-\theta_5}\\
    \sum_{j=1}^{n + N}\rbr{\hat{\theta_6}^{(j)}-\theta_6}
    \end{array}\right) = \left(\begin{array}{c}
    \hat{\theta_1}-\theta_1\\ 
    \hat{\theta_2}-\theta_2\\ 
    \hat{\theta_3}-\theta_3\\ 
    \hat{\theta_4}-\theta_4\\ 
    \hat{\theta_5}-\theta_5\\
    \hat{\theta_6}-\theta_6
    \end{array}\right) 
\end{align*}

For simplicity, we write $\Psi \left(\theta; Z_i, X_i, Y_i, D_i, S_i, \right)$ as $\Psi^{(i)}(\theta).$ Note that $\theta=\hat{\theta}$ is a solution to 
$\frac{1}{n + N}\sum_{i=1}^{n + N} \Psi^{(i)}(\theta) = 0.$ Denote $\theta^{*}$ be the solution of the population estimating equation 
\begin{align*}
    \EE[\Psi \left(\theta; Z, X, Y, D, S \right)]=0
\end{align*}
and $\frac{\partial \Psi^{(i)}(\theta)}{\partial \theta}$ be the derivative matrix of $\Psi^{(i)}(\theta),$ i.e., $\rbr{\frac{\partial \Psi^{(i)}(\theta)}{\partial \theta}}_{jk}=\frac{\partial \Psi_{j}^{(i)}(\theta)}{\partial \theta_k}.$ We define the matrices:
\begin{align*}
    A_{n + N}(\theta) &= \frac{1}{n + N} \sum_{i=1}^{n + N} \EE\sbr{\frac{\partial \Psi^{(i)}(\theta)}{\partial \theta}} \text{ and } B_{n + N}(\theta) = \frac{1}{n + N}\sum_{i=1}^{n + N}\text{cov}(\Psi^{(i)}(\theta)).
\end{align*}
We denote $A(\theta) = \text{lim}_{n + N \rightarrow \infty} A_{n + N}(\theta)$ and $B(\theta) = \text{lim}_{n + N \rightarrow \infty} B_{n + N}(\theta).$ Define $\Sigma_{\theta} = A^{-1}(\theta)B(\theta)A^{-T}(\theta).$ Then
\begin{align} \label{eq: covariance matrix known weights}
    \Sigma_{\theta^*} = A^{-1}(\theta^*)B(\theta^*)A^{-T}(\theta^*)
\end{align}
is the covariance matrix with $\theta$ replaced by the solution to the population estimation equation, $\theta^{*}.$
By the theory of estimating equation \citep{buchanan2018generalizing, carroll2006measurement, huber1967behavior} (see Theorem 9.3 of \cite{li2019graduate} for details), we derive that $\sqrt{n + N}\rbr{\hat{\theta} - \theta^{*}}\rightarrow N(0, \Sigma_{\theta^*}).$  We now apply the delta method, which leads us to $\sqrt{n + N}\rbr{g(\hat{\theta}) - g\rbr{\theta^{*}}}\rightarrow N(0, \nabla g ^T \Sigma_{\theta^*} \nabla g).$ We note that $g(\hat{\theta}) = \hat{\tau}_{w}.$ By Theorem, $g(\theta^{*}) = \tau_\tca.$  Therefore, we complete the proof. 
\end{proof}

\subsection{Theorem~\ref{thm: cre logistic model asymptotics Hajek} (Asymptotic Distribution of the Weighted Estimator for Unknown Selection Mechanism)} 
\begin{proof} \label{pro: randomization Hajek Estimator asymptotic unknown}
    The proof is similar to that of Theorem~\ref{thm: cre known model asymptotics Hajek}, except that we need to take the unknown $\text{dim}X-$dimensional parameter $\beta$ into account. Since $\hat{\beta}$ is the maximum likelihood estimator, it satisfies the $\text{dim}X-$dimensional estimating equation:
\begin{align*}
    \frac{1}{n + N} \sum_{i=1}^{n + N} \psi (\beta; X_i, S_i) = \frac{1}{n + N} \sum_{i=1}^{n + N} X_i (S_i - \sigma(\beta^ T X_i)) = 0.
\end{align*}
We define $\hat{\theta}$ the same as that in the proof of Theorem~\ref{thm: cre known model asymptotics Hajek}. Therefore, $(\hat{\theta}, \hat{\beta})$ solves the estimation equation:
\begin{align*}
    \frac{1}{n + N}\sum_{i=1}^{n + N} \Phi \left(\theta, \beta; Z_i, X_i, Y_i, D_i, S_i\right) = \frac{1}{n + N} \left(\begin{array}{c}\sum_{i=1}^{n + N} \Psi^{(i)}(\theta) \\ \sum_{i=1}^{n + N} \psi (\beta; X_i, S_i)\end{array}\right) = 0,
\end{align*}
where $\Psi^{(i)}(\theta)$ is defined in the proof of Theorem~\ref{thm: cre known model asymptotics Hajek}. For simplicity, we write $\Phi \left(\theta, \beta; Z_i, X_i, Y_i, D_i, S_i\right)$ as $\Phi^{(i)}(\theta, \beta).$ We define the matrices 
\begin{align*}
    C_{n + N}(\theta, \beta)= \frac{1}{n + N} \sum_{i=1}^{n + N} \EE[\frac{\partial \Phi^{(i)}(\theta, \beta)}{\partial (\theta, \beta)}] \text{ and } D_{n + N}(\theta, \beta) &= \frac{1}{n + N}\sum_{i=1}^{n + N}\text{cov}(\Phi^{(i)}(\theta, \beta)).
\end{align*}
Define $C(\theta, \beta) = \text{lim}_{n + N \rightarrow \infty} C_{n + N}(\theta, \beta)$ and $D(\theta, \beta) = \text{lim}_{n + N \rightarrow \infty} D_{n +N}(\theta, \beta).$ Define $\Sigma_{\theta, \beta} = C^{-1}(\theta, \beta)D(\theta, \beta)C^{-T}(\theta, \beta).$ Let $(\theta^{*}, \beta^{*})$ be the solution of the population estimating equation $\EE[\Phi \left(\theta, \beta; Z, X, Y, D, S\right) = 0.$ Then
\begin{align} \label{eq: weighted covariance matrix unknown weights}
    \Sigma_{\theta^*, \beta^*} = C^{-1}(\theta^*, \beta^*)D(\theta^*, \beta^*)C^{-T}(\theta^*, \beta^*)
\end{align}
is the covariance matrix with $(\theta, \beta)$ replaced by the solution to the population estimation equation, $(\theta^*, \beta^*).$ By the theory of estimating equations (see Theorem 9.3 of \cite{li2019graduate} for details), we derive that 
\begin{align*}
    \sqrt{n + N}\left(\begin{array}{c}\hat{\theta} - \theta^{*} \\ \hat{\beta} - \beta^* \end{array}\right) \rightarrow N(0, \Sigma_{\theta^*, \beta^*}).
\end{align*}
By applying the delta method and the consistency Theorem~\ref{thm: randomization Hajek Estimator consistency}, we conclude that $\sqrt{n+N}(\hat{\tau}_{w} - \tau_\tca) \overset{d}{\rightarrow} N\rbr{0, \nabla g ^T \Sigma_{\theta^*, \beta^*} \nabla g},$ and thus complete the proof.
\end{proof}

\subsection{Theorem \ref{thm: randomization wls Estimator consistency} (Consistency of the Weighted Least Squares Estimator)} 
\begin{proof} \label{pro: randomization wls Estimator consistency}
   By \cite{freedman2008weighting}, we can rewrite the weighted least squares estimator $\hat{\tau}_{\text{wls}}$ in the inverse weighting form:
    \begin{align*}
        \hat{\tau}^{Y}_{\text{wls}} =  \frac{\sum_{i: S_i=1} \hat{w}_i(X_i) Z_i(Y_i - X_i ^ T \hat{\gamma}^Y)} {\sum_{i: S_i=1} \hat{w}_i(X_i) Z_i} - \frac{\sum_{i: S_i=1} \hat{w}_i(X_i) (1 - Z_i)(Y_i -  X_i ^ T \hat{\gamma}^Y)} {\sum_{i: S_i=1} \hat{w}_i(X_i) (1 - Z_i)}.
    \end{align*}
    Rearranging, we obtain

    \begin{align*}
        \hat{\tau}^{Y}_{\text{wls}} = \hat{\tau}^{Y}_{w} -  \Bigg[\frac{\sum_{i: S_i=1} \hat{w}_i(X_i) Z_i(X_i ^ T \hat{\gamma}^Y)} {\sum_{i: S_i=1} \hat{w}_i(X_i) Z_i} - \frac{\sum_{i: S_i=1} \hat{w}_i(X_i) (1 - Z_i)(X_i ^ T \hat{\gamma}^Y)} {\sum_{i: S_i=1} \hat{w}_i(X_i) (1 - Z_i)} \Bigg].
    \end{align*}

    By examing the proof of Theorem~\ref{thm: randomization Hajek Estimator consistency}, $\hat{\tau}^{Y}_{w} \overset{p}{\rightarrow} \EE_{X \given S=0}[\mu_{y1}(X)-\mu_{y0}(X)].$ Denote $\gamma^Y = \text{lim}_{N \rightarrow \infty} \hat{\gamma}^Y.$ By the law of large numbers and the continuous mapping theorem,
   \begin{align} 
        &\frac{\sum_{i: S_i=1} \hat{w}_i(X_i) Z_i(X_i ^ T \hat{\gamma}^Y)} {\sum_{i: S_i=1} \hat{w}_i(X_i) Z_i} - \frac{\sum_{i: S_i=1} \hat{w}_i(X_i) (1 - Z_i)(X_i ^ T \hat{\gamma}^Y)} {\sum_{i: S_i=1} \hat{w}_i(X_i) (1 - Z_i)} \nonumber \\
        &\overset{p}{\rightarrow} \frac{\EE[S w_1(X) Z X^T]\gamma^Y}{\EE[S w_1(X) Z]} - \frac{\EE[S w_0(X) (1 - Z) X^T]\gamma^Y}{\EE[S w_0(X) (1 - Z)]}.\label{eq: consistency wls}
    \end{align}
    Looking at the numerator of the first term on the right-hand side,
    \begin{align*}
        \EE[Sw_1(X) Z X^T]\gamma^Y &= \EE_X\{\EE[Sw_1(X) Z X^T \given X]\}\gamma^Y \nend
        &= \EE_X\{\EE[w_1(X) X^T \given X] \prob(Z=1 \given X) \prob(S=1 \given X) \}\gamma^Y \nend
        &= \EE_X\{\EE[w_1(X) X^T  \prob(Z=1 \given X) \prob(S=1 \given X) \given X]\}\gamma^Y \nend
        &= \EE_X\{\EE[w_1(X) X^T  \prob(Z=1 \given X) \given X]\}\gamma^Y\nend
        &= \EE_X\{\EE[\prob(S=0 \given X) X^T  \given X]\}\gamma^Y \nend
        &= \EE[\prob(S=0 \given X) X^T]\gamma^Y.
    \end{align*}
    Similarly, $\EE[Sw_2(X) (1 - Z) X^T]\gamma^Y = \EE[\prob(S=0 \given X) X^T]\gamma^Y$ and 
    \begin{align*}
        \EE[Sw_1(X) Z] = \EE[Sw_2(X) (1 - Z)] = \EE[\prob(S=0 \given X)].
    \end{align*}
    Therefore, the right-hand side of \eqref{eq: consistency wls} is zero. Consequently, 
    \begin{align*}
        \hat{\tau}^{Y}_{\text{wls}} = \hat{\tau}^{Y}_{w} + o_p(1) \overset{p}{\rightarrow} \EE_{X \given S=0}[\mu_{y1}(X)-\mu_{y0}(X)].
    \end{align*}
    We can use the same technique to show that 
    \begin{align*}
        \hat{\tau}^{D}_{\text{wls}} = \hat{\tau}^{D}_{w} + o_p(1) \overset{p}{\rightarrow} \EE_{X \given S=0}[\mu_{d1}(X)-\mu_{d0}(X)].
    \end{align*}
    By the continuous mapping theorem, we complete the proof by invoking Theorem~\ref{thm: randomization identification tcace}:
    \begin{align*}
        \hat{\tau}_{\text{wls}} = \frac{\hat{\tau}^{Y}_{\text{wls}}}{\hat{\tau}^{D}_{\text{wls}}} \overset{p}{\rightarrow} \frac{ \EE_{X \given S=0}[\mu_{y1}(X)-\mu_{y0}(X)]}{\EE_{X \given S=0}[\mu_{d1}(X)-\mu_{d0}(X)]} = \tau_\tca.
    \end{align*}
\end{proof}

\subsection{Theorem \ref{thm: cre logistic model asymptotics distribution wls} (Asymptotic Distribution of the Weighted Least Squares Estimator for Unknown Selection Mechanism)} 
\begin{proof} \label{pro: cre logistic model asymptotics distribution wls}
    The proof is similar to that of Theorem~\ref{thm: cre logistic model asymptotics Hajek}, except that we need to take the estimated parameter $\hat{\theta}_{\text{wls}}=
    (\hat{\tau}_{\text{wls}}^{Y}, \hat{\tau}_{\text{wls}}^{D}, \hat{\gamma}^Y, \hat{\gamma}^D)$ into account. By taking the partial derivative, we establish the estimating equation for $\hat{\theta}_{\text{wls}} \in \RR^{(2 + 2\text{dim}X)}$ as:
    \begin{align*}
    \frac{1}{n + N}\sum_{i=1}^{n + N} \Psi_{\text{wls}} \left(\theta_{\text{wls}}; Z_i, X_i, Y_i, D_i, S_i \right) = \frac{1}{n + N} \left(\begin{array}{c}
    \sum_{i=1}^{n + N} 2S_i w_i(X_i) (Y_i - \tau_{\text{wls}}^Y Z_i - {\gamma^Y}^T X_i) (-Z_i)\\ 
    \sum_{i=1}^{n + N} 2S_i w_i(X_i) (D_i - \tau_{\text{wls}}^D Z_i - {\gamma^D}^T X_i) (-Z_i)\\ 
    \sum_{i=1}^{n + N} 2S_i w_i(X_i) (Y_i - \tau_{\text{wls}}^Y Z_i - {\gamma^Y}^T X_i) (-X_i)\\ 
    \sum_{i=1}^{n + N} 2S_i w_i(X_i) (D_i - \tau_{\text{wls}}^D Z_i - {\gamma^D}^T X_i) (-X_i)
    \end{array}\right),
\end{align*}
and denote $\theta_{\text{wls}} = (\tau_{\text{wls}}^{Y}, \tau_{\text{wls}}^{D}, \gamma^Y, \gamma^D).$ Note that $\theta_{\text{wls}} = \hat{\theta}_{\text{wls}}$ solves the equations.
Moreover, as in \S\ref{pro: randomization Hajek Estimator asymptotic unknown}, $\hat{\beta}$ satisfies the $\text{dim}X-$dimensional estimating equation:
\begin{align*}
    \frac{1}{n + N} \sum_{i=1}^{n + N} \psi (\beta; X_i, S_i) = \frac{1}{n + N} \sum_{i=1}^{n + N} \tilde{X_i} (S_i - \sigma(\beta^ T \tilde{X_i})) = 0,
\end{align*}
here $\tilde{X}_i$ denotes $X_i$ augmented with a leading column of ones. Therefore, $\rbr{\hat{\theta}_{\text{wls}}, \hat{\beta}}$ solves the estimation equation:
\begin{align*}
    \frac{1}{n + N}\sum_{i=1}^{n + N} \Phi_{\text{wls}} \left(\theta_{\text{wls}}, \beta; Z_i, X_i, Y_i, D_i, S_i\right) = \frac{1}{n + N} \left(\begin{array}{c}\sum_{i=1}^{n + N} \Psi_{\text{wls}}^{(i)}(\theta) \\ \sum_{i=1}^{n + N} \psi (\beta; X_i, S_i)\end{array}\right) = 0,
\end{align*}
For simplicity, we write $\Phi_{\text{wls}} \left(\theta, \beta; Z_i, X_i, Y_i, D_i, S_i\right)$ as $\Phi_{\text{wls}}^{(i)}(\theta, \beta).$ We define the matrices 
\begin{align*}
    E_{n + N}(\theta_{\text{wls}}, \beta)= \frac{1}{n + N} \sum_{i=1}^{n + N} \EE\sbr{\frac{\partial \Phi_{\text{wls}}^{(i)}(\theta_{\text{wls}}, \beta)}{\partial (\theta_{\text{wls}}, \beta)}} \text{ and } F_{n+N}(\theta_{\text{wls}}, \beta) &= \frac{1}{n + N}\sum_{i=1}^{n + N}\text{cov}\rbr{\Phi_{\text{wls}}^{(i)}(\theta_{\text{wls}}, \beta)}.
\end{align*}
Define $E(\theta_{\text{wls}}, \beta) = \text{lim}_{n + N \rightarrow \infty} E_{n + N}(\theta_{\text{wls}}, \beta)$ and $F(\theta_{\text{wls}}, \beta) = \text{lim}_{n + N \rightarrow \infty} F_{n + N}(\theta_{\text{wls}}, \beta).$ Define $\Sigma_{\theta_{\text{wls}}, \beta} = E^{-1}(\theta_{\text{wls}}, \beta)F(\theta_{\text{wls}}, \beta)E^{-T}(\theta_{\text{wls}}, \beta).$ Let $(\theta_{\text{wls}}^{*}, \beta^{*})$ be the solution of the population estimating equation $\EE\sbr{\Phi_{\text{wls}} \rbr{\theta_{\text{wls}}, \beta; Z, X, Y, D, S}} = 0.$ Then
\begin{align} \label{eq: wls covariance matrix unknown weights}
    \Sigma_{\theta_{\text{wls}}^*, \beta^*} = E^{-1}(\theta_{\text{wls}}^*, \beta^*)F(\theta_{\text{wls}}^*, \beta^*)E^{-T}(\theta_{\text{wls}}^*, \beta^*)
\end{align}
is the covariance matrix with $(\theta_{\text{wls}}, \beta)$ replaced by the solution to the population estimation equation, $(\theta_{\text{wls}}^*, \beta^*).$ By the theory of estimating equations (see Theorem 9.3 of \cite{li2019graduate} for details), we derive that 
\begin{align*}
    \sqrt{n + N}\left(\begin{array}{c}\hat{\theta}_{\text{wls}} - \theta_{\text{wls}}^{*} \\ \hat{\beta} - \beta^* \end{array}\right) \rightarrow N\rbr{0, \Sigma_{\theta_{\text{wls}}^*, \beta_{\text{wls}}^*}},
\end{align*}
Let $g_{\text{wls}}(\cdot):\mathbb{R}^{(2 + 2\text{dim}X)}\rightarrow \mathbb{R}$ be defined as $g_{\text{wls}}(\theta_{\text{wls}}) ={{\theta_{\text{wls}}}_1}/{{\theta_{\text{wls}}}_2}.$ By applying the delta method to $g_{\text{wls}}$ and Theorem \ref{thm: randomization wls Estimator consistency}, we conclude that $\sqrt{n + N}(\hat{\tau}_{\text{wls}} - \tau_\tca) \overset{d}{\rightarrow} N\rbr{0, \nabla g_{\text{wls}} ^T \Sigma_{\theta_{\text{wls}}^*, \beta^*} \nabla g_{\text{wls}}},$ and thus complete the proof.
\end{proof}

\subsection{Theorem~\ref{thm: consistency doubly robust estimator} (Consistency Properties of the Multiply Robust Estimator)} 
\begin{proof} \label{pro: consistency doubly robust estimator}
    We now proceed to analyze the two conditions stated in Theorem~\ref{thm: consistency doubly robust estimator} separately. For each case, we provide a detailed explanation of why $\hat{\tau}_{\text{mr}}^{Y} \overset{p}{\rightarrow} \tau_{w}^{Y}.$  $\hat{\tau}_{\text{mr}}^{D} \overset{p}{\rightarrow} \tau_{w}^{D}$ follows similarly. Finally, the overall consistency result is established using the continuous mapping theorem.
    \paragraph{Case 1 when Condition (i) is satisfied.}
    We first consider the case where the model of the study sample selection process is correctly specified. To prove $\hat{\tau}_{\text{mr}}^{Y} \overset{p}{\rightarrow} \tau_{w}^{Y},$ by Theorem~\ref{thm: randomization Hajek Estimator consistency}, it suffices to show that 
\begin{align} \label{eq: consistency sample correctly specified equation 1}
    \frac{\sum_{i: S_i=0} (\hat{\mu}_{y1}(X_i) - \hat{\mu}_{y0}(X_i))} {N} -  \frac{\sum_{i: S_i=1} \hat{w}_1(X_i) Z_i\hat{\mu}_{y1}(X_i)} {\sum_{i: S_i = 1} \hat{w}_1(X_i) Z_i} \nend
    +\frac{ \sum_{i: S_i = 1} \hat{w}_0(X_i) (1-Z_i)\hat{\mu}_{y0}(X_i)} {\sum_{i: S_i = 1} \hat{w}_0(X_i) (1-Z_i)} \overset{p}{\rightarrow} 0.
\end{align}
    By the convergence condition and the law of large numbers, the left-hand side of (\ref{eq: consistency sample correctly specified equation 1}) converges to 
\begin{align} \label{eq: consistency sample correctly specified equation 2}
     \frac{\EE[(1-S)(\tilde{\mu}_{y1}(X) - \tilde{\mu}_{y0}(X))]} {\prob(S=0)} -  \frac{\EE[w_1(X)ZS\tilde{\mu}_{y1}(X)]}{\EE[w_1(X)ZS]} + \frac{\EE[w_0(X)(1-Z)S\tilde{\mu}_{y0}(X)]}{\EE[w_0(X)(1-Z)S]}.
\end{align}
By applying the law of total probability on the support of $S$, the first term of \eqref{eq: consistency sample correctly specified equation 2} converges in probability to $\EE_{X \given S=0}[\tilde{\mu}_{y1}(X)-\tilde{\mu}_{y0}(X)].$ In the proof of Theorem \ref{thm: randomization identification tcace IPW}, we show that \eqref{eq: ipw outcome identification 1} is true, i.e.,
\begin{align*}
    \mathbb{E} \left[ w_1(X)ZSY \right] = \mathbb{P}(S=0) \cdot \mathbb{E}_{X|S=0}\{\mathbb{E} [Y\given Z=1, S=1, X]\}.
\end{align*}
We can show in the same way that 
\begin{align*}
    \EE[w_1(X)ZS\tilde{\mu}_{y1}(X)] &= \mathbb{P}(S=0) \cdot \mathbb{E}_{X|S=0}\{\mathbb{E} [\tilde{\mu}_{y1}(X)\given Z=1, S=1, X]\} \nend
     &= \mathbb{P}(S=0) \cdot \mathbb{E}_{X|S=0}[\tilde{\mu}_{y1}(X)].
\end{align*}
We also have 
\begin{align*}
    \EE[w_0(X)(1-Z)S\tilde{\mu}_{y0}(X)] &= \mathbb{P}(S=0) \cdot \mathbb{E}_{X|S=0}\{\mathbb{E} [\tilde{\mu}_{y0}(X)\given Z=0, S=1, X]\} \nend
     &= \mathbb{P}(S=0) \cdot \mathbb{E}_{X|S=0}[\tilde{\mu}_{y0}(X)].
\end{align*}
In addition, we have 
\begin{align*}
    \EE[w_1(X)ZS] = \EE[w_0(X)(1-Z)S] = \mathbb{P}(S=0)
\end{align*}
from \eqref{eq: ipw outcome identification 2}. Consequently, we prove that \eqref{eq: consistency sample correctly specified equation 2} equals 0, and hence \eqref{eq: consistency sample correctly specified equation 1} is true. As a result, $\hat{\tau}_{\text{mr}}^{Y} \overset{p}{\rightarrow} \tau_{w}^{Y}.$ $\hat{\tau}_{\text{mr}}^{D} \overset{p}{\rightarrow} \tau_{w}^{D}$ follows in a similar way. We conclude that $\hat{\tau}_{\text{mr}} \overset{p}{\rightarrow}  \tau_{w}^{Y}/{\tau_{w}^{D}} = \tau_\tca$ using Corollary \ref{thm: randomization identification tcace IPW} and the continuous mapping theorem.

\paragraph{Case 2 when Condition (ii) is satisfied.}
Condition (ii) states that both the outcome and treatment received models are correctly specified. Then by the law of large numbers, we have 
\begin{align*}
    \frac{ \sum_{i: S_i=0} (\hat{\mu}_{y1}(X_i) - \hat{\mu}_{y0}(X_i))} {N} \overset{p}{\rightarrow} \EE_{X \given S=0}[\mu_{y1}(X)-\mu_{y0}(X)] = \tau_{w}^{Y},
\end{align*}
where the equality holds by Theorem~\ref{thm: randomization identification tcace}. Therefore, to prove $\hat{\tau}_{\text{mr}}^{Y} \overset{p}{\rightarrow} \tau_{w}^{Y},$ it suffices to show that 

\begin{align} \label{eq: consistency outcome correctly specified equation 1}
   \frac{\sum_{S_i: i=1} \hat{w}_1(X_i) Z_i(Y_i - \hat{\mu}_{y1}(X_i))} {\sum_{i: S_i=1} \hat{w}_1(X_i) Z_i} - \frac{\sum_{S_i: i=1} \hat{w}_0(X_i) (1-Z_i)(Y_i - \hat{\mu}_{y0}(X_i))} {\sum_{S_i: i=1} \hat{w}_0(X_i) (1-Z_i)} \overset{p}{\rightarrow} 0.
\end{align}

We now show that the first term of \eqref{eq: consistency outcome correctly specified equation 1} converges to zero in probability. It is then easy to see that the second term converges to zero in probability in a similar way. By Condition (ii) and the law of large numbers, we have
\begin{align} \label{eq: consistency outcome correctly specified equation 2}
 \frac{1}{n + N} \sum_{S_i: i=1} \hat{w}_1(X_i) Z_i(Y_i - \hat{\mu}_{y1}(X_i)) {\rightarrow} \EE[\tilde{w}_1(X)ZS(Y - \mu_{y1}(X))].
\end{align}
By the same steps to derive \eqref{eq: ipw outcome identification 1}, we can rewrite the right-hand side of \eqref{eq: consistency outcome correctly specified equation 2} as:

\begin{align*}
    \EE[\tilde{w}_1(X)ZS(Y - \mu_{y1}(X))] &= \mathbb{P}(S=0) \cdot \mathbb{E}_{X|S=0}\{\mathbb{E} [ Y- \mu_{y1}(X)\given Z=1, S=1, X]\},
\end{align*}
which is zero by definition of $\mu_{y1}(X).$ Therefore, 
\begin{align*} 
 \frac{1}{n + N} \sum_{S_i: i=1} \hat{w}_1(X_i) Z_i(Y_i - \hat{\mu}_{y1}(X_i)) {\rightarrow} 0
\end{align*}
by \eqref{eq: consistency outcome correctly specified equation 2}. Hence the first term of \eqref{eq: consistency outcome correctly specified equation 1} converges to zero in probability. Similarly, the second term of \eqref{eq: consistency outcome correctly specified equation 1} converges to zero in probability. Consequently, we show that \eqref{eq: consistency outcome correctly specified equation 1} holds, and $\hat{\tau}_{\text{mr}}^{Y} \overset{p}{\rightarrow} \tau_{w}^{Y}.$ Since $\hat{\tau}_{\text{mr}}^{D} \overset{p}{\rightarrow} \tau_{w}^{D}$ follows in a similar way, we conclude the proof using Corollary \ref{thm: randomization identification tcace IPW} and the continuous mapping theorem.
\end{proof}

\section{Variance Estimators and Confidence Intervals for the Weighted Estimator}\label{app: cre confidence intervals}

\subsection{Construction of the Variance Estimator}
We give the details of the construction of the sandwich-type variance estimator and the Wald-type confidence intervals. Recall from \S\ref{pro: randomization Hajek Estimator asymptotic unknown}, the asymptotic variance is 
\begin{align} \label{eq: cre asy variance unknown}
    \nabla g ^T \Sigma_{\theta^*, \beta^*} \nabla g = \nabla g ^T C^{-1}(\theta^*, \beta^*)D(\theta^*, \beta^*)C^{-T}(\theta^*, \beta^*) \nabla g.
\end{align}
Since $(\theta^*, \beta^*)$ is unknown, we estimate the right-hand side of \eqref{eq: cre asy variance unknown} term by term, using their consistent empirical estimators.

\paragraph{$\mathbf{\nabla g ^T}$:} By definition, 
\begin{align*}
    g(\theta, \beta) = \frac{\theta_1 \theta_3^{-1} - \theta_2 \theta_4^{-1}}{\theta_5 \theta_3^{-1} - \theta_6 \theta_4^{-1}}.
\end{align*}
Denote $\Delta_g^{(1)} = \theta_1 \theta_3^{-1} - \theta_2 \theta_4^{-1}$ and $\Delta_g^{(2)} = \theta_5 \theta_3^{-1} - \theta_6 \theta_4^{-1}.$ By direct calculation, \begin{align*}
    \nabla g = \left(\begin{array}{c}
    \theta_3^{-1}/\Delta_g^{(2)}\\ 
    -\theta_4^{-1}/\Delta_g^{(2)}\\ 
    \rbr{-\Delta_g^{(2)} \theta_1 \theta_3^{-2} + \Delta_g^{(1)} \theta_5 \theta_3^{-2}}/\rbr{\Delta_g^{(2)}}^2\\ 
    \rbr{\Delta_g^{(2)} \theta_2 \theta_4^{-2} - \Delta_g^{(1)} \theta_6 \theta_4^{-2}}/\rbr{\Delta_g^{(2)}}^2\\ 
    -\Delta_g^{(1)} \theta_3^{-1} / \rbr{\Delta_g^{(2)}}^2\\
    \Delta_g^{(1)} \theta_4^{-1} / \rbr{\Delta_g^{(2)}}^2
    \end{array}\right).
\end{align*}

\paragraph{$\boldsymbol{C}(\boldsymbol{\theta}^*, \boldsymbol{\beta}^*)$:}
Since $C(\theta^*, \beta^*)=  \text{lim}_{{n + N} \rightarrow \infty}\frac{1}{n + N} \sum_{i=1}^{n + N} \EE\sbr{\frac{\partial \Phi^{(i)}(\theta^*, \beta^*)}{\partial (\theta^*, \beta^*)}},$ we use 
\begin{align*}
    \hat{C}_{n + N}(\hat{\theta}, \hat{\beta})= \frac{1}{n + N} \sum_{i=1}^{n + N} \frac{\partial \Phi^{(i)}\rbr{\hat{\theta}, \hat{\beta}}}{\partial \rbr{\hat{\theta}, \hat{\beta}}}
\end{align*}
to estimate it. The definition of $\Phi^{(i)}$ can be found at \S\ref{pro: randomization Hajek Estimator asymptotic} and \S\ref{pro: randomization Hajek Estimator asymptotic unknown}. By taking the partial derivatives, we know for any $i \in [n], j \in [\text{dim}X_i], k \in [\text{dim}X_i],$ we have
\begin{align*}
    \frac{\partial X_{ij}(S_i - \sigma(\beta^T X_i))}{\partial \beta_k} &= \sigma(\beta^TX_i)(\sigma(\beta^TX_i)-1)X_{ij}X_{ik}, \text{ and}\nend
  \frac{\partial(1 - \sigma(\beta^TX_i))/\sigma(\beta^TX_i)}{\partial \beta_k} &= (1 - \sigma^{-1}(\beta^TX_i))X_{ik}.
\end{align*}
As a result, we deduce that
\begin{align*}
\hat{C}_{n + N}(\hat{\theta}, \hat{\beta}) &= \frac{1}{n + N} \sum_{i=1}^{n + N} \begin{pmatrix}
    - I_{6 \times 6} & \Theta^{(i)} \\
    0_{p \times 6} & K^{(i)}
\end{pmatrix},
\end{align*}
where the matrices $\Theta^{(i)} \in \RR^{6 \times \text{dim}X_i}$ and $K^{(i)} \in \RR^{\text{dim}X_i \times \text{dim}X_i}$ are defined as 
\begin{align*}
K_{k,j}^{(i)} &= \sbr{\sigma\rbr{\beta^T X_i^{(j)}} \rbr{\sigma\rbr{\beta^T X_i^{(j)}} - 1}} X_{ik} X_{ij}, \nend
\Theta^{(i)}_{1k} &=\sum_{i: S_i=1} \frac{Z_i  Y_i}{\prob(Z_i = 1 \given S_i=1, X_i)} \cdot \rbr{1 - \sigma\rbr{\beta^T X_i}} X_{ik}, \nend
\Theta^{(i)}_{2k} &=\sum_{i: S_i=1} \frac{(1-Z_i) Y_i}{\prob(Z_i = 0 \given S_i=1, X_i)} \cdot \rbr{1 - \sigma\rbr{\beta^T X_i}} X_{ik}, \nend
\Theta^{(i)}_{3k} &=\sum_{i: S_i=1} \frac{Z_i}{\prob(Z_i = 1 \given S_i=1, X_i)} \cdot\rbr{1 - \sigma\rbr{\beta^T X_i}} X_{ik}, \nend
\Theta^{(i)}_{4k} &=\sum_{i: S_i=1} \frac{(1-Z_i)}{\prob(Z_i = 0 \given S_i=1, X_i)} \cdot \rbr{1 - \sigma\rbr{\beta^T X_i}} X_{ik}, \nend
\Theta^{(i)}_{5k} &=\sum_{i: S_i=1} \frac{Z_iD_i}{\prob(Z_i = 1 \given S_i=1, X_i)} \cdot \rbr{1 - \sigma\rbr{\beta^T X_i}} X_{ik}, \nend
\Theta^{(i)}_{6k} &=\sum_{i: S_i=1} \frac{(1-Z_i) D_i}{\prob(Z_i = 0 \given S_i=1, X_i)} \cdot \rbr{1 - \sigma\rbr{\beta^T X_i}} X_{ik}.
\end{align*}

\paragraph{$\boldsymbol{D}(\boldsymbol{\theta}^*, \boldsymbol{\beta}^*)$:}
Recall $D(\theta^*, \beta^*)=  \text{lim}_{ {n+N} \rightarrow \infty}\frac{1}{n + N} \sum_{i=1}^{n + N} \text{cov}\rbr{\Phi^{(i)}(\theta^*, \beta^*)}.$ Define the within-sample means
\begin{align*}
    \bar{\Phi}_{1}(\hat{\theta},\hat{\beta})
    &:=
    \frac{1}{n}\sum_{i:S_i=1}
    \Phi^{(i)}(\hat{\theta},\hat{\beta}),
    \\
    \bar{\Phi}_{0}(\hat{\theta},\hat{\beta})
    &:=
    \frac{1}{N}\sum_{i:S_i=0}
    \Phi^{(i)}(\hat{\theta},\hat{\beta}).
\end{align*}
Then we estimate \(D(\theta^*,\beta^*)\) by
\begin{align*}
    \hat{D}_{n+N}(\hat{\theta},\hat{\beta})
    &:=
    \frac{1}{n+N}
    \sum_{i:S_i=1}
    \left\{
    \Phi^{(i)}(\hat{\theta},\hat{\beta})
    -
    \bar{\Phi}_{1}(\hat{\theta},\hat{\beta})
    \right\}
    \left\{
    \Phi^{(i)}(\hat{\theta},\hat{\beta})
    -
    \bar{\Phi}_{1}(\hat{\theta},\hat{\beta})
    \right\}^{T}
    \\
    &\quad+
    \frac{1}{n+N}
    \sum_{i:S_i=0}
    \left\{
    \Phi^{(i)}(\hat{\theta},\hat{\beta})
    -
    \bar{\Phi}_{0}(\hat{\theta},\hat{\beta})
    \right\}
    \left\{
    \Phi^{(i)}(\hat{\theta},\hat{\beta})
    -
    \bar{\Phi}_{0}(\hat{\theta},\hat{\beta})
    \right\}^{T}.
\end{align*}

Combine all terms, a sandwich-type variance estimator is given by 
\begin{align} \label{eq: cre sandwich estimator}
    \nabla g ^T \Sigma_{\theta^*, \beta^*} \nabla g \approx \nabla g ^T \hat{C}_{n + N}^{-1}(\hat{\theta}, \hat{\beta})\hat{D}_{n + N}(\hat{\theta}, \hat{\beta})\hat{C}_{n + N}^{-T}(\hat{\theta}, \hat{\beta}) \nabla g.
\end{align}

Denote $\hat{\Sigma}_{\hat{\theta}, \hat{\beta}} = \hat{C}_{n + N}^{-1}(\hat{\theta}, \hat{\beta})\hat{D}_{n + N}(\hat{\theta}, \hat{\beta})\hat{C}_{n + N}^{-T}(\hat{\theta}, \hat{\beta}),$ a $\alpha \%$ Wald-type confidence interval is given by
\begin{align*}
    \sbr{\hat{\tau}_{w} - z_{1- \alpha/2} \sqrt{\frac{1}{n + N} \nabla g^\top \hat{\Sigma}_{\hat{\theta}, \hat{\beta}} \nabla g},\quad \hat{\tau}_{w} + z_{1- \alpha/2} \sqrt{\frac{1}{n + N} \nabla g^\top \hat{\Sigma}_{\hat{\theta}, \hat{\beta}} \nabla g} },
\end{align*}
where $z_{1- \alpha/2}$ stands for the $1-\alpha/2$-th quantile of a standard Gaussian distribution.

\subsection{Consistency of the Sandwich-Type Variance Estimator}

We now show that the sandwich-type variance estimator in
\eqref{eq: cre sandwich estimator} is consistent for the asymptotic variance in
\eqref{eq: cre asy variance unknown}. Let
\begin{align*}
    \eta := (\theta^\top,\beta^\top)^\top,
    \qquad
    \eta^* := (\theta^{*\top},\beta^{*\top})^\top .
\end{align*}
Write \(\Phi_i(\eta)\) for the stacked estimating function
\(\Phi^{(i)}(\theta,\beta)\) used in the proof of Theorem~\ref{thm: cre logistic model asymptotics Hajek}. Thus
\begin{align*}
    \frac{1}{n+N}\sum_{i=1}^{n+N}\Phi_i(\hat\eta)=0,
    \qquad
    \hat\eta=(\hat\theta^\top,\hat\beta^\top)^\top.
\end{align*}

Recall that the \(n\) observations with \(S_i=1\)
and the \(N\) observations with \(S_i=0\) are independent samples from
possibly different distributions. Define
\begin{align*}
    \bar\Phi_{1}(\eta)
    &:=
    \frac{1}{n}\sum_{i:S_i=1}\Phi_i(\eta),
    \qquad
    \bar\Phi_{0}(\eta)
    :=
    \frac{1}{N}\sum_{i:S_i=0}\Phi_i(\eta).
\end{align*}

\begin{assumption}[Regularity Conditions for Sandwich Consistency under the Two-Sample Design]
\label{ass: sandwich consistency}
In addition to the assumptions of Theorem~\ref{thm: cre logistic model asymptotics Hajek}, suppose there exist integrable envelope functions \(M_0(O_i)\), \(M_1(O_i)\),
    and \(M_2(O_i)\) such that, for all \(\eta\in\mathcal N\),
    \begin{align*}
        ||\Phi_i(\eta)||
        &\leq M_0(O_i),
        \\
        \Bigl\lVert \frac{\partial \Phi_i(\eta)}{\partial\eta^\top}\Bigr\rVert
        &\leq M_1(O_i),
        \\
        \nbr{
        \Phi_i(\eta)\Phi_i(\eta)^\top}
        &\leq M_2(O_i),
    \end{align*}
    with
    \begin{align*}
        \EE\sbr{M_k(O_i)\mid S_i=1}<\infty,
        \qquad
        \EE\sbr{M_k(O_i)\mid S_i=0}<\infty,
        \qquad
        k=0,1,2.
    \end{align*}
    Here \(O_i\) denotes the observed tuple
    \((S_i,X_i,S_iZ_i,S_iD_i,S_iY_i)\).
\end{assumption}

\begin{theorem}[Consistency of the Sandwich-Type Variance Estimator]
\label{thm: sandwich consistency}
Define
\begin{align*}
    \hat V_{n+N}
    :=
    \nabla g(\hat\theta)^\top
    \hat C_{n+N}^{-1}(\hat\theta,\hat\beta)
    \hat D_{n+N}(\hat\theta,\hat\beta)
    \hat C_{n+N}^{-T}(\hat\theta,\hat\beta)
    \nabla g(\hat\theta),
\end{align*}
Under Assumption~\ref{ass: sandwich consistency},
\begin{align*}
    \hat V_{n+N}
    \overset{p}{\rightarrow}
    V
    :=
    \nabla g(\theta^*)^\top
    C^{-1}(\theta^*,\beta^*)
    D(\theta^*,\beta^*)
    C^{-T}(\theta^*,\beta^*)
    \nabla g(\theta^*),
\end{align*}
Consequently,
\begin{align*}
    \frac{\hat V_{n+N}}{n+N}
\end{align*}
is a consistent estimator of the asymptotic variance of
\(\hat\tau_w=g(\hat\theta)\).
\end{theorem}

\begin{proof}
The proof proceeds by showing consistency of each component of the sandwich
expression. First, by Theorem~\ref{thm: cre logistic model asymptotics Hajek},
\begin{align*}
    \hat\eta \overset{p}{\rightarrow}\eta^*.
\end{align*}
Therefore, for any open neighborhood \(\mathcal N\) of \(\eta^*\),
\begin{align*}
    P(\hat\eta\in\mathcal N)\to 1.
\end{align*}

We first record the two-sample uniform law of large numbers that will be used
repeatedly. Let \(h_i(\eta)\) denote any entry of
\(\Phi_i(\eta)\), of
\(\partial\Phi_i(\eta)/\partial\eta^\top\), or of
\(\Phi_i(\eta)\Phi_i(\eta)^\top\). Under
Assumption~\ref{ass: sandwich consistency},
\begin{align}
    \sup_{\eta\in\mathcal N}
    \nbr{\frac{1}{n+N}\sum_{i=1}^{n+N} h_i(\eta)
    -
    \frac{n}{n+N}\,\EE\sbr{h_i(\eta)\mid S_i=1}
    -
    \frac{N}{n+N}\,\EE\sbr{h_i(\eta)\mid S_i=0}}
    \overset{p}{\rightarrow}0.
    \label{eq: two sample ulln}
\end{align}
To see this, write the left-hand side as the weighted sum of the two empirical
processes from the study sample and the target-population sample:
\begin{align*}
    &\frac{1}{n+N}\sum_{i=1}^{n+N} h_i(\eta)
    -
    \frac{n}{n+N}\,\EE\sbr{h_i(\eta)\mid S_i=1}
    -
    \frac{N}{n+N}\,\EE\sbr{h_i(\eta)\mid S_i=0}
    \\
    &=
    \frac{n}{n+N}
    \left[
    \frac{1}{n}\sum_{i:S_i=1} h_i(\eta)
    -
    \EE\sbr{h_i(\eta)\mid S_i=1}
    \right]
    \\
    &\quad+
    \frac{N}{n+N}
    \left[
    \frac{1}{N}\sum_{i:S_i=0} h_i(\eta)
    -
    \EE\sbr{h_i(\eta)\mid S_i=0}
    \right].
\end{align*}
Each term converges uniformly to zero by the uniform laws of large numbers (see Lemma~2.4 of
\citet{newey1994large}), applied separately to the two i.i.d. samples.
This proves \eqref{eq: two sample ulln}.

We first prove that
\begin{align*}
    \hat C_{n+N}(\hat\theta,\hat\beta)
    \overset{p}{\rightarrow}
    C(\theta^*,\beta^*).
\end{align*}
By definition,
\begin{align*}
    \hat C_{n+N}(\theta,\beta)
    =
    \frac{1}{n+N}
    \sum_{i=1}^{n+N}
    \frac{\partial\Phi_i(\eta)}{\partial\eta^\top}.
\end{align*}
Applying \eqref{eq: two sample ulln} to the entries of
\(\partial\Phi_i(\eta)/\partial\eta^\top\) gives
\begin{align*}
    \sup_{\eta\in\mathcal N}
    \nbr{
    \hat C_{n+N}(\theta,\beta)
    -
    C(\theta,\beta)}
    \overset{p}{\rightarrow}0.
\end{align*}
On the event \(\{\hat\eta\in\mathcal N\}\), we therefore have
\begin{align*}
    &
    \nbr{
    \hat C_{n+N}(\hat\theta,\hat\beta)
    -
    C(\theta^*,\beta^*)}
    \\
    &\leq
    \sup_{\eta\in\mathcal N}
    \nbr{
    \hat C_{n+N}(\theta,\beta)
    -
    C(\theta,\beta)}
    +
    \nbr{
    C(\hat\theta,\hat\beta)
    -
    C(\theta^*,\beta^*)}.
\end{align*}
The first term is \(o_p(1)\) by the two-sample uniform law of large numbers.
The second term is \(o_p(1)\) by
\(\hat\eta\overset{p}{\rightarrow}\eta^*\) and the continuity of
\((\theta,\beta)\mapsto C(\theta,\beta)\). Hence
\begin{align*}
    \hat C_{n+N}(\hat\theta,\hat\beta)
    \overset{p}{\rightarrow}
    C(\theta^*,\beta^*).
\end{align*}
Since \(C(\theta^*,\beta^*)\) is nonsingular, the continuous mapping theorem
implies
\begin{align*}
    \hat C_{n+N}^{-1}(\hat\theta,\hat\beta)
    \overset{p}{\rightarrow}
    C^{-1}(\theta^*,\beta^*),
    \qquad
    \hat C_{n+N}^{-T}(\hat\theta,\hat\beta)
    \overset{p}{\rightarrow}
    C^{-T}(\theta^*,\beta^*).
\end{align*}

Next, we prove that
\begin{align*}
    \hat D_{n+N}(\hat\theta,\hat\beta)
    \overset{p}{\rightarrow}
    D(\theta^*,\beta^*).
\end{align*}
For \(s\in\{0,1\}\), define
\begin{align*}
    \mu_s(\eta)
    &:=
    \EE\sbr{\Phi_i(\eta)\mid S_i=s},
    \\
    Q_s(\eta)
    &:=
    \EE\sbr{\Phi_i(\eta)\Phi_i(\eta)^\top\mid S_i=s}.
\end{align*}
Also define the corresponding empirical quantities
\begin{align*}
    \hat Q_1(\eta)
    &:=
    \frac{1}{n}\sum_{i:S_i=1}
    \Phi_i(\eta)\Phi_i(\eta)^\top,
    \qquad
    \hat Q_0(\eta)
    :=
    \frac{1}{N}\sum_{i:S_i=0}
    \Phi_i(\eta)\Phi_i(\eta)^\top .
\end{align*}
Then we have
\begin{align*}
    \hat D_{n+N}(\theta,\beta)
    &=
    \frac{n}{n+N}
    \left\{
    \hat Q_1(\eta)
    -
    \bar\Phi_1(\eta)\bar\Phi_1(\eta)^\top
    \right\}
    \\
    &\quad+
    \frac{N}{n+N}
    \left\{
    \hat Q_0(\eta)
    -
    \bar\Phi_0(\eta)\bar\Phi_0(\eta)^\top
    \right\}.
\end{align*}
By the two-sample uniform law of large numbers, applied separately to the entries
of \(\Phi_i(\eta)\) and \(\Phi_i(\eta)\Phi_i(\eta)^\top\),
\begin{align*}
    \sup_{\eta\in\mathcal N}
    \nbr{
    \bar\Phi_s(\eta)-\mu_s(\eta)}
    &\overset{p}{\rightarrow}0,
    \\
    \sup_{\eta\in\mathcal N}
      \nbr{
    \hat Q_s(\eta)-Q_s(\eta)
    }
    &\overset{p}{\rightarrow}0,
    \qquad s\in\{0,1\}.
\end{align*}
Therefore, on the event \(\{\hat\eta\in\mathcal N\}\),
\begin{align*}
    \hat Q_s(\hat\eta)
    -
    \bar\Phi_s(\hat\eta)\bar\Phi_s(\hat\eta)^\top
    \rightarrow{p}
    Q_s(\eta^*)-\mu_s(\eta^*)\mu_s(\eta^*)^\top,
    \qquad s\in\{0,1\},
\end{align*}
where we also used \(\hat\eta\overset{p}{\rightarrow}\eta^*\) and the continuity of
\(\mu_s(\eta)\) and \(Q_s(\eta)\), which follows from dominated convergence
under Assumption~\ref{ass: sandwich consistency}. Since
\begin{align*}
    Q_s(\eta^*)-\mu_s(\eta^*)\mu_s(\eta^*)^\top
    =
    \operatorname{cov}\{\Phi_i(\eta^*)\mid S_i=s\},
\end{align*}
we obtain
\begin{align*}
    \hat D_{n+N}(\hat\theta,\hat\beta)
    &\overset{p}{\rightarrow}
    \frac{n}{n+N}\,
    \operatorname{cov}\{\Phi_i(\theta^*,\beta^*)\mid S_i=1\}
    \\
    &\quad+
    \frac{N}{n+N}\,
    \operatorname{cov}\{\Phi_i(\theta^*,\beta^*)\mid S_i=0\}
    \\
    &=
    D(\theta^*,\beta^*).
\end{align*}

Finally, since \(\hat\theta\overset{p}{\rightarrow}\theta^*\),
\begin{align*}
    \nabla g(\hat\theta)
    \overset{p}{\rightarrow}
    \nabla g(\theta^*).
\end{align*}
Combining the preceding convergences and applying the continuous mapping
theorem gives
\begin{align*}
    \hat V_{n+N}
    &=
    \nabla g(\hat\theta)^\top
    \hat C_{n+N}^{-1}(\hat\theta,\hat\beta)
    \hat D_{n+N}(\hat\theta,\hat\beta)
    \hat C_{n+N}^{-T}(\hat\theta,\hat\beta)
    \nabla g(\hat\theta)
    \\
    &\overset{p}{\rightarrow}
    \nabla g(\theta^*)^\top
    C^{-1}(\theta^*,\beta^*)
    D(\theta^*,\beta^*)
    C^{-T}(\theta^*,\beta^*)
    \nabla g(\theta^*)
    =
    V.
\end{align*}

Since Theorem~\ref{thm: cre logistic model asymptotics Hajek} gives
\begin{align*}
    \sqrt{n+N}\rbr{\hat\tau_w-\tau_{\mathrm{T\text{-}CACE}}}
    \xrightarrow{d}
    N(0,V),
\end{align*}
we conclude that $\frac{\hat V_{n+N}}{n+N}$ consistently estimates the asymptotic variance of. Although the proof of
consistency is given for the weighted estimator under an unknown selection
mechanism, the result for the weighted estimator under a known selection
mechanism follows in a closely analogous manner, with the estimating equation
for \(\beta\) removed.
\end{proof}

\section{Proofs in Section~\ref{sec: practical considerations}} 
\subsection{Identification of $\hat{\tau}_{\text{w-pc}}$}\label{app: partial compliance information identification}

\begin{assumption}[Assumptions for Identification of  $\hat{\tau}_{\text{w-pc}}$] \label{asp: partial compliance information}
\mbox{}
\begin{enumerate} 
    \item[(a)] $Z \indep \{X, Y(0), Y(1), D(0), D(1)\}.$ (i.e., randomized treatment assignment)
    \item[(b)] $\prob(D(0) = 1) = 0.$ (i.e., no always takers)
    \item[(c)] Mean exchangeability of selection and treatment effect heterogeneity (i.e., Assumption~\ref{asp: cond_ign_selection}) holds.
    \item[(d)] The monotonicity and valid instrument assumptions (i.e., Assumption~\ref{asp: IV} (a)-(b)) hold.
    \item[(g)] Overlap (i.e., Assumption~\ref{asp: cre overlap}) holds.
    \item[(h)] $\mathbb{P}(S = 1) = \frac{n}{n+N}$ is fixed, so that the ratio between $n$ and $N$ remains constant as $n \to \infty$ and $N \to \infty$.
\end{enumerate}
\end{assumption}
\begin{theorem}[Consistency of $\hat{\tau}_{\text{w-pc}}$]\label{thm: consistency partial information}
Under Assumption \ref{asp: partial compliance information}:
\begin{align*}
    \hat{\tau}_{\text{w-pc}} \overset{p}{\rightarrow} \tau_\tca.
\end{align*}

\begin{proof}
Recall that we define 

\begin{align*} 
    \hat{\tau}_{\text{w-pc}} = \hat{\tau}_{w}^{Y} \times \frac{\sum_{i: S_i =0}Z_i} {\sum_{i: S_i =0}D_i}.
\end{align*}

By inspecting the proof of Theorem 3.1, Corollary \ref{thm: randomization identification tcace IPW} and Theorem~\ref{thm: randomization Hajek Estimator consistency}, we see that $ \hat{\tau}_{w}^{Y} \overset{p}{\rightarrow} \EE\sbr{Y(1) - Y(0)\given S=0, C = 1}\prob(C=1 \given S=0).$ We also have $\prob(C=1 \given S=0) =  \prob(C=1 \given S=0, Z=1)$ by randomized treatment assignment. Then, by the weak law of large numbers, we know $\sum_{i: S_i =0}Z_i/ {\sum_{i: S_i =0}D_i} \overset{p}{\rightarrow} 1/ \prob(C=1 \given S=0, Z=1) = 1/ \prob(C=1 \given S=0).$ We conclude the proof using the continuous mapping theorem.
\end{proof}
\end{theorem}

\subsection{Consistency of the Weighted Estimator with Unmeasured Confounders} 
\begin{corollary} \label{cor: randomization Hajek Estimator consistency unmeasured covariates}
Assume that $\mathbb{P}(S = 1) = \frac{n}{n+N}$ is fixed, so that the ratio between $n$ and $N$ remains constant as $n \to \infty$ and $N \to \infty$. Suppose that treatment ignorability (i.e., Assumption~\ref{asp: randomization}) and monotonicity and valid instrument (i.e., Assumption~\ref{asp: IV} (a)-(b)) hold. Suppose Assumption~\ref{asp: cond_ign_selection unmeasured covariates} and Assumption~\ref{asp: randomized treatment received exchangeability unmeasured covariates} hold. Denote $\hat{w}^{*}_1(x, u) = \frac{\hat{\prob}(S = 0 \given x, u)}{\hat{\prob}(S = 1 \given x, u)\hat{\prob}(Z = 1 \given S=1, x, u)}$ and $\hat{w}^{*}_0(x, u) = \frac{\hat{\prob}(S = 0 \given x, u)}{\hat{\prob}(S = 1 \given x, u)\hat{\prob}(Z = 0 \given S=1, x, u)}$ be the weights that account for the unmeasured covariates. Let $w^{*}_1(x, u)$ and $w^{*}_0(x, u)$ be their population counterparts. Suppose $\sup_{x,u \in \cX \times \cU} |\hat{w}^{*}_1(x, u) - w^{*}_1(x, u)| = o_p(1)$ and $\sup_{x,u \in \cX \times \cU} |\hat{w}^{*}_0(x, u) - w^{*}_0(x, u)| = o_p(1).$ We have
\begin{align*}
    \hat{\tau}(\hat{w}^{*}) \overset{p}{\rightarrow} \tau_\tca.
\end{align*}
    \begin{proof}
        The proof is identical to that of Theorem~\ref{thm: randomization Hajek Estimator consistency}.

    \end{proof}
\end{corollary}

\subsection{Sensitivity Analysis Optimization via Linear Programming} \label{app: sensitivity}
In general, it is implausible to directly test the mean exchangeability assumptions (i.e., Assumption~\ref{asp: cond_ign_selection} and Assumption~\ref{asp: randomized treatment received exchangeability}). The challenges in doing so are similar to those involved in developing tests for treatment ignorability (i.e., Assumption~\ref{asp: randomization}). In both cases, the assumptions impose restrictions on unobserved potential outcomes and potential treatment received rather than observed data. These difficulties are even more severe for the exchangeability assumptions. First, the target population contains only pre-treatment covariates, so treatment assignment, treatment received, and outcomes are not observed in the target population. Second, the first-stage exchangeability assumption involves latent compliance behavior, which is not directly observed for any individual and is especially difficult to compare across populations. For this reason, we focus primarily on a sensitivity analysis to evaluate the robustness of the underlying results to potential violations in mean exchangeability. 

For a fixed $\Gamma$ value, we can bound the range of possible values of the T-CACE:
\begin{equation} 
\tau_{\textsc{T-CACE}} \in \sbr{\min_{\tilde w \in \varepsilon(\Gamma)} \tau(\tilde w), \ \  \max_{\tilde w \in \varepsilon(\Gamma)} \tau(\tilde w)},
\label{eqn:partial_id}
\end{equation}
where $\tau(\tilde w)$ represents the population-level estimator $\tau(\tilde w) = \frac{\E(\tilde w SY \mid Z =1)/\E(\tilde w S\mid Z =1) - \E(\tilde w S Y \mid Z = 0)/\E(\tilde w S \mid Z = 0)}{\E(\tilde w S D \mid Z = 1)/\E(\tilde w S \mid Z = 1) - \E(\tilde w S D \mid Z = 0)/\E(\tilde w S \mid Z = 0)}$. We use $\tau^Y(\tilde w)$ to represent the numerator, and $\tau^D(\tilde w)$ to represent the denominator. We define the empirical counterpart of $\tau^Y(\tilde w)$ 
\begin{align} \label{eq: proof IPW in ratio q 1}
    \hat{\tau}^Y(r) = \frac{\sum_{i:S_i=1, Z_i=1} r_i \hat{w}_i(X_i) Y_i} {\sum_{i:S_i=1, Z_i=1}  r_i \hat{w}_i(X_i) } - \frac{\sum_{i:S_i=1, Z_i=0}  r_i \hat{w}_i(X_i) Y_i} {\sum_{i:S_i=1, Z_i = 0}  r_i \hat{w}_i(X_i)},
\end{align}
and $ \cR_{\Gamma} = \left\{  r ~ : ~ \Gamma^{-1} \leq r_i \leq \Gamma \text{ for all } i \right\}.$

Consequently, $\min / \max_{r \in \cR_{\Gamma}}\hat{\tau}^Y(r)$ can be achieved by separately solving one minimization/maximization problem for $\{r_i: S_i =1, Z_i = 1 \}$ and one minimization/maximization for $\{r_i: S_i=1, Z_i = 0\}$ over the two terms in \eqref{eq: proof IPW in ratio q 1}. For the first term, we introduce a new variable $t>0,$ and let $\bar{r} = tr.$ We further impose the constraint $\sum_{i: S_i = 1, Z_i=1}\bar{r}_i\hat{w}_i=1.$ It is direct to check that 
\begin{align*}
    \min / \max_{r \in \cR_{\Gamma}} \frac{\sum_{i: Z_i=1} r_i \hat{w}_i Y_i} {\sum_{i: Z_i=1} r_i \hat{w}_i}
\end{align*}
is equivalent to the following linear programming:
\begin{align*}
    \text{minimize or maximize }&\sum_{i: S_i=1, Z_i=1} \bar{r}_i \hat{w}_i Y_i, \nend
    \text{ subject to } & t > 0, \Gamma^{-1} \leq r_i \leq \Gamma \text{ for all $i$ and} \nend
    & \sum_{i: S_i=1, Z_i=1}\bar{r}_i\hat{w}_i=1.
\end{align*}

We can similarly reformulate the minimization and maximization problem of the second term in $\hat{\tau}^{Y}(r)$ as a linear programming problem in a similar fashion. Thus, solving $\min_{r \in \cR_{\Gamma}}\hat{\tau}^Y(r)$ and $\max_{r \in \cR_{\Gamma}}\hat{\tau}^Y(r)$ reduces to solving four linear programming problems. The same approach applies to $\hat{\tau}^{D}(r).$ Define
\[
    \hat{\tau}^{Y}_{L}=\min_{r \in \cR_{\Gamma}}\hat{\tau}^Y(r), 
    \qquad
    \hat{\tau}^{Y}_{U}=\max_{r \in \cR_{\Gamma}}\hat{\tau}^Y(r),
\]
and
\[
    \hat{\tau}^{D}_{L}=\min_{r \in \cR_{\Gamma}}\hat{\tau}^D(r), 
    \qquad
    \hat{\tau}^{D}_{U}=\max_{r \in \cR_{\Gamma}}\hat{\tau}^D(r).
\]
Assuming for all $\tilde w \in \varepsilon(\Gamma)$, ${\tau}^D(\tilde w) >0$, and its empirical counterpart is bounded away from zero over $\cR_{\Gamma}$, so that $\hat{\tau}^{D}_{L}>0$, then the interval
\begin{align} \label{eq: sensitivity interval}
    \sbr{
    \min_{\substack{a \in \{\hat{\tau}^{Y}_{L},\hat{\tau}^{Y}_{U}\} \\
                    d \in \{\hat{\tau}^{D}_{L},\hat{\tau}^{D}_{U}\}}}
    \frac{a}{d},
    \ 
    \max_{\substack{a \in \{\hat{\tau}^{Y}_{L},\hat{\tau}^{Y}_{U}\} \\
                    d \in \{\hat{\tau}^{D}_{L},\hat{\tau}^{D}_{U}\}}}
    \frac{a}{d}
    }
\end{align}
can be obtained by solving a total of eight linear programming problems. Regarding computational complexity, Proposition 4.5 of \cite{zhao2019sensitivity} shows that each $r_i$ that solves the linear programming problem takes a value of either $\Gamma$ or $\Gamma^{-1}.$ Furthermore, they demonstrate that the magnitude ordering of the solution $\{r_i\}$ must either match or be the reverse of the ordering of $\{Y_i\}$ or $\{D_i\}.$ These properties enable each linear programming problem to be solved in $\cO(n).$ We refer readers to their proof for further details.
\clearpage 

\section{Extended Simulation Study Results} \label{app: other simulation study results}
\subsection{Details of Simulation Set-Up and Full Results} \label{app: detailed standard simulation setup}
To generate the compliance status of each unit in the simulation, we first create a matrix of random coefficients with dimensions $2 \times (10+1)$, where each coefficient is drawn from $\text{Unif}(-1, 1)$. We then compute linear predictors by multiplying the covariate matrix (augmented with an intercept term) by the transpose of the random coefficient matrix. These linear predictors are transformed into probabilities using the softmax function. Specifically, let $\beta_{1j}, \beta_{2j} \sim \text{Unif}(-1, 1)$ for $j=0,1,\ldots, 10$ and $X_i^* = (1, X_i)$ be the augmented covariate vector, the probability of unit $i$ belonging to each compliance type is calculated as:
\begin{align*}
    \prob(C_i = 1) &= \frac{3}{3 + \exp(\beta_1^T X_i^* ) + \exp(\beta_2^T X_i^*)}, \\
    \prob(\text{unit $i$ is a never-taker}) &= \frac{\exp(\beta_1^T X_i^*)}{3 + \exp(\beta_1^T X_i^*) + \exp(\beta_2^T X_i^*)}, \\
    \prob(\text{unit $i$ is an always-taker}) &= \frac{\exp(\beta_2 ^ T X_i^* )}{3 + \exp(\beta_1 ^ T X_i^*) + \exp(\beta_2 ^ T X_i^*)}.
\end{align*}

\begin{figure}[htbp]
\centering
\includegraphics[width=0.8\textwidth]{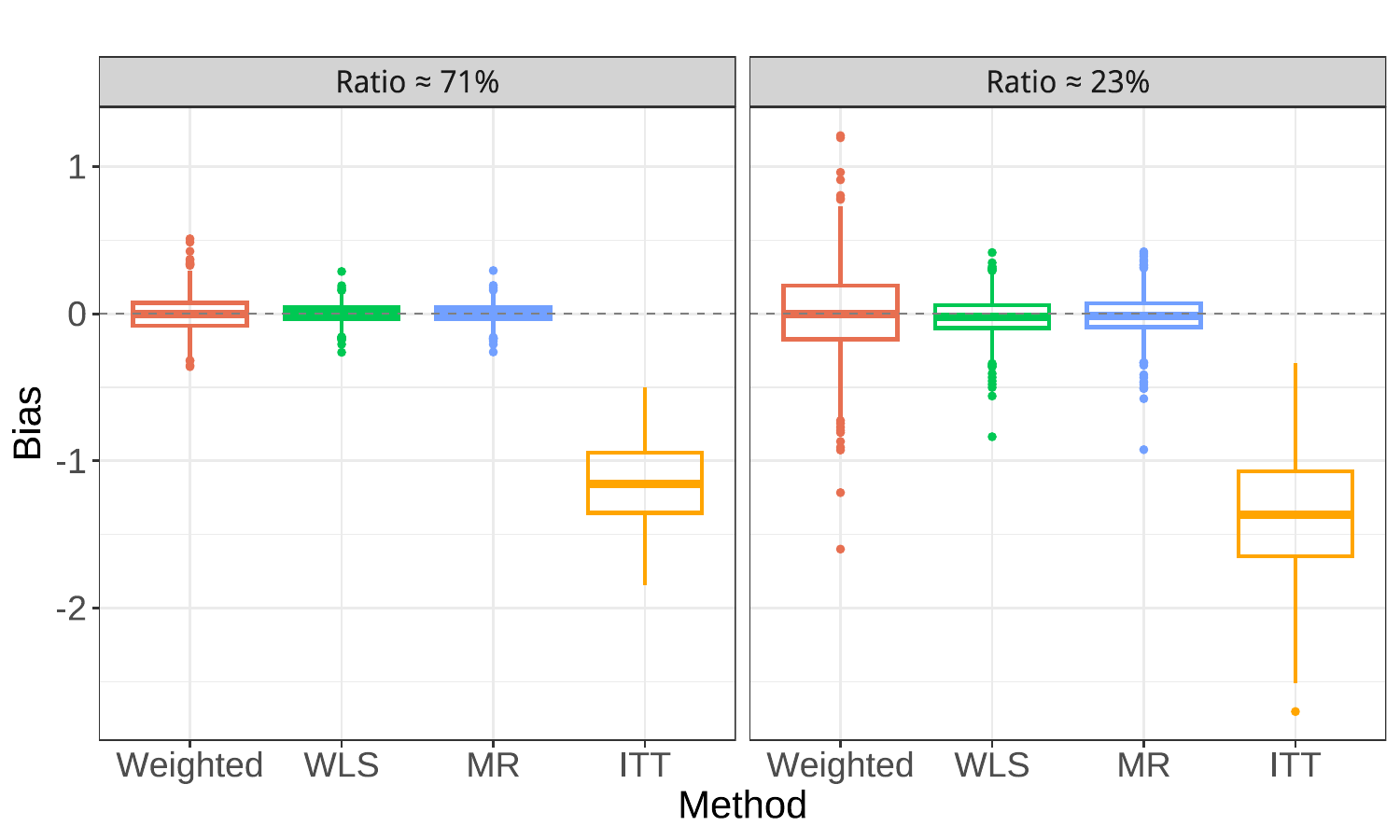}
\caption{Boxplots of the bias of the weighted estimator, WLS estimator, multiply robust estimator, and the weighted ITT estimator. The results are based on 1,000 trials with a total sample size of $N + n = 5000$. The left plot represents $n/(n + N) \approx 0.71$, while the right plot corresponds to $n/(n + N) \approx 0.23$.}
\label{fig: bias four estimators box plot}
\end{figure}

 Note that  $\hat{\tau}_{w}^Y$ is the numerator of $\hat{\tau}_{w}$ and generalizes the ITT from the study sample to the target population. We use logistic regression to estimate $\hat{\prob}(S_i = s_i \given X_i)$ for 4 estimators. For $\hat{\tau}_{\text{mr}},$ we use ordinary linear regression models for both $\hat{\mu}_{yz}(X)$ and $\hat{\mu}_{dz}(X).$ We compute the variance of $\hat{\tau}_{\text{mr}}$ using the nonparametric bootstrap method. The number of bootstrap samples is 500. Figure \ref{fig: bias four estimators box plot} displays the results for the four estimators when $N + n = 5000$, under two settings where $n/(n + N) \approx 0.71$ or $n/(n + N) \approx 0.23$. Table \ref{table:full_combined_coverage} presents the full results. We compare the coverage of the different estimators across the simulation settings. The coverage of the multiply robust estimator fluctuates around the nominal level, echoing the poor coverage observed in \citet{clark2024transportability} under the misspecification of either the outcome or treatment received models. Since variance is estimated via nonparametric bootstrap, we conjecture that an additional calibration step — such as those proposed in \citet{carlin1990approaches, chen2025empirical} — could help achieve accurate coverage.   

\begin{landscape}
\begin{table}[htbp]
  \centering
  \small 
  \caption{Simulation results for 4 estimators over 1000 trials. The approximate ratio measures the ratio of study sample size to the total sample size. The mean error is defined as the mean of the difference between the estimated T-CACE and the ground truth T-CACE. The empirical standard deviation of each estimator is provided in parentheses. We also provide the coverage of the $95\%$ confidence intervals.}
  \label{table:full_combined_coverage}
  \begin{tabular}{cccccccccc} 
    \toprule
    $N$ & \begin{tabular}[c]{@{}c@{}}Approximate \\ Ratio\end{tabular} & \multicolumn{2}{c}{Weighted Estimator} & \multicolumn{2}{c}{WLS Estimator} & \multicolumn{2}{c}{Multiply Robust} & \multicolumn{2}{c}{Weighted (T-ITT)} \\
    \cmidrule(lr){3-4} \cmidrule(lr){5-6} \cmidrule(lr){7-8} \cmidrule(lr){9-10}
     &  & \begin{tabular}[c]{@{}c@{}}Mean Error\end{tabular} & \begin{tabular}[c]{@{}c@{}}Coverage \\ (\%)\end{tabular} & \begin{tabular}[c]{@{}c@{}}Mean Error\end{tabular} & \begin{tabular}[c]{@{}c@{}}Coverage \\ (\%)\end{tabular} & \begin{tabular}[c]{@{}c@{}}Mean Error\end{tabular} & \begin{tabular}[c]{@{}c@{}}Coverage \\ (\%)\end{tabular} & \begin{tabular}[c]{@{}c@{}}Mean Error\end{tabular} & \begin{tabular}[c]{@{}c@{}}Coverage \\ (\%)\end{tabular} \\
    \midrule
    1500 & 0.71 & 0.01 (0.21) & 93.9 & 0.02 (0.10) & 94.6 & 0.01 (0.10) & 91.6 & -1.13 (0.30) & 0.0 \\
    1500 & 0.55 & 0.01 (0.17) & 94.3 & 0.00 (0.09) & 96.1 & 0.00 (0.09) & 92.4 & -1.26 (0.33) & 0.0 \\
    1500 & 0.40 & -0.01 (0.21) & 95.0 & -0.00 (0.10) & 98.1 & -0.00 (0.10) & 94.5 & -1.32 (0.34) & 0.0 \\
    1500 & 0.23 & -0.04 (0.54) & 93.4 & -0.06 (0.21) & 93.6 & -0.04 (0.23) & 91.2 & -1.36 (0.51) & 10.1 \\
    \midrule
    5000 & 0.71 & -0.00 (0.12) & 93.6 & 0.00 (0.06) & 92.8 & 0.00 (0.06) & 87.6 & -1.15 (0.27) & 0.0 \\
    5000 & 0.55 & -0.00 (0.10) & 92.8 & -0.00 (0.06) & 94.6 & -0.00 (0.06) & 87.0 & -1.28 (0.31) & 0.0 \\
    5000 & 0.40 & 0.00 (0.11) & 94.9 & 0.00 (0.06) & 96.9 & 0.00 (0.06) & 91.9 & -1.34 (0.34) & 0.0 \\
    5000 & 0.23 & 0.00 (0.30) & 95.2 & -0.02 (0.13) & 96.3 & -0.01 (0.13) & 93.4 & -1.37 (0.40) & 1.0 \\
    \midrule
    10000 & 0.71 & 0.00 (0.08) & 93.6 & 0.00 (0.05) & 92.6 & 0.00 (0.05) & 88.6 & -1.15 (0.27) & 0.0 \\
    10000 & 0.55 & -0.00 (0.07) & 91.5 & -0.00 (0.04) & 92.0 & -0.00 (0.04) & 84.7 & -1.28 (0.31) & 0.0 \\
    10000 & 0.40 & 0.00 (0.08) & 93.6 & 0.00 (0.05) & 92.1 & 0.00 (0.05) & 85.2 & -1.34 (0.33) & 0.0 \\
    10000 & 0.23 & -0.00 (0.21) & 95.4 & -0.01 (0.09) & 95.4 & -0.01 (0.09) & 92.5 & -1.38 (0.37) & 0.0 \\
    \bottomrule
  \end{tabular}
\end{table}
\end{landscape}

\subsection{Instrumental Variable vs Principal Stratification} \label{subsec: IV vs PS}
\citet{clark2024transportability} is a recent study that also proposes an identification formula for T-CACE. Instead of exclusion restriction (i.e., Assumption~\ref{asp: IV}-(b)), their result is based on principal ignorability:
\begin{assumption} [Principal Ignorability] \label{asp: principal ignorability} The following relationships hold: 
\begin{itemize}
    \item [(i).] $\EE[Y(1) \given D(1)=0, D(0)=0, S=1, X] = \EE[Y(1) \given D(1)=1, S=1, X].$
    \item [(ii).] $\EE[Y(0) \given  D(1)=0, D(0)=0, S=1, X] = \EE[Y(0) \given D(0)=0, S=1, X].$
\end{itemize}
\end{assumption}
In this section, we compare the performance of the proposed WLS estimator with that of the inverse probability estimator defined in Theorem 2 of \citet{clark2024transportability}, under two settings: (1) when principal ignorability (i.e., Assumption \ref{asp: principal ignorability}) is violated; and (2) when exclusion restriction (i.e., Assumption~\ref{asp: IV}-(b)) is violated.

\subsubsection{Violation of Principal Ignorability}

\begin{figure}[htbp]
\centering
\includegraphics[width=1.0\textwidth, height=0.6\textwidth]{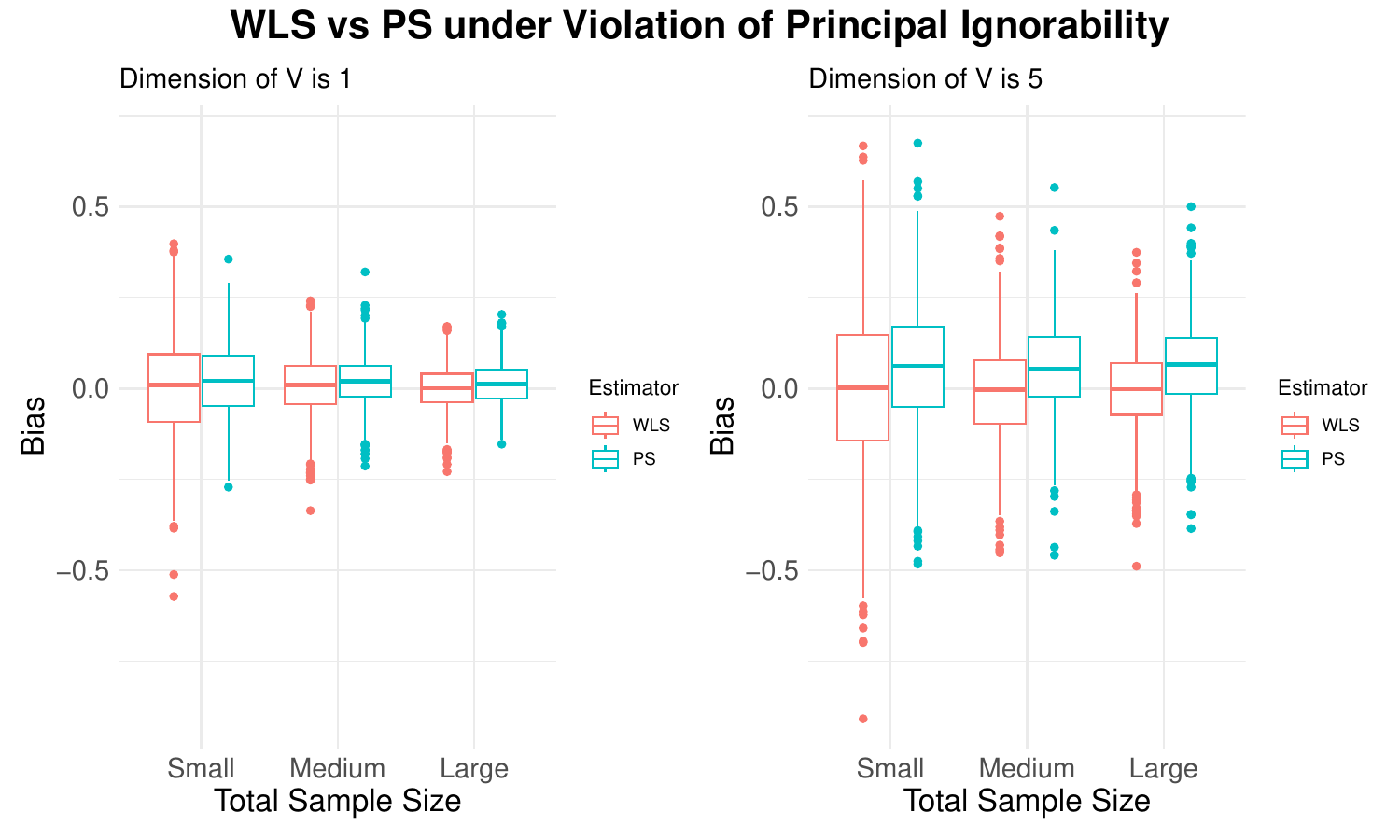}
\caption{Comparison of bias of the WLS estimator and the PS estimator when principal ignorability is violated. The plot at the left is when the unmeasured confounder $V \in \RR^{1}.$ The plot at the right is when $V \in \RR^{5}.$ There are $1000$ trials for each sample size. For the label of the $x$-axis “small” corresponds to  $N=1500$, “medium” to $N = 5000$, and “large” to $N=10000.$}
\label{fig: WLS PS CRE PS violate}
\end{figure}

We focus on scenarios where principal ignorability (i.e., Assumption \ref{asp: principal ignorability}) is violated. We modify the data generating process detailed in Section~\ref{sec: simulation study} and \S\ref{app: detailed standard simulation setup} by setting the confounders $V^{j}_i \overset{i.i.d.} {\sim} N(0, 1)$ with $j \in [\text{dim}V]$ and $i \in [n + N]$. Hence, $V_i \in \RR^{\text{dim}V}$ for each $i.$, we vary $\text{dim}V$ between 1 and 5. The pre-treatment observed covariates $X_i = (X^{1}_i, ..., X^{10}_i) \in \RR^{10}$ are generated independently as $X^{j}_i \overset{i.i.d.} {\sim} \text{Unif(-0.3, 0.5)}.$  We consider a completely randomized treatment indicator $Z_i \sim \text{Bernoulli}(0.5)$. The compliance type $C_i$ for each unit $i$ is based on a randomly generated multinomial logit model on $X_i$ and $V_i.$ We create a matrix of random coefficients for $X$ with dimensions $2 \times (10+1)$, where each coefficient is drawn from $\text{Unif}(-1, 1)$. We also create a matrix of random coefficients for $V$ with dimensions $2 \times \text{dim}V$, where each coefficient is drawn from $\text{Unif}(-0.5, 1).$ We then compute linear predictors by multiplying the covariate matrix (augmented with an intercept term) and the unmeasured confounder matrix by the transpose of their corresponding random coefficient matrices. These linear predictors are transformed into probabilities using the softmax function. Specifically, denote $\beta_{1j}, \beta_{2j} \sim \text{Unif}(-1, 1)$ for $j=0,1,\ldots, 10,$ $\beta_{3j}, \beta_{4j} \sim \text{Unif}(-0.5, 1)$ for $j=0,1,\ldots, \text{dim}V,$ and $X_i^* = (1, X_i)$ the augmented covariate vector. The probability of unit $i$ belonging to each compliance type is calculated as:
\begin{align*}
    \prob(C_i = 1) &= \frac{3}{3 + \exp(\beta_1^T X_i^* ) + \exp(\beta_2^T X_i^*) + 1.5\exp(\beta_3^T V_i) + 1.5\exp(\beta_4^T V_i)}, \\
    \prob(\text{unit $i$ is a never-taker}) &= \frac{\exp(\beta_1^T X_i^*) + 1.5\exp(\beta_3^T V_i)}{3 + \exp(\beta_1^T X_i^*) + \exp(\beta_2^T X_i) + 1.5\exp(\beta_3^T V_i) + 1.5\exp(\beta_4^T V_i)}, \\
    \prob(\text{unit $i$ is an always-taker}) &= \frac{\exp(\beta_1^T X_i^*) + 1.5\exp(\beta_4^T V_i)}{3 + \exp(\beta_1^T X_i^*) + \exp(\beta_2^T X_i) + 1.5\exp(\beta_3^T V_i) + 1.5\exp(\beta_4^T V_i)}.
\end{align*}
We generate the sample selection indicator as $S_i\given X_i \sim \text{Bernoulli}\cbr{\sigma\rbr{\sum_{j=1}^{10}X^{j}_i}}.$ We generate the outcome 
\begin{align*}
    Y_i = 2D_i + \sum_{j=1}^{10}X^{j}_i + \sum_{j=1}^{\text{dim}V}V^{j}_i + D_i \times \rbr{\sum_{j=1}^{10}X^{j}_i + 1.5 \sum_{j=1}^{\text{dim}V}V^{j}_i}  +  \epsilon,
\end{align*}
where $\epsilon \sim N(0, 0.5).$  We assume that $V$ is unobserved, so that $D$ is endogenous and the principal ignorability assumption fails. We run experiments with 1000 trials for each combination of $N \in \{1500, 5000, 10000\}$ and $\text{dim}V \in \{1, 5\}.$ In Figure \ref{fig: WLS PS CRE PS violate}, we visualize the comparison of the biases for the proposed WLS estimator and the principal stratification (PS estimator). Table \ref{table:IV_vs_PS} records the numerical results of the experiment. There is a systematic bias of the PS estimator regardless of the sample sizes. When $\text{dim}V$ increases from $1$ to $5,$ the bias of the principal stratification estimator increases, while that of the WLS estimator remains robust. This implies that the performance of the PS estimator is severely affected when there exist unmeasured confounders. Moreover, despite the common criticism that instrumental variable-based estimators have higher variance compared to those based on principal ignorability \citep{hartman2023improving}, the WLS estimator exhibits a comparable standard deviation to the PS estimator, indicating an improvement in precision.

\begin{table}[htbp]
  \centering
  \begin{tabular}{cccc}
    \toprule
    $N$ & Estimator & dim$V$ & Mean Error\\
    \midrule
    1500 & WLS & 1 & 0.00 (0.14) \\
    1500 & PS  & 1 & 0.02 (0.10) \\
    1500 & WLS & 5 & -0.00 (0.22) \\
    1500 & PS  & 5 & 0.06 (0.17) \\
    \midrule
    5000 & WLS & 1 & 0.01 (0.08) \\
    5000 & PS  & 1 & 0.02 (0.07) \\
    5000 & WLS & 5 & -0.01 (0.14) \\
    5000 & PS  & 5 & 0.06 (0.13) \\
    \midrule
    10000 & WLS & 1 & -0.00 (0.06) \\
    10000 & PS  & 1 & 0.01 (0.06) \\
    10000 & WLS & 5 & -0.01 (0.11) \\
    10000 & PS  & 5 & 0.06 (0.12) \\
    \bottomrule
  \end{tabular}
  \vspace{0.5em}
  \caption{Simulation results comparing WLS and PS estimators across different sample sizes under the violation of principal ignorability. The mean error is defined as the difference between the T-CACE estimate and the ground truth T-CACE. Values in parentheses represent standard deviations.}
  \label{table:IV_vs_PS}
\end{table}

\subsubsection{Violation of Exclusion Restriction}
\begin{figure}[htbp]
\centering
\includegraphics[width=1.0\textwidth, height=0.6\textwidth]{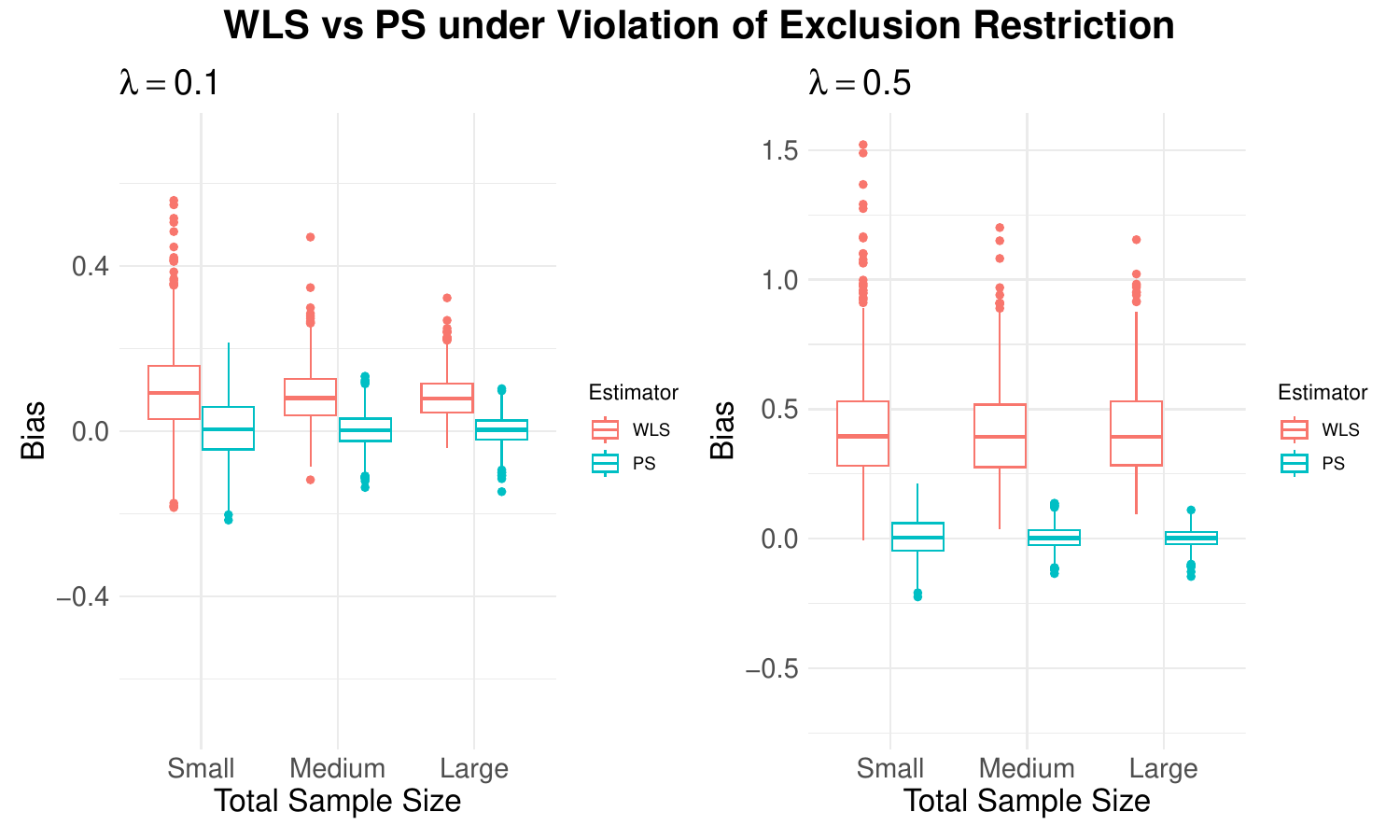}
\caption{Comparison of bias of the WLS estimator and the PS estimator when exclusion restriction is violated. The left plot corresponds to $\lambda = 0.1$, representing a smaller violation, while the right plot corresponds to $\lambda = 0.5$, representing a more severe violation. Each setting includes 1000 trials. For the $x$-axis labels, “small” corresponds to $N = 1500$, “medium” to $N = 5000$, and “large” to $N = 10000$.}
\label{fig: WLS PS CRE IV violate}
\end{figure}

\begin{table}[htbp]
  \centering
  \begin{tabular}{cccc}
    \toprule
    $N$ & Estimator & $\lambda$ & Mean Error\\
    \midrule
    1500 & WLS & 0.1 & 0.10 (0.11) \\
    1500 & PS  & 0.1 & 0.01 (0.07) \\
    1500 & WLS & 0.5 & 0.42 (0.20) \\
    1500 & PS  & 0.5 & 0.01 (0.08) \\
    \midrule
    5000 & WLS & 0.1 & 0.08 (0.07) \\
    5000 & PS  & 0.1 & 0.00 (0.04) \\
    5000 & WLS & 0.5 & 0.41 (0.18) \\
    5000 & PS  & 0.5 & 0.00 (0.04) \\
    \midrule
    10000 & WLS & 0.1 & 0.08 (0.05) \\
    10000 & PS  & 0.1 & 0.00 (0.04) \\
    10000 & WLS & 0.5 & 0.42 (0.17) \\
    10000 & PS  & 0.5 & 0.00 (0.04) \\
    \bottomrule
  \end{tabular}
  \vspace{0.5em}
  \caption{Simulation results comparing WLS and PS estimators across different sample sizes under the violation of the exclusion restriction. The mean error is defined as the difference between the T-CACE estimate and the ground truth T-CACE. Values in parentheses represent standard deviations.}
  \label{table:IV_vs_PS_ER}
\end{table}

We focus on scenarios in which exclusion restriction (i.e., Assumption~\ref{asp: IV}-(b)) is violated. To do so, we modify the data-generating process described in Section~\ref{sec: simulation study} and \S\ref{app: detailed standard simulation setup} by allowing the outcome to be directly influenced by the treatment assignment:
\begin{align*}
    Y_i = 2D_i + \sum_{j=1}^{10}X^{j}_i + D_i \times \sum_{j=1}^{10}X^{j}_i + \lambda Z_i + \epsilon_i.
\end{align*}
Here, $\lambda$ governs the degree to which the exclusion restriction is violated: the larger the value of $\lambda$, the stronger the direct effect of treatment assignment $Z_i$ on the outcome $Y_i$, and thus the greater the violation. For each $\lambda \in \cbr{0.1, 0.5}$, we run 1000 trials to compare the bias of the WLS and PS estimators, as shown in Figure \ref{fig: WLS PS CRE IV violate}. Table \ref{table:IV_vs_PS_ER} shows the full numerical results. The WLS estimator exhibits a non-negligible bias. As $\lambda$ increases from $0.1$ to $0.5$, the bias of the WLS estimator grows, highlighting the importance of the exclusion restriction in consistently estimating the T-CACE.

\subsection{Observational Study} \label{subsec: sim obs study}
We conduct a simulation study in a setting where randomized treatment assignment, \\ $Z \indep \{X, Y(0), Y(1), D(0), D(1)\} \given \{S=1\},$ fails, but unconfoundedness, $Z \indep \{Y(1), Y(0),D(0), D(1)\} \given \{S=1\}, X$, holds. We use the same setting as the simulation set-up detailed in Section~\ref{sec: simulation study}, except that $Z_i\given X_i \sim \text{Bernoulli}\cbr{0.2 \cdot \sigma\rbr{\sum_{j=1}^{5}X^{j}_i}}.$ Table \ref{table:combined_coverage_obs} shows the results for the weighted estimator, the WLS estimator, and the multiply robust estimator, respectively.

\begin{table}[htbp]
  \centering
  \footnotesize
  \caption{Simulation results for 4 estimators over 1000 trials under the observational setting. The approximate ratio measures the ratio of study sample size to the total sample size. The mean error is defined as the mean of the difference between the estimated T-CACE and the ground truth T-CACE. The empirical standard deviation of each estimator over 1000 trials is provided in parentheses. We also provide the coverage of the $95\%$ confidence intervals.}
  \label{table:combined_coverage_obs}
  \resizebox{\textwidth}{!}{%
  \begin{tabular}{cccccccccc}
    \toprule
    $N$ & \begin{tabular}[c]{@{}c@{}}Approximate \\ Ratio\end{tabular} & \multicolumn{2}{c}{Weighted Estimator} & \multicolumn{2}{c}{WLS Estimator} & \multicolumn{2}{c}{Multiply Robust} & \multicolumn{2}{c}{Weighted (T-ITT)} \\
    \cmidrule(lr){3-4} \cmidrule(lr){5-6} \cmidrule(lr){7-8} \cmidrule(lr){9-10}
     &  & \begin{tabular}[c]{@{}c@{}}Mean Error\end{tabular} & \begin{tabular}[c]{@{}c@{}}Coverage \\ (\%)\end{tabular} & \begin{tabular}[c]{@{}c@{}}Mean Error\end{tabular} & \begin{tabular}[c]{@{}c@{}}Coverage \\ (\%)\end{tabular} & \begin{tabular}[c]{@{}c@{}}Mean Error\end{tabular} & \begin{tabular}[c]{@{}c@{}}Coverage \\ (\%)\end{tabular} & \begin{tabular}[c]{@{}c@{}}Mean Error\end{tabular} & \begin{tabular}[c]{@{}c@{}}Coverage \\ (\%)\end{tabular} \\
    \midrule
    1500 & 0.71 & 0.00 (0.21) & 93.8 & 0.01 (0.10) & 95.5 & 0.01 (0.10) & 92.8 & -1.13 (0.30) & 0.0 \\
    1500 & 0.55 & 0.00 (0.17) & 93.9 & 0.00 (0.09) & 97.0 & 0.00 (0.09) & 93.5 & -1.25 (0.32) & 0.0 \\
    1500 & 0.40 & 0.01 (0.20) & 95.1 & -0.00 (0.11) & 97.4 & -0.00 (0.11) & 93.9 & -1.32 (0.36) & 0.0 \\
    1500 & 0.23 & -0.01 (0.56) & 92.7 & -0.07 (0.21) & 92.6 & -0.04 (0.24) & 91.2 & -1.34 (0.51) & 11.1 \\
    \midrule
    5000 & 0.71 & 0.00 (0.12) & 94.3 & 0.00 (0.06) & 94.7 & 0.00 (0.06) & 90.6 & -1.17 (0.27) & 0.0 \\
    5000 & 0.55 & 0.00 (0.10) & 93.6 & 0.00 (0.06) & 94.4 & 0.00 (0.06) & 87.9 & -1.30 (0.31) & 0.0 \\
    5000 & 0.40 & 0.00 (0.11) & 95.4 & 0.00 (0.06) & 97.1 & -0.00 (0.06) & 91.2 & -1.36 (0.33) & 0.0 \\
    5000 & 0.23 & 0.01 (0.31) & 94.8 & -0.03 (0.13) & 96.7 & -0.02 (0.14) & 91.9 & -1.38 (0.39) & 0.5 \\
    \midrule
    10000 & 0.71 & -0.00 (0.08) & 95.5 & -0.00 (0.04) & 93.2 & -0.00 (0.04) & 88.9 & -1.15 (0.27) & 0.0 \\
    10000 & 0.55 & -0.00 (0.07) & 93.1 & -0.00 (0.04) & 93.0 & -0.00 (0.04) & 81.9 & -1.28 (0.30) & 0.0 \\
    10000 & 0.40 & -0.00 (0.08) & 92.7 & -0.00 (0.05) & 94.2 & -0.00 (0.05) & 86.5 & -1.35 (0.32) & 0.0 \\
    10000 & 0.23 & 0.01 (0.22) & 96.0 & -0.02 (0.09) & 96.2 & -0.01 (0.10) & 92.8 & -1.37 (0.37) & 0.1 \\
    \bottomrule
  \end{tabular}%
  }
\end{table}

\subsection{Sensitivity Analysis} \label{subsec: sensitivity analysis simulation}
We conduct sensitivity analysis under the setting when there exists an unmeasured confounder $U$ such that mean exchangeability (i.e., Assumption~\ref{asp: cond_ign_selection} and Assumption~\ref{asp: randomized treatment received exchangeability}) fails.  Let $N=1500$, the covariates $X$ follow a uniform distribution (i.e., $X_i \in \RR^{5}$, where $X^{j}_i \overset{i.i.d.} {\sim} \text{Unif(-0.3, 0.5)}$), and the unmeasured confounder is generated as $U_i \sim \text{Unif(-0.1, 0.5)}.$ We generate the sample selection indicator as
\begin{align*}
   S_i = 0 \given X_i, U_i \sim \text{Bernoulli}\rbr{\frac{\ \exp(\beta_0 + \sum_{j=1}^{5}X^{j}_i + \kappa U_i)}{1 + \exp(\beta_0 + \sum_{j=1}^{5}X^{j}_i + \kappa U_i)}},
\end{align*}

where $\kappa$ is a hyperparameter controlling the extent of how $U$ affects the experiment participation. We vary $\kappa$ over the set $\{0.1, 0.3, 0.7, 1.0\}$ in the simulation. We consider a completely randomized treatment indicator $Z_i \sim \text{Bernoulli}(0.5)$. The compliance type $C_i$ for each unit $i$ is based on a randomly generated multinomial logit model on $X_i$ and $U_i.$ By “randomly generated,” we mean that we first create a matrix of random coefficients with dimensions $2 \times (6+1)$, where each coefficient is drawn from $\text{Unif}(-1, 1)$. We then compute linear predictors by multiplying the covariate matrix (augmented with an intercept term) by the transpose of the random coefficient matrix. These linear predictors are transformed into probabilities using the softmax function. Specifically, if $\beta_{1j}, \beta_{2j} \sim \text{Unif}(-1, 1)$ for $j=0,1,\ldots, 6$ and $X_i^* = (1, X_i, U_i)$ is the augmented covariate vector, the probability of unit $i$ belonging to each compliance type is calculated as:
\begin{align*}
    \prob(C_i = 1) &= \frac{3}{3 + \exp(\beta_1^T X_i^* ) + \exp(\beta_2^T X_i^*)}, \\
    \prob(\text{unit $i$ is a never-taker}) &= \frac{\exp(\beta_1^T X_i^*)}{3 + \exp(\beta_1^T X_i^*) + \exp(\beta_2^T X_i)}, \\
    \prob(\text{unit $i$ is an always-taker}) &= \frac{\exp(\beta_2 ^ T X_i^* )}{3 + \exp(\beta_1 ^ T X_i^*) + \exp(\beta_2 ^ T X_i^*)}.
\end{align*}

Finally, the outcome is a linear combination of the treatment received, the covariates and the interaction terms: 
\begin{align*}
    Y_i = 2D_i + \sum_{j=1}^{5}X^{j}_i + U_i + D_i \times \rbr{\sum_{j=1}^{5}X^{j}_i + 2U_i} + \epsilon,
\end{align*}
where $\epsilon \sim N(0, 0.5).$ The oracle $\tau_\tca$ can be computed using the Monte Carlo Method. Define the weight ratio that accounts for the unmeasured confounder compared to the one that does not
\begin{align*}
    r_{i}^{*} = \frac{w^*(X_i,U_i)}{w(X_i)} = \frac{\prob(S_i=0 \given X_i, U_i)}{\prob(S_i=1 \given X_i, U_i) } \times \frac{\prob(S_i=1 \given X_i)}{\prob(S_i=0 \given X_i)}.
\end{align*}

Since for each $x_i,$ $\prob(S_i=0 \given X_i = x_i) = \int_{\cU} \prob(S_i=0 \given X_i=x_i, U_i = u) p_{U}(u) du,$ there exists a $\tilde{u_i} \in [-0.1, 0.5]$ such that $\prob(S_i = 1 \given X_i=x_i) \leq \prob(S_i = 1 \given X_i=x_i, U_i = \tilde{u_i}).$ Then

\begin{align*}
    r_{i}^{*} \given \{X_i = x_i\} &= \frac{\prob(S_i=0 \given X_i=x_i, U_i)}{\prob(S_i=1 \given X_i=x_i, U_i) } \times \frac{\prob(S_i=1 \given X_i=x_i)}{\prob(S_i=0 \given X_i=x_i)} \nend 
    & \leq \frac{\prob(S_i=0 \given X_i=x_i, U_i)}{\prob(S_i=1 \given X_i=x_i, U_i) } \times \frac{\prob(S_i=1 \given X_i=x_i, U_i = \tilde{u_i})}{\prob(S_i=0 \given X_i=x_i, U_i = \tilde{u_i})} \nend 
    &= \exp\{\kappa(U_i - \tilde{u_i})\} \nend 
    & \leq e^{0.6\kappa}.
\end{align*}

Similarly, $r_{i}^{*} \given \{X_i = x_i\} \geq e^{-0.6\kappa}.$ Therefore, the marginal sensitivity model assumption (i.e., Assumption~\ref{asp: marginal sensitivity model}) is satisfied with $\Gamma = e^{0.6\kappa}.$ 

For each of $\gamma \in \{1.06, 1.11, 1.16, 1.21, 1.26\},$ we compute the interval defined in \eqref{eq: sensitivity interval} for 1000 trials. In Table \ref{table:sensitivity_analysis_combine_coverage}, we record the empirical coverage in $\%$ of the sensitivity intervals for each $\gamma$: 

\begin{align*}
\frac{1}{1000} \sum_{k=1}^{1000} \mathds{1}\{ 
\tau_\tca \in I^k\},
\end{align*}

where $I^k$ denotes the interval for the $k$-th simulation trial, as defined in \eqref{eq: sensitivity interval}. Note that the empirical coverage increases with $\gamma$, reaching approximately $100\%$ when $\gamma \approx  \Gamma = e^{0.6\kappa}.$

\begin{figure}[htbp]
\centering
\includegraphics[width=0.6\textwidth]{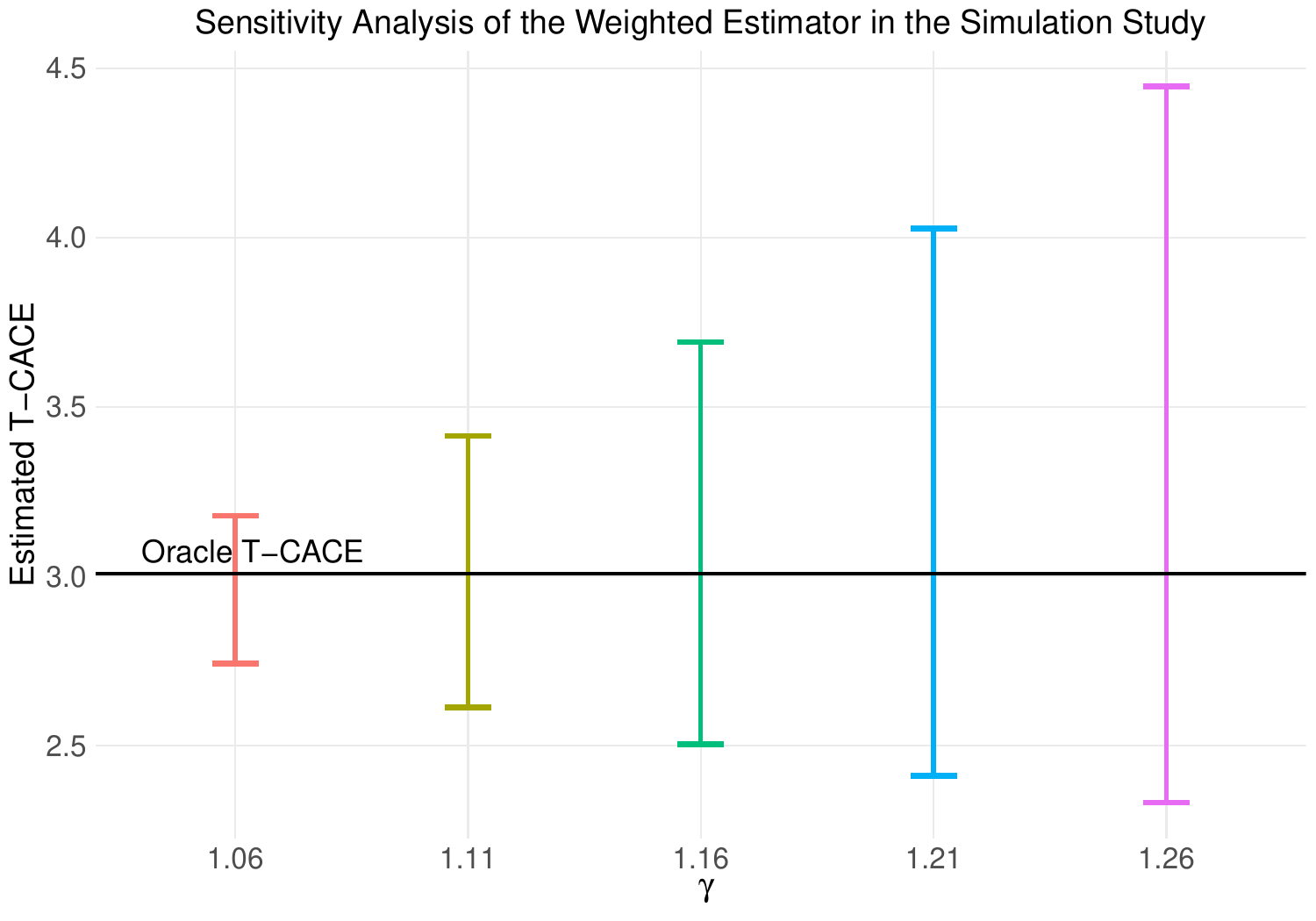}
\caption{Plot of the estimated T-CACE with respect to the sensitivity parameter $\gamma$ when $\kappa=1.0$. The solid error bars are the means of the upper ends and lower ends of the partial identification interval of the weighted estimator, computed over 1000 simulation trials for each $\gamma$, (i.e., $\sbr{\min_{r \in \cR_{\gamma}}\hat{\tau}(r), \max_{r \in \cR_{\gamma}}\hat{\tau}(r)}).$ The black solid lines represents the oracle T-CACE under the simulation setting.}
\label{fig: simulation study sensitivity intervals}
\end{figure}

To examine the widths of the intervals, Figure \ref{fig: simulation study sensitivity intervals} displays the mean of the upper bounds, ${\max_{r \in \cR_{\gamma}}\hat{\tau}^Y(r)}/{\min _{r \in \cR_{\gamma}}\hat{\tau}^D(r)},$ and the lower bounds, ${\min_{r \in \cR_{\gamma}}\hat{\tau}^Y(r)}/{\max _{r \in \cR_{\gamma}}\hat{\tau}^D(r)}$ computed over 1000 simulation trials for each $\gamma$ when $\kappa = 1$. When $\gamma=1,$ the unmeasured confounder is not accounted for. We conclude that even when the chosen $\gamma$ is conservative (e.g., 1.11),  the partially identified intervals can capture the oracle T-CACE with high probability, while avoiding excessively wide intervals.

\begin{table}[htbp]
  \centering
  \begin{tabular}{ccccc}
    \toprule
    $\kappa$ & $\Gamma = e^{0.6\kappa}$ & Mean of $\hat{\Gamma}$ & $\gamma$ & Coverage (\%) \\
    \midrule
    0.1 & 1.06 & 1.09 (0.08) & 1.06 & 72.2 \\
    0.1 & 1.06 & 1.09 (0.08) & 1.11 & 94.9 \\
    0.1 & 1.06 & 1.09 (0.08) & 1.16 & 98.6 \\
    0.1 & 1.06 & 1.09 (0.08) & 1.21 & 99.7 \\
    0.1 & 1.06 & 1.09 (0.08) & 1.26 & 100 \\
    \midrule
    0.3 & 1.19 & 1.13 (0.09) & 1.06 & 72.4 \\
    0.3 & 1.19 & 1.13 (0.09) & 1.11 & 94.0 \\
    0.3 & 1.19 & 1.13 (0.09) & 1.16 & 98.5 \\
    0.3 & 1.19 & 1.13 (0.09) & 1.21 & 99.7 \\
    0.3 & 1.19 & 1.13 (0.09) & 1.26 & 100 \\
    \midrule
    0.7 & 1.52 & 1.26 (0.13) & 1.06 & 69.3 \\
    0.7 & 1.52 & 1.26 (0.13) & 1.11 & 93.5 \\
    0.7 & 1.52 & 1.26 (0.13) & 1.16 & 98.7 \\
    0.7 & 1.52 & 1.26 (0.13) & 1.21 & 99.8 \\
    0.7 & 1.52 & 1.26 (0.13) & 1.26 & 99.9 \\
    \midrule
    1.0 & 1.82 & 1.39 (0.15) & 1.06 & 67.0 \\
    1.0 & 1.82 & 1.39 (0.15) & 1.11 & 93.4 \\
    1.0 & 1.82 & 1.39 (0.15) & 1.16 & 98.6 \\
    1.0 & 1.82 & 1.39 (0.15) & 1.21 & 99.8 \\
    1.0 & 1.82 & 1.39 (0.15) & 1.26 & 99.9 \\
    \bottomrule
  \end{tabular}
  \caption{Coverage (\%) of the sensitivity intervals in 1000 trials for each $\gamma$ and $\kappa$ value. Here, $\Gamma$ is the theoretical bound defined in the marginal sensitivity model (i.e., Assumption~\ref{asp: marginal sensitivity model}). For each simulation trial, we also compute $\hat{\Gamma}$, the value of $\max \left\{ \max_{i: S_i = 1} \frac{\hat{w}^*(X_i,U_i)}{\hat{w}(X_i)}, \left(\min_{i \in [n]}\frac{\hat{w}^*(X_i,U_i)}{\hat{w}(X_i)}\right)^{-1} \right\}.$}
  \label{table:sensitivity_analysis_combine_coverage}
\end{table}

\section{Assessing Instrumental Variable Validity}
\label{app: IV_test}

\subsection{Behavioral Content of Monotonicity.}
 Monotonicity (i.e., Assumption~\ref{asp: IV}-(a)) is a sign-uniformity condition on how the instrument affects treatment received. This condition does not require identical compliance behavior, since always-takers, never-takers, and compliers may all be present. It rules out defiers, whose treatment received would move opposite to the treatment assignment. To see why this is needed, let \(F\) denote defiers. Without monotonicity, the target Wald ratio equals
\[
\frac{p_C E[Y_i(1)-Y_i(0)\mid C,S=0]-p_F E[Y_i(1)-Y_i(0)\mid F,S=0]}{p_C-p_F},
\]
where \(p_C=P(C\mid S=0)\) and \(p_F=P(F\mid S=0)\). Thus the ratio is generally a net complier-defier contrast, not the T-CACE, unless \(p_F=0\) or additional restrictions are imposed. This is the sense in which monotonicity gives the Wald estimand its LATE interpretation in \citet{imbens1994identification}.

\subsection{Tests for Instrument Validity}
There is a stream of literature \citep{kitagawa2015test, mourifie2017testing, yu2025binary} that develops tests for instrument validity. We remark that these tests are not directly applicable to our problem without further assumptions. Our identification results require monotonicity and valid instrument (i.e. Assumption~\ref{asp: IV}) to hold in the target population, where we only observe the pre-treatment covariates, making the tests infeasible. Nevertheless, applied researchers may still use these tests if they believe the assignment-specific observed response distributions can be generalized from the study sample to the target population:

\begin{assumption}[Sample Assignment Ignorability for IV Tests] \label{asp: IV generalizability} 
For each $z \in \{0,1\}$, assume
\begin{align}
\rbr{Y(z,D(z)),D(z)} \indep S \given Z,X. \label{eq:sample_assignment_ignorability_iv}
\end{align}
In addition, assume that the conditional first-stage contrast is invariant across the study sample and the target population:
\begin{align}
\EE\sbr{D(1) - D(0) \mid X, S = 1} 
=
\EE\sbr{D(1) - D(0) \mid X, S = 0}. \label{eq:first_stage_transport}
\end{align}
\end{assumption} 

Assumption \ref{asp: IV generalizability} states that, conditional on the treatment assignment and pre-treatment covariates, the assignment-specific observed response pair $\rbr{Y(z,D(z)),D(z)}$ has the same distribution in the study sample and the target population. This assumption is plausible in social science applications where the study sample and target population are embedded in the same institutional environment, the instrument is generated by the same assignment or encouragement mechanism, and differences between the two populations are believed not to arise from different assignment-specific response distributions. For the remainder of this section, we assume Assumption \ref{asp: IV generalizability} holds, and therefore, tests of the study-sample observable implications of instrument validity can be interpreted as tests of the corresponding target-population observable implications.

First, for instrument relevance (i.e., Assumption~\ref{asp: IV}-(c)), Assumption \ref{asp: IV generalizability} implies
\begin{align}
\EE\sbr{D(1)-D(0) \mid S=0}
&=
\EE\sbr{
    \EE\sbr{D(1)-D(0) \mid X,S=0}
    \mid S=0
} \nonumber \\
&=
\EE\sbr{
    \EE\sbr{D(1)-D(0) \mid X,S=1}
    \mid S=0
} \nonumber \\
&=
\EE\sbr{
    \EE\sbr{D \mid Z=1,S=1,X}
    -
    \EE\sbr{D \mid Z=0,S=1,X}
    \mid S=0
}, \label{eq:target_first_stage_test}
\end{align}
where the second equality follows from \eqref{eq:first_stage_transport}, and the last equality follows from treatment ignorability in the study sample (i.e., Assumption 1). Therefore, the target population instrument relevance condition can be assessed by estimating the first-stage contrast in \eqref{eq:target_first_stage_test}. A statistically significant estimate of this contrast, together with the corresponding \(t\)-statistic or \(F\)-statistic, provides support for the instrument relevance condition in the target population. 

\citet{kitagawa2015test} shows that for every Borel set \(B \subseteq \mathcal{Y}\), where $\mathcal{Y}$ is the support of the outcome variable, the original IV test implication in the target population is
\begin{align} 
\prob\left(Y \in B, D = 1 \mid Z = 1, S = 0, X\right)
-
\prob\left(Y \in B, D = 1 \mid Z = 0, S = 0, X\right)
&\geq 0, \label{eq:target_kitagawa_test_0} \\
\prob\left(Y \in B, D = 0 \mid Z = 0, S = 0, X\right)
-
\prob\left(Y \in B, D = 0 \mid Z = 1, S = 0, X\right)
&\geq 0. \label{eq:target_kitagawa_test_1}
\end{align}   

Assumption \ref{asp: IV generalizability} transports the target population implications \eqref{eq:target_kitagawa_test_0}--\eqref{eq:target_kitagawa_test_1} into study-sample implications. To see this, fix $z \in \{0,1\}$, $d \in \{0,1\}$, and a Borel set $B \subseteq \mathcal{Y}$. By consistency,
\begin{align}
\prob\left(Y \in B, D=d \mid Z=z,S=0,X\right)
&=
\prob\left(Y(z,D(z)) \in B, D(z)=d \mid Z=z,S=0,X\right). \label{eq:consistency_target}
\end{align}
Then, by Assumption \ref{asp: IV generalizability},
\begin{align}
\prob\left(Y(z,D(z)) \in B, D(z)=d \mid Z=z,S=0,X\right)
&=
\prob\left(Y(z,D(z)) \in B, D(z)=d \mid Z=z,S=1,X\right). \label{eq:transport_observable_law}
\end{align}
Applying consistency again gives
\begin{align}
\prob\left(Y(z,D(z)) \in B, D(z)=d \mid Z=z,S=1,X\right)
&=
\prob\left(Y \in B, D=d \mid Z=z,S=1,X\right). \label{eq:consistency_study}
\end{align}
Combining \eqref{eq:consistency_target}--\eqref{eq:consistency_study}, we obtain
\begin{align}
\prob\left(Y \in B, D=d \mid Z=z,S=0,X\right)
=
\prob\left(Y \in B, D=d \mid Z=z,S=1,X\right). \label{eq:target_study_law_equivalence}
\end{align}
Substituting \eqref{eq:target_study_law_equivalence} into the target-population implications \eqref{eq:target_kitagawa_test_0}--\eqref{eq:target_kitagawa_test_1} yields
\begin{align} 
\prob\left(Y \in B, D = 1 \mid Z = 1, S = 1, X\right)
-
\prob\left(Y \in B, D = 1 \mid Z = 0, S = 1, X\right)
&\geq 0, \label{eq:kitagawa_test_0} \\
\prob\left(Y \in B, D = 0 \mid Z = 0, S = 1, X\right)
-
\prob\left(Y \in B, D = 0 \mid Z = 1, S = 1, X\right)
&\geq 0. \label{eq:kitagawa_test_1}
\end{align}   

Therefore, under Assumption \ref{asp: IV generalizability}, the study-sample inequalities \eqref{eq:kitagawa_test_0}--\eqref{eq:kitagawa_test_1} are equivalent to the target-population inequalities \eqref{eq:target_kitagawa_test_0}--\eqref{eq:target_kitagawa_test_1}. Thus, applying the Kitagawa test in the study sample provides a valid falsification test for the target population observable implications of monotonicity and exclusion restriction.

\citet{kitagawa2015test} further shows that this implication is sharp: no other feature of the observed data can provide additional information for ruling out invalid instruments. Under Assumption \ref{asp: IV generalizability}, this sharpness statement applies to the target population observable implications. \citet{kitagawa2015test} then develops a variance-weighted Kolmogorov–Smirnov test statistic to measure the magnitude of violations of \eqref{eq:kitagawa_test_0} and \eqref{eq:kitagawa_test_1}. \citet{mourifie2017testing} proposed another, easier-to-implement test by rewriting the same implication as conditional moment inequalities and applying the intersection bounds framework of \citet{chernozhukov2013intersection}. These tests are available in the R package \texttt{ivcheck} \citep{coverdale2026ivcheck}. We implement the test for instrument relevance based on \eqref{eq:target_first_stage_test} and the test proposed by \citep{mourifie2017testing} for the deep canvassing application in \S \ref{subsec:empirical_IV_test}.

\subsection{What If the Instrument Validity Test Fails?}

If the sharp IV validity tests reject in the study sample, we recommend that the researchers only report the target intent-to-treat effect (i.e., $\tau_{\textsc{T-ITT}}$) \citep{buchanan2018generalizing}. The reason is that such a rejection indicates that the study sample data are not consistent with the conditions listed in study-to-target generalizability of IV assumptions (i.e., Assumption \ref{asp: IV generalizability}), and therefore, there is no empirical basis for the identification of T-CACE. 

If researchers still wish to report a complier-relevant quantity, two possible ways are examined in \citet{de2017tolerating} and \citet{liao2025extending}. \citet{de2017tolerating} shows that the Wald estimand can still have a CACE interpretation in the presence of defiers under additional restrictions. In our setting, an analogous argument would require assuming that in the target population, any defiers can be offset by a comparable subgroup of compliers. Under this interpretation, the generalized Wald ratio

\begin{align}
\label{eq:generalized Wald ratio}
    \frac{\EE_{X \given S=0} \sbr{\mu_{y1}(x) - \mu_{y0}(x)}}{\EE_{X \given S=0} \sbr{\mu_{d1}(x) - \mu_{d0}(x)}}
\end{align}
should not be viewed as the effect for all target compliers, but rather as the effect for a remaining subpopulation of target compliers.

A second approach is motivated by \citet{liao2025extending}, who proposes replacing a refuted IV model with a non-refutable relaxed model that deviates as little as possible from the original assumptions. Applied to our setting, this would mean imposing a minimal-deviation relaxation in the target population, such as allowing the smallest amount of defiance or the smallest departure from treatment ignorability to make the observed data compatible with the relaxed assumptions. In this case, \eqref{eq:generalized Wald ratio} can be interpreted as a target relaxed average complier treatment effect: the CACE under the closest admissible latent IV structure that explains the data. Importantly, this quantity does not a same interpretation of the original T-CACE unless the original IV assumptions are not refuted; rather, it should be reported as a robustness estimand whose interpretation depends on the chosen relaxation criterion.

In either case, these additional assumptions must hold along with the mean exchangeability assumptions (i.e., Assumption~\ref{asp: cond_ign_selection} and Assumption~\ref{asp: randomized treatment received exchangeability}). As a result, we believe researchers should consider these estimates only if they have an exceptionally strong reason to justify the additional assumptions.
\section{Extended Results for Empirical Application}
\label{app: empirical}
\subsection{Illustrating the Sensitivity Analysis}
\label{app: illustrating sensitivity analysis}
\begin{figure}[htbp]
\centering
\includegraphics[width=0.8\textwidth]{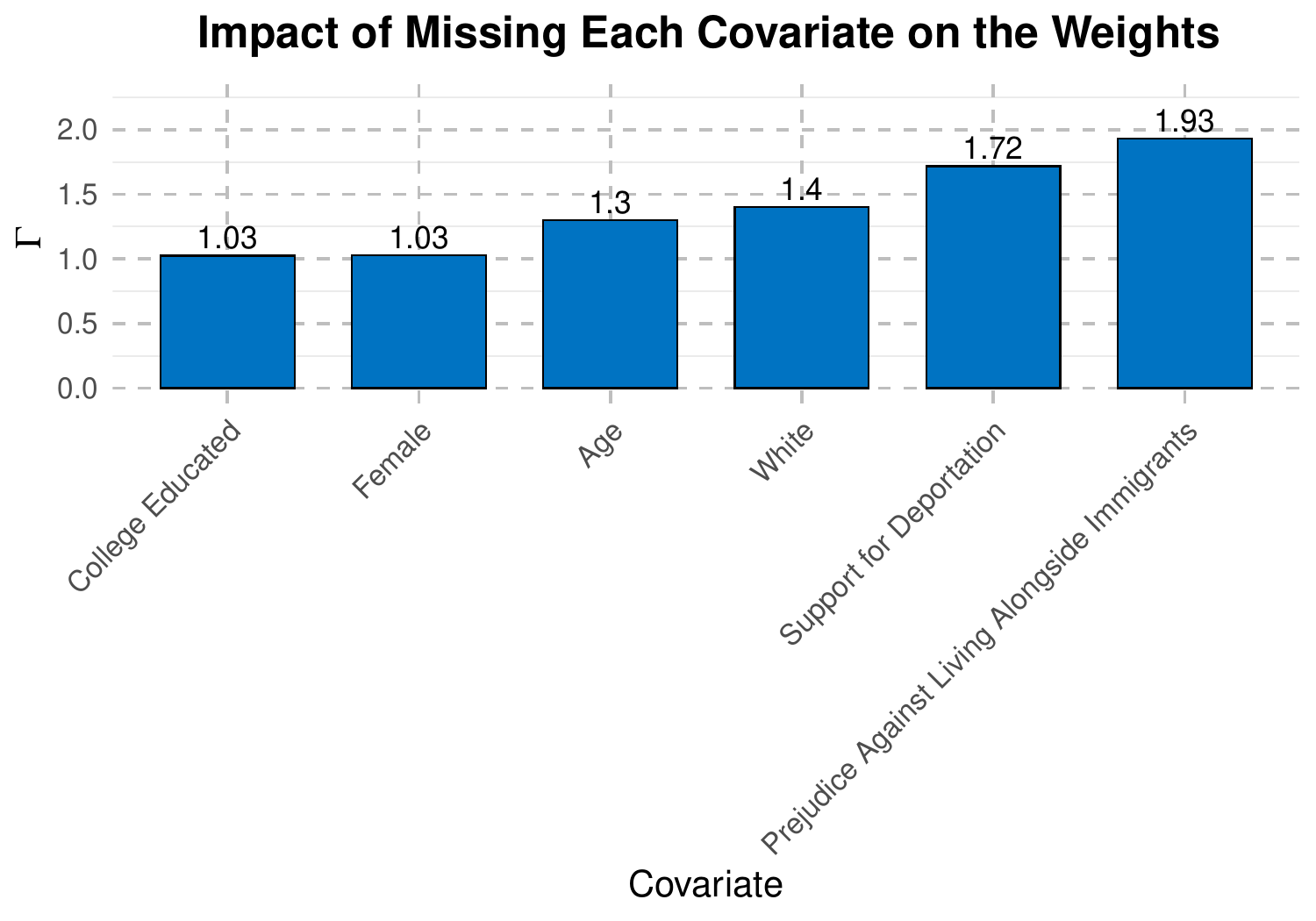}
\caption{Plot of \({\Gamma}\) for each of the 6 selected covariates, arranged in increasing order. The covariates \textbf{Female} and \textbf{Age} represent the voter’s gender and age, respectively. \textbf{White} and \textbf{College Educated} are binary variables denoting whether the voter identifies as white and whether they have attended college. The variables \textbf{Support for Deportation} and \textbf{Prejudice Against Living Alongside Immigrants} are measured on a Likert scale, reflecting the voter's stance on deporting all undocumented immigrants and their comfort with living near undocumented immigrants, respectively.}
\label{fig: tilde gamma}
\end{figure}

\begin{figure}[htbp]
\centering
\includegraphics[width=0.6\textwidth]{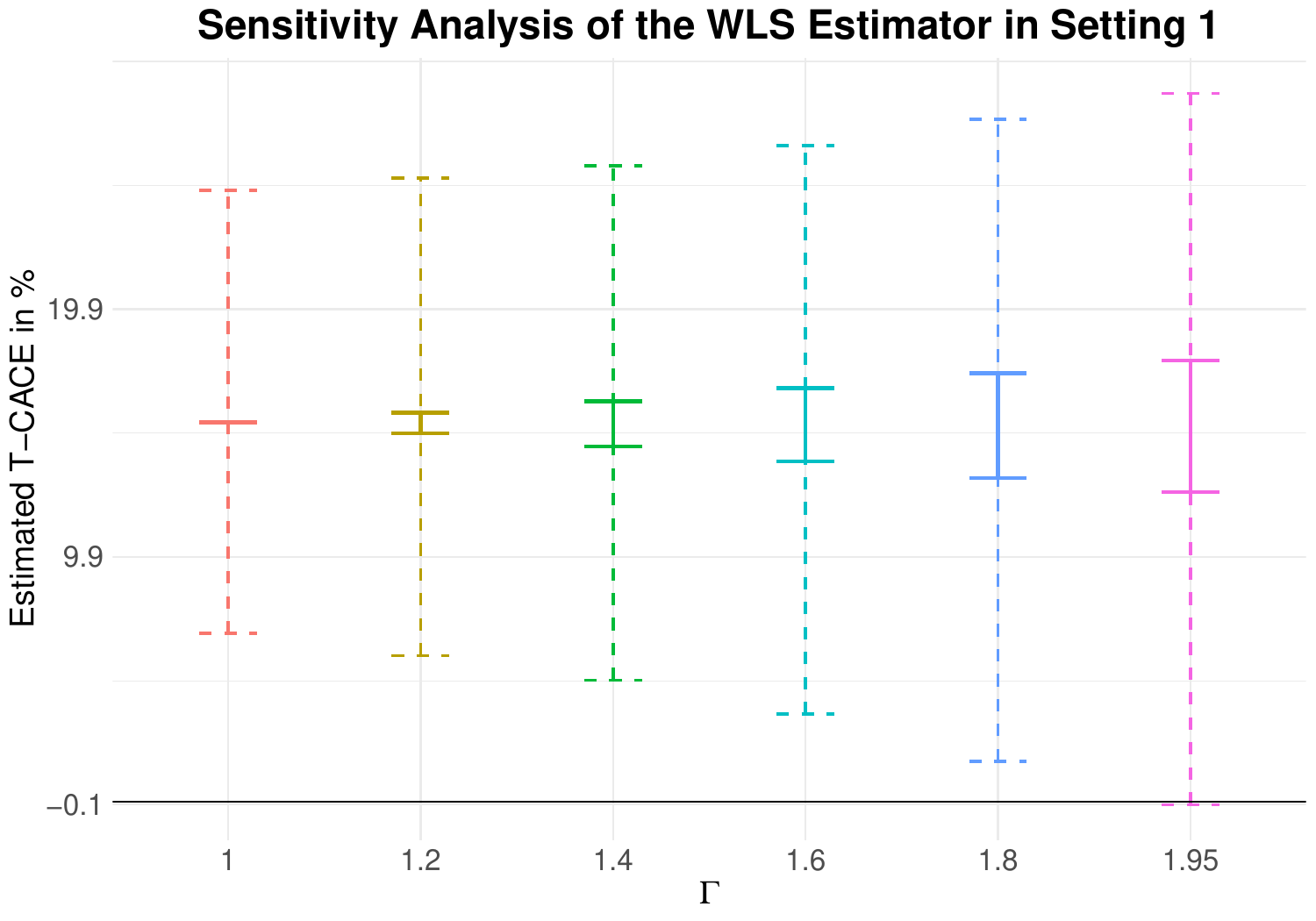}
\caption{Plot of the estimated T-CACE with respect to the sensitivity parameter $\gamma.$ The solid error bars are the range of point estimates of the WLS estimator, i.e., $\sbr{\min_{\tilde w \in \varepsilon(\Gamma)} \tau(\tilde w), \ \  \max_{\tilde w \in \varepsilon(\Gamma)} \tau(\tilde w)}.$ The dashed error bars are the 95\% percentile bootstrap confidence intervals. The black horizontal line indicates $y$-axis value 0.}
\label{fig: empirical whole sensitivity analysis}
\end{figure}

To evaluate the sensitivity of our results to the potential omission of a key moderator for participants who did not complete the follow-up survey, as defined in Setting 1, we apply our proposed sensitivity analysis. We begin with a benchmarking approach, comparing the weight ratio before and after omitting an observed covariate to estimate a plausible range for $\Gamma$. Figure \ref{fig: tilde gamma} displays the $\Gamma$ values for 6 selected covariates. Across all 33 covariates, the $\Gamma$ values fall within the range $[1.03, 1.96]$. This implies that $\Gamma$ is unlikely to substantially exceed 1.96 if we believe the omitted confounder does not have a much stronger moderating effect than the observed covariates. In Figure \ref{fig: empirical whole sensitivity analysis}, we vary $\Gamma \geq 1$ and observe that the percentile bootstrap sensitivity analysis confidence intervals cross zero when $\Gamma^* = 1.95.$ 

The highest $\Gamma$ obtained by omitting a single covariate is 1.96. This indicates that $\Gamma$ would reach a similar value as $\Gamma^*$ only if the unmeasured confounder $U$ induces bias comparable to the worst-case bias from omitting an observed covariate. We conclude that only an exceptionally strong moderator could result in the T-CACE estimate statistically insignificant.

\subsection{Additional tables} \label{app: empirical addtional tables}
We summarize the mean and standard deviation of the covariates available in both settings of the deep canvassing application in Table \ref{tab: summary_covariates}. Table \ref{table: exclusionary attitudes} presents the numerical values of different estimators as well as the estimated within-sample effects.
\begin{table}[ht] 
\centering 
\scriptsize
\begin{tabular}{lcccccc}   
\toprule 
Covariate & Sample Mean & Sample SD & Target 1 Mean & Target 1 SD & Target 2 Mean & Target 2 SD \\    
\midrule 
Imm Better Worse & 0.74 & 1.13 & 0.58 & 1.18 & — & — \\    
Imm Police & 0.12 & 1.57 & 0.22 & 1.55 & — & — \\    
Imm Driverslicense & 0.34 & 1.59 & 0.46 & 1.54 & — & — \\    
Imm Daca & 0.82 & 1.36 & 0.75 & 1.39 & — & — \\    
Imm Citizenship & 0.35 & 1.44 & 0.34 & 1.43 & — & — \\    
Imm Deportall & -0.26 & 1.50 & -0.14 & 1.47 & — & — \\    
Imm Attorney & 0.13 & 1.51 & 0.04 & 1.53 & — & — \\    
Imm Prej Living & 0.42 & 1.36 & 0.30 & 1.37 & — & — \\    
Imm Prej Neighbor & -0.96 & 1.22 & -0.86 & 1.24 & — & — \\    
Imm Prej Speaking & 1.13 & 1.23 & 1.05 & 1.25 & — & — \\    
Imm Prej Workethic & -1.18 & 1.00 & -1.10 & 1.04 & — & — \\    
Imm Prej Fit & -0.37 & 1.41 & -0.30 & 1.38 & — & — \\    
Imm Know & 0.19 & 0.39 & 0.17 & 0.38 & — & — \\    
Social Distance Immigrant & 3.05 & 1.88 & 3.23 & 1.91 & — & — \\    
Therm Illegal Immigrant & 48.55 & 28.58 & 46.96 & 28.29 & — & — \\    
Therm Legal Immigrant & 83.51 & 20.55 & 82.77 & 20.76 & — & — \\    
College Educ & 0.60 & 0.49 & 0.58 & 0.49 & — & — \\    
Asian & 0.06 & 0.24 & 0.08 & 0.27 & — & — \\    
Latino & 0.11 & 0.31 & 0.12 & 0.32 & — & — \\    
Black & 0.03 & 0.16 & 0.04 & 0.20 & — & — \\    
White & 0.79 & 0.41 & 0.74 & 0.44 & — & — \\    
Born In Us & 0.94 & 0.23 & 0.92 & 0.27 & — & — \\    
Factor Undoc Immigrant & 0.06 & 0.98 & -0.02 & 0.97 & — & — \\    
Factor Lgbt & 0.04 & 0.93 & -0.01 & 0.93 & — & — \\    
Factor Trump & 0.04 & 0.98 & -0.01 & 0.98 & — & — \\    
Age & 52.21 & 16.78 & 49.44 & 16.84 & 49.05 & 17.66 \\    
Voted 08 & 0.71 & 0.46 & 0.65 & 0.48 & 0.59 & 0.49 \\    
Voted 10 & 0.61 & 0.49 & 0.54 & 0.50 & 0.44 & 0.50 \\    
Voted 12 & 0.75 & 0.43 & 0.70 & 0.46 & 0.62 & 0.48 \\    
Voted 14 & 0.62 & 0.48 & 0.53 & 0.50 & 0.40 & 0.49 \\    
Voted 16 & 0.90 & 0.30 & 0.84 & 0.37 & 0.74 & 0.44 \\    
Female & 0.52 & 0.50 & 0.52 & 0.50 & 0.51 & 0.50 \\    
Site & 2.07 & 0.81 & 2.08 & 0.86 & 2.10 & 0.85 \\    
\bottomrule 
\end{tabular} 
\caption{Summary of the mean and standard deviation of the covariates available in the study sample, along with the corresponding statistics for the target populations. \textbf{Target 1} represents participants who did not complete the follow-up survey, while \textbf{Target 2} represents participants who did not participate in the experiment. For the precise definitions of each covariate, refer to the online appendix of \cite{kalla2020reducing}.}  
\label{tab: summary_covariates} 
\end{table}

\paragraph{Variance Estimator Adjusted for Household-Level Clustering}
In the deep canvassing application, treatment assignment is conducted at the household level. Therefore, although the estimating equations in Theorem~\ref{thm: cre logistic model asymptotics Hajek} and Theorem \ref{thm: cre logistic model asymptotics distribution wls} are written as sums of individual-level contributions, the empirical covariance matrix in the sandwich estimator is adjusted to allow for arbitrary dependence among observations within the same household. Let $h(i)\in\{1,\ldots,G\}$ denote the household containing unit $i$. For the weighted estimator in Theorem~\ref{thm: cre logistic model asymptotics Hajek}, let $\widehat{\Phi}_i=\Phi^{(i)}(\widehat{\theta},\widehat{\beta})$ denote the stacked estimating-function contribution for $(\theta,\beta)$ evaluated at the estimated parameters. We replace the individual-level matrix $\widehat D_{n+N}$ with the household-level analogue
\begin{align*}
    \widehat D^{\mathrm{hh}}_{n+N}
    =
    \frac{1}{n+N}
    \sum_{h=1}^{G}
    \left(
        \widehat U_h-\bar U
    \right)
    \left(
        \widehat U_h-\bar U
    \right)^{T},
    \qquad
    \widehat U_h
    :=
    \sum_{i:h(i)=h}
    \widehat{\Phi}_i,
    \qquad
    \bar U
    :=
    \frac{1}{G}\sum_{h=1}^{G}\widehat U_h .
\end{align*}
That is, we first sum the individual estimating-function contributions within each household, then center the household-level sums, and finally compute their cross product. The bread matrix $\widehat C_{n+N}$ is unchanged, so the household-adjusted sandwich estimator for the weighted estimator is
\begin{align*}
    \nabla g(\widehat\theta)^T
    \widehat C_{n+N}^{-1}
    \widehat D^{\mathrm{hh}}_{n+N}
    \widehat C_{n+N}^{-T}
    \nabla g(\widehat\theta).
\end{align*}
We make the same modification for the weighted least squares estimator in Theorem~\ref{thm: cre logistic model asymptotics distribution wls}. Let $\widehat{\Phi}^{\mathrm{wls}}_i$ denote the stacked WLS estimating-function contribution for $(\tau^Y_{\mathrm{wls}},\tau^D_{\mathrm{wls}},\gamma^Y,\gamma^D,\beta)$, evaluated at the corresponding estimates. The household-level meat matrix is
\begin{align*}
    \widehat F^{\mathrm{hh}}_{n+N}
    =
    \frac{1}{n+N}
    \sum_{h=1}^{G}
    \left(
        \widehat U^{\mathrm{wls}}_h-\bar U^{\mathrm{wls}}
    \right)
    \left(
        \widehat U^{\mathrm{wls}}_h-\bar U^{\mathrm{wls}}
    \right)^{T},
    \qquad
    \widehat U^{\mathrm{wls}}_h
    :=
    \sum_{i:h(i)=h}
    \widehat{\Phi}^{\mathrm{wls}}_i .
\end{align*}
The derivative matrix $\widehat E_{n+N}$ remains unchanged, and hence the WLS covariance estimator becomes
\begin{align*}
    \widehat \Sigma^{\mathrm{hh}}_{\mathrm{wls}}
    =
    \widehat E_{n+N}^{-1}
    \widehat F^{\mathrm{hh}}_{n+N}
    \widehat E_{n+N}^{-T}.
\end{align*}
For the WLS T-CACE estimator, we apply the delta method with the gradient of
$g_{\mathrm{wls}}(\theta)=\theta_1/\theta_2$, where
$\theta_1=\tau^Y_{\mathrm{wls}}$ and $\theta_2=\tau^D_{\mathrm{wls}}$. For the WLS T-ITT estimator, the estimand is simply the first component $\tau^Y_{\mathrm{wls}}$, so the corresponding gradient is the first basis vector. Thus its variance is the $(1,1)$ entry of $\widehat \Sigma^{\mathrm{hh}}_{\mathrm{wls}}$.

The consistency argument for the sandwich-type variance estimator follows the same logic as Theorem~\ref{thm: sandwich consistency}, with households replacing individuals as the independent sampling units. Under the regularity conditions used there, together with independent households and suitably bounded household sizes, the household-level sums $\widehat U_h$ satisfy a law of large numbers for the cluster-level covariance. Since the point estimators remain consistent and $\widehat C_{n+N}$ and $\widehat E_{n+N}$ are unchanged and consistently estimate their population analogues, the continuous mapping argument used in Theorem~\ref{thm: sandwich consistency} continues to apply after replacing the individual-level matrices by the household-level matrices. Finally, for the weighted estimator based on the principal score, we estimate uncertainty using a nonparametric household bootstrap: households are sampled with replacement, all observations belonging to sampled households are retained, and the nuisance models and the PS estimator are recomputed in each bootstrap sample.

\begin{table}
  \centering
  \begin{tabular}{lcccc}
    \toprule
     Estimator &Point Estimate & Standard Deviation & $95\%$ Confidence Interval\\
    \midrule
    Within-Sample \\
    S-ITT &8.8 &2.4 &[4.2, 13.5] \\ 
    S-CACE &13.2 &3.5 &[6.2, 20.1] \\ \midrule 
    \textbf{Setting 1:} \\
    T-ITT (WLS) &10.3 &4.0 &[2.4, 18.2] \\ 
    T-CACE \\
    $\quad$ Weighted &20.5 &10.2 &[0.4, 40.6] \\
    $\quad$ Weighted Least Squares &15.8 &5.5 &[4.9, 26.6] \\ 
    $\quad$ Weighted (PS) &12.4 &5.1 &[2.8, 23.3] \\\midrule 
    \textbf{Setting 2:} \\ 
     T-ITT (WLS) &10.8 &7.5 &[-3.8, 25.5] \\ 
     T-CACE \\ 
      $\quad$ Weighted &12.8 &10.9 &[-8.6, 34.1] \\ 
      $\quad$ Weighted Least Squares  &16.3 &11.3 &[-5.8, 38.3] \\
      $\quad$ Weighted (PS) &19.2 &8.0  &[3.5, 35.0] \\  
    \bottomrule
  \end{tabular} \vspace{2mm} 
  \caption{Estimated impact of deep canvassing on support for immigration-related policies. All values in this table are expressed as percentages.}
  \label{table: exclusionary attitudes}
\end{table}

\subsection{A Test for Instrument Validity}
\label{subsec:empirical_IV_test}
 We implement the t-test for instrument relevance (i.e., Assumption~\ref{asp: IV}-(c)) using $D_i = \alpha + \beta_z Z_i + \beta_x ^T X_i + \varepsilon_i$. We then implement the test for monotonicity and exclusion restriction (i.e., Assumption~\ref{asp: IV}-(a) and Assumption~\ref{asp: IV}-(b)) proposed by \citep{mourifie2017testing} available in the R package \texttt{ivcheck} \citep{coverdale2026ivcheck}. Because \texttt{ivcheck} only allows the conditioning of one covariate, we use a scalar propensity-score summary of the pretreatment covariates. Specifically, we estimate
\begin{align*}
    {e}_D(X) := \prob(D_i=1 \given Z_i=1, X_i, S_i=1)
\end{align*}
using a logistic regression of treatment received on the pretreatment covariates among units assigned to treatment, and use $\widehat{e}_D(X_i) : =\widehat \prob(D_i=1 \given Z_i=1, X_i, S_i=1)$ as the conditioning variable in the \texttt{ivcheck} implementation.

This choice is motivated by both validity and power considerations. For validity, if the testable inequalities implied by monotonicity and exclusion hold conditional on the full covariate vector \(X\), then they also hold conditional on any measurable scalar function of \(X\), including $\widehat{e}_D(X_i)$. Thus, replacing \(X\) by a one-dimensional summary does not create a false testable implication under the null. For power, the compliance propensity score is a natural summary because, under monotonicity, \eqref{eq:kitagawa_test_0} becomes
\begin{align*}
&\prob(Y\in B,D=1\mid Z=1,X=x,S=1) \\
&\quad =
\prob(Y(1)\in B,D(1)=1\mid Z=1,X=x,S=1) \\
&\quad =
\prob(Y(1)\in B,C\mid Z=1,X=x,S=1) \\
&\quad =
\prob(Y(1)\in B,C\mid X=x,S=1) \\
&\quad =
\prob(C\mid X=x,S=1)
\prob(Y(1)\in B\mid C,X=x,S=1).
\end{align*}
and \eqref{eq:kitagawa_test_1} similarly becomes
\begin{align*}
\prob(Y\in B,D=0\given Z=0,X=x, S=1) &- \prob(Y\in B,D=0\given Z=1,X=x, S=1)\\
&= \prob(C\given X=x, S=1)\prob(Y(0)\in B\given C,X=x, S=1).
\end{align*}
When monotonicity holds, $\prob(C\mid X=x, S=1)={e}_D(x)$. Hence strata with larger ${e}_D(X)$ contain more of the variation relevant for the test, whereas strata with ${e}_D(X)$ contribute little to the test statistic. Conditioning on $\widehat e_D(X)$ therefore keeps units with similar compliance-related signal together, while avoiding the curse of dimensionality that would arise from conditioning flexibly on the full covariate vector. 

Table \ref{tab:IV_test} presents the results for both the instrument relevance test and the test of \citet{mourifie2017testing}. The first-stage coefficient on $Z$ is large and statistically significant, and the first-stage $F$-statistic is well above the conventional rule-of-thumb threshold of 10, providing strong evidence of instrument relevance. The \citet{mourifie2017testing} test yields a statistic of 31 with a $p$-value of 0.55, so we fail to reject the sharp observable implications of monotonicity and exclusion at the 5\% level. Together with the t-test, we conclude that there is no evidence against the IV validity conditions (i.e. Assumption \ref{asp: IV generalizability}).

\begin{table}[htbp]
\centering
\caption{Instrument Relevance and Validity Tests}
\label{tab:IV_test}
\begin{tabular}{llcc}
\hline
Test & Quantity & Estimate/Statistic & $p$-value \\
\hline
Instrument Relevance & First-Stage Coefficient on $Z$ & 0.67 & $6.02 \times 10^{-165}$ \\
                     & Standard Error & 0.02 &  \\
                     & 95\% Confidence Interval & $[0.63, 0.71]$ &  \\
                     & $t$-Statistic & 33.1 &  \\
                     & First-Stage $F$-Statistic & 1096.4 & $6.02 \times 10^{-165}$ \\
\hline
\citet{mourifie2017testing} Test & Sample Size & 1079 &  \\
               & Test Statistic & 31 & 0.55 \\
               & Bootstrap Replications & 1000 &  \\
\hline
\end{tabular}
\begin{flushleft}
\footnotesize
\textit{Notes:} The first-stage regression is $D_i = \alpha + \beta_z Z_i + \beta_x ^T X_i + \varepsilon_i$.
\end{flushleft}
\end{table}

\end{document}